\documentclass[nofootinbib, aps, twocolumn, superscriptaddress, longbibliography]{revtex4-1}
\usepackage{amsmath}
\usepackage{amssymb}
\usepackage{graphicx}
\usepackage[usenames,dvipsnames]{xcolor}
\usepackage{tikz}
\usepackage{pgffor}
\usepackage{verbatim}
\usepackage{bbm}
\usepackage{float}
\usepackage{mathrsfs}
\usepackage{multibib}
\usepackage[colorlinks, breaklinks=true,linkcolor=red, citecolor=blue, linktocpage=true]{hyperref}
\usepackage{color,colortbl}

\newcommand{\modulo}[1]{\ ({\rm mod} \ #1)}
\newcommand{\lcm}{{\rm lcm}}
\definecolor{Gray}{gray}{0.85}
\definecolor{LightCyan}{rgb}{0.88,1,1}

\begin{document}

\title{Interacting fermionic symmetry-protected topological phases in two dimensions}

\author{Chenjie Wang}
\affiliation{Perimeter Institute for Theoretical Physics, Waterloo, Ontario N2L 2Y5, Canada}

\author{Chien-Hung Lin}
\affiliation{Department of Physics, University of Alberta, Edmonton, Alberta T6G 2E1, Canada}

\author{Zheng-Cheng Gu}
\affiliation{Department of Physics, The Chinese University of Hong Kong, Shatin, New Territories, Hong Kong}
\affiliation{Perimeter Institute for Theoretical Physics, Waterloo, Ontario N2L 2Y5, Canada}

\date{\today}

\begin{abstract}
We classify and construct models for two-dimensional (2D) interacting fermionic symmetry-protected topological (FSPT) phases with general finite Abelian unitary symmetry $G_f$.  To obtain the classification, we couple the FSPT system to a dynamical discrete gauge field with gauge group $G_f$ and study braiding statistics in the resulting gauge theory. Under reasonable assumptions, the braiding statistics data allows us to infer a potentially complete classification of 2D FSPT phases with Abelian symmetry.  The FSPT models that we construct are simple stacks of the following two kinds of existing models: (i) free-fermion models and (ii) models obtained through embedding of bosonic symmetry-protected topological (BSPT) phases. Interestingly, using these two kinds of models, we are able to realize almost all FSPT phases in our classification, except for one class. We argue that this exceptional class of FSPT phases can never be realized through models (i) and (ii), and therefore can be thought of as intrinsically interacting and intrinsically fermionic. The simplest example of this class is associated with $\mathbb Z_4^f\times\mathbb Z_4\times\mathbb Z_4$ symmetry. We show that all 2D FSPT phases with a finite Abelian symmetry of the form $ \mathbb Z_2^f\times G$ can be realized through the above models (i), or (ii), or a simple stack of them. Finally, we study the stability of BSPT phases when they are embedded into fermionic systems.
\end{abstract}

\maketitle


\section{Introduction}

Recently much attention has been attracted to the so-called symmetry-protected topological (SPT) phases \cite{gu09, pollmann10, fidkowski11,chen11a, chen11b, schuch11,chen13}. A gapped quantum many-body system  is said to belong to a nontrivial SPT phase if it satisfies the following conditions: First, the Hamiltonian is invariant under certain global symmetries, which are not spontaneously broken in the ground state. Second, the ground state is short-range entangled. That is, it can be continuously connected to a product state (for bosonic systems) or an atomic insulator (for fermionic systems) through a local unitary transformation.  Third, it is impossible to connect the ground state to a product state or an atomic insulator without closing the energy gap or breaking one of the symmetries. The product state and atomic insulator are said to be the trivial SPT phases. Two nontrivial SPT phases are said to be inequivalent if they cannot be smoothly connected without closing the energy gap or breaking one of the symmetries. Famous examples of nontrivial SPT phases include the 2D and 3D topological insulators, which are protected by the charge conservation symmetry and time-reversal symmetry \cite{hasan10,qi11}.

One of the main themes in the theoretical study of SPT phases is the classification of SPT phases in a given spatial dimension $d$ and for a given symmetry group $G$. So far, complete classification is only rigorously obtained for free fermion systems\cite{ff1,ff2} and 1D interacting bosonic and fermionic systems\cite{fidkowski11,chen11a,schuch11}. For higher-dimensional interacting systems, various classification methods\cite{chen13,kapustin14a,freed14,kitaevsre} have been proposed, but all under assumptions of some kind. Perhaps the most influential method so far is the group cohomology classification scheme for bosonic SPT (BSPT) phases, proposed by Chen, et al in Ref.~\onlinecite{chen13}. They systematically constructed a class of BSPT models, each labeled by an element of the cohomology group $H^{d+1}[G, U(1)]$. Under the assumption that these models exhaust all possible SPT phases, it is claimed that BSPT phases in $d$ spatial dimension with symmetry $G$ is classified by $H^{d+1}[G, U(1)]$.  It turns out that this classification works very well. In 2D and 3D, the only known example beyond this classification is an SPT phase of 3D time-reversal symmetric bosonic systems\cite{vishwanath13}.

While the group cohomology classification greatly advances our understanding of BSPT phases,  strongly interacting fermionic SPT (FSPT) phases in higher dimensions are much less understood.  One direction that has obtained fruitful results is the study on reduction of the free-fermion classification under the effect of strong interaction\cite{fidkowski10, gu14b, ryu12,yao13,qi13,wangc-science, fidkowski13,metlitski14,wangc14,you14,morimoto15,queiroz16,witten15}. However, these works miss those FSPT phases that can be realized {\it only} in interacting systems. The first attempt to classify interacting FSPT phases in general dimensions and with general symmetry was taken by Gu and Wen.\cite{gu-super} They follow a similar idea behind the group cohomology models and generalize these models to the so-called group super-cohomology models for FSPT phases. However, unlike its bosonic counterpart, the super-cohomology classification only gives rise to a subset of FSPT phases in 2D and 3D. Many known FSPT phases are beyond the super-cohomology classification. More recently, several other attempts have been made, and some of them obtain more complete classification for interacting FSPT phases.\cite{kapustin14,freed14, cheng15,lu12,lan15,lan16} For 2D  FSPT phases, Ref.~\onlinecite{cheng15} obtained a fairly complete classification, by studying topological properties of external symmetry defects. That work focuses on onsite unitary symmetry $G_f$ of the form $\mathbb Z_2^f\times G$, where $\mathbb Z_2^f$ is the fermion parity group. This classification is recently supported  by Refs.~\onlinecite{tarantino16} and \onlinecite{gaiotto16}, where commuting projector Hamiltonian models are constructed for each FSPT phase in the classification of Ref.~\onlinecite{cheng15}. For 3D FSPT phases, less is known for general symmetries (see some results in Refs.~\onlinecite{kapustin14,freed14}).

In this work, we go beyond the previous works and study the classification of 2D interacting FSPT phases with general finite Abelian symmetry in the form
\begin{equation}
G_f = \mathbb  Z_{N_0}^f \times \prod_{i=1}^{K} \mathbb Z_{N_i} \label{group}
\end{equation}
where $N_0=2m$ is an even positive integer, and $K, N_i$ are positive integers. We also assume that the symmetry is {\it onsite} (internal) and {\it unitary}. Here, the notation ``$\mathbb Z_{N_0}^f$'' is used to indicate that the fermion parity group $\mathbb Z_2^f$ is a subgroup of $\mathbb Z_{N_0}^f$.  Such a symmetry group $G_f$ goes beyond the previous studies, because $G_f$ can be a {\it nontrivial} $\mathbb Z_2^f$ extension of $G\equiv G_f/\mathbb Z_2^f$ while the previous works focus on the cases that $G_f$ is a {\it trivial} extension of $G$, i.e., $G_f$ is a direct product of $ \mathbb Z_2^f$ and $ G$.\footnote{When $m$ is odd,  $G_f$ in (\ref{group}) is still a trivial $\mathbb Z_2^f$ extension of $G$, due to the isomorphism $\mathbb Z_{2m}^f = \mathbb Z_2^f\times\mathbb Z_m$. When $m$ is even, no such isomorphism exists.} Several simple cases where $G_f$ is a nontrivial extension, such as $\mathbb Z_4^f$ or more generally $\mathbb Z_{2m}^f$, were considered in Ref.~\onlinecite{kapustin14, wangcj16,lan16}. Here, we provide a more systematic classification for general finite Abelian groups.

Besides classification, another important motivation of this work is as follows. Interacting FSPT phases can be divided into three qualitatively different kinds:
\begin{enumerate}
\item The first kind are those that admit a free-fermion realization, i.e., those that can be adiabatically connected to the free-fermion FSPT phases.

\item The second kind can be thought of as ``relatives'' of those BSPT phases that are protected by the symmetry $G=G_f/\mathbb Z_2^f$. These FSPT phases can be obtained by first putting fermions into strongly bound pairs, then letting the pairs form a BSPT state with symmetry $G$. We call this way of obtaining FSPT phases {\it BSPT embedding}, and call the corresponding phases {\it BSPT-embedded phases}.

\item The third kind are any phases other than the first kind, the second kind, and a simple stack of them. One might consider this kind as {\it intrinsically interacting} and {\it intrinsically fermionic}. These phases can only be realized in strongly interacting fermionic systems.
\end{enumerate}
Note that the first and second kinds of FSPT phases are not exclusive from one another. Some FSPT phases admit both a free-fermion realization and a realization through BSPT embedding.

Our motivation is to seek for the third kind of FSPT phases.  Reference \onlinecite{kapustin14} discusses the possibility of finding such phases in 6D and 7D fermionic systems. However, there is no confirmative realization of these FSPT phases. In this work, we look for the third kind of FSPT phases in two-dimensional fermionic systems with general finite Abelian symmetry (\ref{group}). Within our classification scheme, we find that the third kind of FSPT phases indeed can be supported by certain finite Abelian symmetry, with the simplest one being $\mathbb Z_4^f\times\mathbb Z_4\times\mathbb Z_4$ symmetry (see Sec.~\ref{sec:exception}).\footnote{Motivated by these 2D FSPT phases, we also find that one-dimensional fermionic systems with $\mathbb Z_4^f\times\mathbb Z_4$ symmetry can support similar FSPT phases of the third kind; see a discussion in Sec.~\ref{sec:1dfspt}.} We note that $\mathbb Z_4^f\times\mathbb Z_4\times\mathbb Z_4$ is a nontrivial $\mathbb Z_2^f$ extension of $\mathbb Z_2\times\mathbb Z_4\times\mathbb Z_4$. In fact, we are able to show that in two dimensions, all finite Abelian symmetry $G_f$ of the form $\mathbb Z_2^f\times G$ {\it cannot} support the third kind of FSPT phases.

\subsection{Main results}

\begin{table*}
\caption{Subgroups $A, B_i, C_{ij}, D_{ijk}$ of the stacking group $H_{\rm stack}$, Eq.~(\ref{stackgroup}), of two-dimensional FSPT phases with arbitrary finite Abelian unitary symmetry of the form (\ref{group}). The notation $N_{0i}$ denotes the greatest common divisor of $N_0$ and $N_i$, and the notation $N_{0ij}$ denotes the greatest common divisor of $N_0, N_i$ and $N_j$. The notation $N_{ij}$ and $N_{ijk}$ are similar. The number $N_0$ is even, and we use the convention $m=N_0/2$.  The last column lists the values of topological invariants (defined in Sec.~\ref{sec:invariants}) for the generating FSPT phases, which correspond to the generators of each cyclic subgroup.
}\label{tab1}
\begin{tabular}{llllcc}
\hline\hline
$\ $&   Cases  & $\quad$Group $\quad$ &  $\quad$Topological invariants of generating phases $\quad$
\\
\hline
$A \ $  & if $m$ is odd    & $\quad\mathbb Z_m$      &  $\quad\Theta_0=2\pi/m$\\
        & if $m$ is even   & $\quad\mathbb Z_{m/2}$  &  $\quad\Theta_0=4\pi/m$\\[1pt]
\hline $B_i \ $ & if $N_i$ is odd                                 & $\quad\mathbb Z_{N_i} \times\mathbb Z_{N_{0i}}$                       &  $\quad(\Theta_{i}, \Theta_{0i}, \Theta_{00i}) = \left(2\pi/N_i, 0, 0\right), \  \left( 0,  2\pi/N_{0i},0\right)$
  \\[1pt]
                & if $m, N_i$ are even                            & $\quad\mathbb Z_{2N_i} \times\mathbb Z_{N_{0i}/2}$                    &  $\quad (\Theta_{i}, \Theta_{0i}, \Theta_{00i})= \left(\pi/N_i, 2\pi/N_{0i},0\right), \  \left(0, 4\pi/N_{0i},0\right)
$  \\[1pt]
                & if $m$ is odd, $N_i =2 \ ({\rm mod}\ 4) $       & $\quad\mathbb Z_{4N_i} \times\mathbb Z_{N_{0i}/2}$                    & $\quad (\Theta_{i}, \Theta_{0i}, \Theta_{00i})= \left(\pi/2N_i, \pm\pi/N_{0i},\pi\right)\footnote{The ``$-$'' sign applies when $m = \frac{N_i}{2} \modulo{4}$, and the ``$+$'' sign applies when $m=\frac{N_i}{2}+2 \modulo{4}$.}, \  \left(0, 4\pi/N_{0i},0\right)$
  \\[1pt]
                & if $m$ is odd, $N_i =0 \ ({\rm mod}\ 4) $        & $\quad\mathbb Z_{2N_i} \times\mathbb Z_{N_{0i}}$    &  $\quad (\Theta_{i}, \Theta_{0i}, \Theta_{00i})= \left(\pi/N_i, 2\pi/N_{0i},0\right), \  \left(0, \lambda2\pi/N_{0i},\pi\right)\footnote{Here, $\lambda=1$ when $N_i = 4 \modulo{8}$, and $\lambda=2$ when $N_i =0 \modulo{8}$.} $\\[1pt]
\hline $C_{ij} \ $ & if $m$ is odd, $N_i,N_j =2 ({\rm mod \ }4) $   &  $\quad \mathbb Z_{2N_{ij}} \times \mathbb Z_{N_{0ij}/2}$   & $\quad(\Theta_{ij}, \Theta_{0ij})  = \left(\pi/N_{ij}, 2\pi/N_{0ij}\right), \ \left(0, 4\pi/N_{0ij}\right)$
   \\[1pt]
                   & otherwise                                     &  $\quad \mathbb Z_{N_{ij}} \times \mathbb Z_{N_{0ij}}$      &  $\quad(\Theta_{ij}, \Theta_{0ij})  = \left(2\pi/N_{ij},0 \right), \  \left(0, 2\pi/N_{0ij}\right)
$  \\[1pt]
\hline
$D_{ijk} \ $  & all cases   & $\quad\mathbb Z_{N_{ijk}}$  & $\quad\Theta_{ijk} = 2\pi/N_{ijk}$  \\[1pt]
\hline\hline
\end{tabular}
\end{table*}

As discussed above, the goal of this work is to classify 2D FSPT phases with general finite Abelian symmetry (\ref{group}) and to look for the third kind of FSPT phases which can only be realized in strongly interacting fermionic systems. Since the paper is long, we summarize the main results and general methodology here.

We obtain a (potentially complete) classification of 2D  FSPT phases with arbitrary finite Abelian symmetry in the form (\ref{group}). We assume that the symmetry is onsite (internal) and unitary. As is well known, FSPT phases form a group, where the group identity corresponds to the trivial phase and the group multiplication corresponds to stacking two FSPT phases.\cite{freed14} We refer to this group as the {\it stacking group} and denote it as $H_{\rm stack}$. For arbitrary finite Abelian unitary symmetry group $G_f$ in (\ref{group}), we show that $H_{\rm stack}$ has the following form
\begin{equation}
H_{\rm stack} = A \times \prod_i  B_i\times \prod_{i<j}  C_{ij}\times \prod_{i<j<k}  D_{ijk} \label{stackgroup}
\end{equation}
where the indices $i,j,k$ take values in $1,2,\dots K$, and $A, B_{i}, C_{ij}, D_{ijk}$ are finite Abelian groups given in Table \ref{tab1}. Every element of $H_{\rm stack}$ corresponds to an FSPT phase, the properties of which will be clear later. The stacking group $H_{\rm stack}$ for several small groups are listed in Table \ref{tab2}.



The approach that we use to obtain the classification was first proposed by Ref.~\onlinecite{levin12} and later developed in Refs.~\onlinecite{gu14b,wangcj15,threeloop}. We study FSPT systems by gauging the $G_f$ symmetry, i.e., by coupling the system to a lattice gauge field of gauge group $G_f$. Then, we study the braiding statistics in the resulting gauge theories. With a proper way of gauging the symmetry\cite{levin12,wangcj15}, the braiding statistics in the resulting gauge theory are guaranteed to be invariant under any smooth deformation of the original FSPT systems, as long as the deformation does not close the energy gap and does not break the symmetry. Hence, braiding statistics can be used to distinguish FSPT phases.

More specifically, we define a set of three tensors $\{\Theta_\mu, \Theta_{\mu\nu}, \Theta_{\mu\nu\lambda}\}$ using braiding statistics between the excitations in the gauged system, where the indices $\mu,\nu,\lambda$ take values in the range $0,1,\dots, K$ (see Sec.~\ref{sec:invariants} for definitions). We call these tensors {\it topological invariants}, following the terminology of Ref.~\onlinecite{wangcj15} where similar quantities are defined for BSPT phases. By studying their physical constraints and further solving the constraints, we obtain all possible values that the topological invariants can take. With this result, we make two crucial assumptions: (i) the set $\{\Theta_\mu, \Theta_{\mu\nu}, \Theta_{\mu\nu\lambda}\}$ is complete in the sense that they distinguish every FSPT phase with finite Abelian unitary symmetry and (ii) every possible value of $\{\Theta_\mu, \Theta_{\mu\nu}, \Theta_{\mu\nu\lambda}\}$ from the solutions of the constraints can be realized in a physical system. The two assumptions lead to a one-to-one correspondence between FSPT phases and values that the topological invariants can take. With this correspondence, we then obtain the classification in Eq.~(\ref{stackgroup}) and Table \ref{tab1} from the solutions of the constraints on the topological invariants.

Certainly, we need to justify the assumptions (i) and (ii). We cannot prove the completeness assumption (i), but  can show  some evidence. The most important evidence is that when $m$ is odd, our classification gives the same counting of FSPT phases as that of Refs.~\onlinecite{cheng15} (see Sec.~\ref{sec:completeness} for discussion). Also, our classification reproduces all known examples\cite{gu14b,cheng15,kapustin14}.  Another support of this assumption is perhaps that the bosonic cousins of the topological invariants, studied in Ref.~\onlinecite{wangcj15}, give an equivalent classification to the group cohomology classification\cite{chen13}.


At the same time, we almost succeed to lift the assumption (ii) by constructing  a physical model for every phase in our classification, with only one class of exceptional FSPT phases for which we are not able to construct models. The recipe of our model construction is simple. We obtain new FSPT phases by stacking two types of existing models: (1) the free-fermion models and (2) models that can be obtained from BSPT embedding. (Details of the two types of models are discussed in Sec.~\ref{sec:model_con}.) With this way of constructing models, we find all FSPT phases of the first and second kinds. The exceptional class of FSPT phases for which we are not able to construct models  are  the third kind of FSPT phases. The simplest symmetry group that the exceptional case occurs is $G_f =\mathbb Z_4^f\times\mathbb Z_4\times\mathbb Z_4$. We argue in Sec.~\ref{sec:exception} that the exceptional FSPT phases are indeed of the third kind and can only be realized in interacting fermionic systems. In passing, we also find that 1D fermionic systems with $\mathbb Z_4^f\times\mathbb Z_4$ symmetry can support similar FSPT phases of the third kind.p

Finally, as an aside, we study stability of BSPT phases when they are embedded into fermionic systems. BSPT phases may be unstable, in the sense that certain nontrivial BSPT phases become trivial after embedding. This issue is discussed in Sec.~\ref{sec:stability}.





\subsection{Organization of the paper}

The rest of the paper is organized as follows. We begin with a discussion on the role of the fermion parity as a symmetry of fermionic systems in Sec.~\ref{sec:symmetry}. In Sec.~\ref{sec:invariants}, we define the topological invariants $\Theta_\mu, \Theta_{\mu\nu}, \Theta_{\mu\nu\lambda}$ and study physical constraints on them. We solve the constraints and obtain a classification of FSPT phases in Sec.~\ref{sec:classification}. In particular, we discuss how the group $H_{\rm stack}$ can be read out from the solutions of constraints in Sec.~\ref{sec:stacking}. Then, we move on to construct models for FSPT phases within our classification in Sec.~\ref{sec:model_con}. Several examples of our models are discussed in detail in Sec.~\ref{sec:examples}. In Sec.~\ref{sec:exception}, we argue that there is no free-fermion or BSPT-embedding realization of the exceptional FSPT phases with $\mathbb Z_4^f\times\mathbb Z_4\times\mathbb Z_4$ symmetry, and show evidence of their existence. We discuss the stability/instability of BSPT phases when they are embedded into FSPT phases in Sec.~\ref{sec:stability}. We conclude in Sec.~\ref{sec:conclusion}. In the appendix \ref{appd_canonical_form}, we prove that general finite Abelian symmetry in fermionic systems can always be written in the form (\ref{group}). In Appendix \ref{app_proof}, we prove the constraints of topological invariants discussed in Sec.~\ref{sec:invariants}.

\begin{table}%
\caption{$H_{\rm stack}$ for several small Abelian groups $G_f$, obtained from the general results in Table \ref{tab1}.} \label{tab2}
\begin{tabular}
[c]{lll}\hline\hline
$G_{f}$ &   \ \quad\quad  & $H_{\rm stack}$\\\hline
$\mathbb{Z}_{2}^{f},\mathbb{Z}_{4}^{f}$ &  \  & $\mathbb Z_1$\\ 
$\mathbb Z_8^f$ & & $\mathbb Z_2$ \\
$\mathbb{Z}_{2}^{f}\times\mathbb{Z}_{2}$ &  \  & $\mathbb{Z}_{8}$\\
$\mathbb{Z}_{2}^{f}\times\mathbb{Z}_{4}$ &    & $\mathbb{Z}_{8}\times
\mathbb{Z}_{2}$\\ 
$\mathbb{Z}_{4}^{f}\times\mathbb{Z}_{2}$ &  \  & $\mathbb{Z}_{4}$\\ 
$\mathbb Z_4^f\times\mathbb Z_4 $ & & $\mathbb Z_8\times\mathbb Z_2$
\\
$\mathbb{Z}_{2}^{f}\times\mathbb{Z}_{2}\times\mathbb{Z}_{2}$ &  \  & $\left(
\mathbb{Z}_{8}\right)  ^{2}\times\mathbb{Z}_{4}$\\ 
$\mathbb Z_2^f\times\mathbb Z_2\times\mathbb Z_4$ & & $\left(\mathbb Z_8\right)^2\times\left(\mathbb Z_2\right)^3$ \\
$\mathbb{Z}_{4}^{f}\times\mathbb{Z}_{4}\times\mathbb{Z}_{4}$ &    & $\left(
\mathbb{Z}_{8}\right)^2 \times\left(\mathbb Z_4\right)^2 \times \left(\mathbb{Z}_{2}\right)  ^{2}$\\
$\mathbb{Z}_{2}^{f}\times\mathbb{Z}_{2}\times\mathbb{Z}_{2}\times
\mathbb{Z}_{2}$ &  \  & $\left(  \mathbb{Z}_{8}\right)  ^{3}\times\left(
\mathbb{Z}_{4}\right)  ^{3}\times\mathbb{Z}_{2}$\\
\hline\hline
\end{tabular}
\end{table}


\section{Symmetries in fermionic systems}
\label{sec:symmetry}

We begin with a discussion on symmetries in fermionic systems. To be specific, we consider fermionic systems defined on a lattice. Unlike bosonic systems,  fermionic systems must respect a special symmetry, the fermion parity $P_f = (-1)^F$, where $F$ is the total fermion number. That is, the Hamiltonian of a fermionic system can be written as a sum of terms, each of which must be a product of an {\it even} number of fermion creation or annihilation operators. A term with an odd number of fermion creation or annihilation operators violates the locality principle. The fermion parity $P_f$ is unitary and Hermitian, and it squares to the identity operator, i.e., $P_f^2=1$.


That being said, to specify the full symmetry of a fermionic system, one needs two pieces of information: (i) a symmetry group $G_f$ that is formed by all symmetry operators; and (ii) a special group element in $G_f$, which corresponds to the fermion parity $P_f$.  Since $P_f$ squares to 1, the order of the fermion-parity element is 2. In general, a symmetry operator respects the fermion parity as well. Hence, the fermion-parity element should be central in $G_f$. Accordingly, the identity and fermion-parity element form a normal subgroup of $G_f$, which is usually denoted as $\mathbb Z_2^f$.

In this work, we study 2D fermionic systems on a lattice with general finite Abelian unitary symmetry
\begin{equation}
G_f = \mathbb Z_{N_0}^f \times \prod_{i=1}^{K} \mathbb Z_{N_i} \label{group1}
\end{equation}
where $N_0\equiv 2m$ is a positive even integer, and $K,N_i$ are positive integers. A group element $a \in G_f$ can be labeled by an integer vector
\begin{equation}
a = (a_0, a_1, \dots, a_K)
\end{equation}
where $a_\mu $ takes values in the range $0, 1, \dots, N_\mu-1$, for $\mu=0, 1,\dots, K$. We will use the ``additive'' notation for group multiplication. The components of $a+b$ are given by $(a+b)_\mu = a_\mu + b_\mu \ ({\rm mod} \ N_\mu)$.

The notation ``$\mathbb Z_{N_0}^f$'' in Eq.~(\ref{group1}) is used to indicate our choice of the fermion-parity group element: we choose $(m, 0, \dots, 0)$ to be the fermion-parity element. In general,  the fermion parity may correspond to any order-of-2 element in the symmetry group. Nevertheless, one can show that any finite Abelian group with a given fermion-parity element is isomorphic to a group in the form (\ref{group1}) with the fermion parity being $(m,0,\dots,0)$ (see Appendix \ref{appd_canonical_form} for a proof). Hence, $G_f$ in Eq.~(\ref{group1}) can be thought of as a {\it canonical form} of the most general finite Abelian symmetry in fermionic systems.

It is worth to point out that two groups with the same group structure, but with different assignments of the fermion-parity element, may represent different symmetries for fermionic systems. For example,  $\mathbb Z_2^f\times \mathbb Z_4$ and $\mathbb Z_4^f \times \mathbb Z_2$ have the same group structure, but the assignments of the fermion-parity element are different and inequivalent. So, they should be considered as different symmetries in fermionic systems.

Finally, a note on our convention: Throughout the paper,  Greek indices $\mu,\nu,\dots$ take values in the range $0,1,\dots, K$, while  Roman indices $i,j,\dots$ take values in the range $1,2,\dots, K$.


\section{Topological invariants}
\label{sec:invariants}


In this section, we define a set of {\it topological invariants} for FSPT phases with symmetry group $G_f$ in the form (\ref{group1}). This set consists of three tensors $\Theta_\mu,\Theta_{\mu\nu}$ and $\Theta_{\mu\nu\lambda}$. Besides the definitions of topological invariants, we also study physical constraints on them. As indicated by the name, the topological invariants are defined in a way that they are constant under any smooth deformation that does not close the energy gap and that does not break the symmetry of the system. Hence, they can be thought of as ``order parameters'' that characterize FSPT phases. These topological invariants are very close to (but not exactly the same as) those defined in Ref.~\onlinecite{wangcj15} for BSPT systems.

\subsection{Gauge theories coupled to fermionic matter}
\label{sec:gauge}

To define the topological invariants, the first step is to gauge the symmetry. That is, for a given FSPT system with symmetry $G_f$, we minimally couple it to a lattice gauge field of gauge group $G_f$ (i.e., we gauge the full symmetry, including the fermion parity). The detailed gauging procedure is not important for our purposes, but we require that the symmetry is gauged in a way such that the resulting gauge theory is gapped and deconfined. One may consult Refs.~\onlinecite{levin12, wangcj15} for a particular gauging procedure where the coupling constant is set to exactly 0. References \onlinecite{levin12, wangcj15} are devoted to bosonic systems, however the gauging procedure there can be easily adapted to fermionic systems.

Why do we gauge the symmetry of FSPT systems? The reason is that after gauging, excitations in the resulting gauge theory exhibit nontrivial braiding statistics. The braiding statistics are the same for two systems that belong to the same FSPT phase: the two systems can be smoothly deformed to each other without closing the energy gap and without breaking the symmetry, thereby we can gauge the whole family of systems along the deformation path. Accordingly, there exists a smooth path connecting the two gauged systems, which leads to the same braiding statistics (known as Ocneanu rigidity\cite{kitaev06}). Therefore, if two FSPT systems have different braiding statistics after gauging the symmetry, they must belong to distinct phases. Nevertheless, it is not obvious that two distinct FSPT phases must lead to distinct braiding statistics after gauging. However, previous studies\cite{levin12, gu14b,wangcj15,threeloop} suggest that the latter statement is also true. In this paper, we will assume that braiding statistics have enough resolution to distinguish all FSPT phases. 

We now study braiding statistics between excitations in the resulting gauge theory. Excitations in the gauge theory can be divided into {\it charges} and {\it vortices}.  Charges carry gauge charge. They can be labeled by
\begin{equation}
q = (q_0, q_1, \dots, q_K)
\end{equation}
where each component $q_\mu$ takes values in the range $0, 1,\dots, N_\mu-1$. Vortices carry gauge flux. The gauge flux of a vortex $\alpha$ can  be labeled by a vector
\begin{equation}
\phi_\alpha = (\phi_{\alpha,0}, \phi_{\alpha,1}, \dots, \phi_{\alpha,K})
\end{equation}
where each component $\phi_{\alpha,\mu}$ is a multiple of $2\pi/N_\mu$. Unlike charges, vortices are not uniquely labeled by their gauge flux. Two vortices that carry the same gauge flux can differ by attaching some charge.

There is a one-to-one correspondence between gauge flux and group elements of $G_f$\footnote{If the gauge group is non-Abelian, the correspondence is between gauge flux and conjugacy classes.}. Since the fermion-parity group element is special in $G_f$, we would like to single out the corresponding gauge flux, i.e., the {\it fermion-parity flux}. For $G_f$ in the form (\ref{group1}),  the fermion-parity flux $\Pi$ is given by
\begin{equation}
\Pi = (\pi, 0, \dots, 0)
\end{equation}
In general, there are many vortices that carry fermion-parity flux $\Pi$.

Next,  we discuss braiding statistics between the excitations. In general, we can imagine three kinds of braiding processes: braiding between two charges, braiding between a charge and a vortex, and braiding between two vortices.  The statistical phase $\theta_{q\alpha}$ associated with braiding a charge $q$ around a vortex $\alpha$ should follow the Aharonov-Bohm law:
\begin{equation}
\theta_{q\alpha} = q\cdot\phi_\alpha \label{abphase}
\end{equation}
where ``$\cdot$'' is the vector inner product. Mutual statistics between two charges should be trivial, because charge excitations corresponds to local excitations from the original FSPT system. Nevertheless, the exchange statistics of a charge $q$ may not be trivial: $q$ can either be a boson or a fermion. More specifically, $q$ is a fermion if it carries odd fermion parity; $q$ is a boson if it carries even fermion parity. The fermion parity carried by $q$ can be read out from the Aharonov-Bohm statistics between $q$ and a vortex carrying the fermion-parity flux $\Pi$. Therefore, the exchange statistics $\theta_q$ is given by
\begin{equation}
\theta_q = q\cdot\Pi = \pi q_0 \label{exchange}
\end{equation}
where $q_0$ is the zeroth component of $q$. The statistics between two vortices may be very complicated, and in general can be non-Abelian. Unlike the charge-charge and charge-vortex statistics which are completely determined by the gauge group,  vortex-vortex statistics varies in different FSPT systems. Accordingly, vortex-vortex statistics contains information of the nature of the underlying FSPT phase. Vortex-vortex statistics is the key to characterize FSPT phases.

It is worth to point out that Eqs. (\ref{abphase}) and (\ref{exchange}) can be considered as the defining properties of our system, $G_f$ gauge theory coupled to fermionic matter: Eq.~(\ref{abphase}) implies that the gauge group if $G_f$, and Eq.~(\ref{exchange}) implies that the matter is fermionic.

Finally, we make a comment. When we compare the braiding statistics in two gauge theories, we need to match two properties: (1)
the algebraic structure associated with the braiding statistics, such as fusion rules, $F$ and $R$ symboles, etc and (2) the gauge flux of excitations. We say that the two theories have the same braiding statistics, only if there exists a one-to-one correspondence between the quasiparticle
excitations such that both properties are matched.

\subsection{Defining the topological invariants}
\label{inv_def}

\begin{figure}
\includegraphics{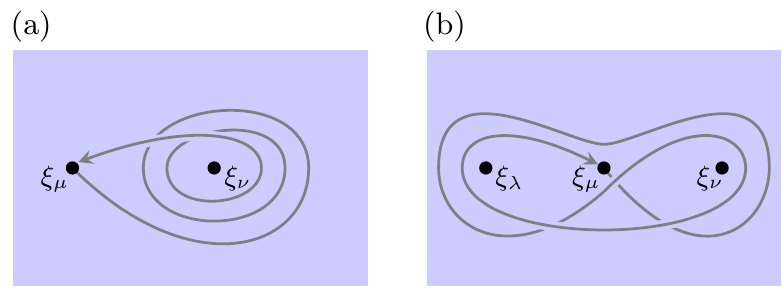}
\caption{Trajectories of $\xi_\mu$ in the braiding processes associated with the topological invariants $\Theta_{\mu\nu}$ (a) and $\Theta_{\mu\nu\lambda}$ (b). }\label{fig1}
\end{figure}

The full set of braiding statistics data is usually complicated in the gauged FSPT systems. In this section, we define a subset of the braiding statistics data, which we call {\it topological invariants}. This set consists of three tensor $\{\Theta_\mu, \Theta_{\mu\nu}, \Theta_{\mu\nu\lambda}\}$, which are defined in terms of braiding statistics between vortices in the gauged FSPT system. One will see that this set captures the essence of the full braiding statistics data.

Let $\xi_\mu$ be a vortex that carries the type-$\mu$ unit flux $\frac{2\pi}{N_\mu}e_\mu$, where $e_\mu = (0, \dots, 1,\dots,0)$ with the $\mu$th entry being 1 and other entries being 0. The topological invariants $\{\Theta_\mu, \Theta_{\mu\nu}, \Theta_{\mu\nu\lambda}\}$  are defined as follows:
\begin{enumerate}
\item $\Theta_0 = 2m \theta_{\xi_0}$, if $m$ is even;\\
$\Theta_0 = m \theta_{\xi_0}$, if $m$ is odd;\\
$\Theta_i = N_i \theta_{\xi_i}$, if $N_i$ is even;\\
$\Theta_i = 2N_i\theta_{\xi_i}$, if $N_i$ is odd, where $i=1, \dots, K$. 

\item $\Theta_{\mu\nu}$ is the Berry phase associated with braiding $\xi_\mu$ around $\xi_\nu$ for $N^{\mu\nu}$ times.

\item $\Theta_{\mu\nu\lambda}$ is the Berry phase associated with the following process: $\xi_\mu$ is first braided around $\xi_\nu$, then around $\xi_\lambda$, then around $\xi_\nu$ in the opposite direction, and finally around $\xi_\lambda$ in the opposite direction.

\end{enumerate}
The braiding processes associated with $\Theta_{\mu\nu}$ and $\Theta_{\mu\nu\lambda}$ are shown in Fig.\ref{fig1}. We have used $N^{\mu\nu}$ to denote the least common multiple of $N_\mu$ and $N_\nu$, for $\mu,\nu=0,1,\dots,K$.  In the definition of $\Theta_\mu$,  the quantity $\theta_{\xi_\mu}$ is the topological spin of the vortex $\xi_\mu$. Usually, $s_\alpha \equiv \theta_\alpha/2\pi$ is denoted as the topological spin of an anyon $\alpha$. In this paper, we use $\theta_\alpha$ as the topological spin instead.

These topological invariants are defined in a very similar way to Ref.~\onlinecite{wangcj15} for BSPT phases. In particular, the invariants $\Theta_{\mu\nu}, \Theta_{\mu\nu\lambda}$ are defined exactly the same as those in Ref.~\onlinecite{wangcj15}. Depending on the parity of $N_\mu$, the definition of $\Theta_\mu$ may differ from its bosonic counterpart by a factor of 2. Such difference is reasonable, since one expects that distinction between BSPT and FSPT shall come from some properties related to exchange statistics/topological spins.

For the above topological invariants to be well defined, we need to show two points: (i) We need to show that the Berry phases associated with the above braiding processes are always Abelian, regardless of the fact that the vortices may be non-Abelian; (ii) These Abelian phases are functions of $\mu,\nu,\lambda$ only and do not depend on the choice of $\xi_\mu,\xi_\nu,\xi_\lambda$ as long as they carry the type-$\mu$, type-$\nu$, and type-$\lambda$ unit flux respectively.  %

The proofs of (i) and (ii) are not particularly relevant to the rest of the paper. Here we only briefly discuss them. For the invariants $\Theta_{\mu\nu}$ and $\Theta_{\mu\nu\lambda}$, points (i) and (ii) can be proven in the same way as in Ref.~\onlinecite{wangcj15} for BSPT phases with no modifications. The fact that we are dealing with fermionic matter does not affect the proofs. For $\Theta_\mu$, point (i) is automatic and we only need to show point (ii). In fact, in defining $\Theta_\mu$, we have chosen proper multiples of $\theta_{\xi_\mu}$ so that point (ii) is satisfied. The proof is again similar to that of Ref.~\onlinecite{wangcj15}, but not exactly the same. Below we show point (ii) for $\Theta_0$; the same argument works well for $\Theta_i$.

Suppose we replace $\xi_0$ by another vortex $\xi_0'$ in the definition of $\Theta_0$, where $\xi_0'$ also carries type-$0$ unit flux. We need to show that $2m\theta_{\xi_0} = 2m \theta_{\xi_0'}$ for even $m$, and $m\theta_{\xi_0} = m \theta_{\xi_0'}$ for odd $m$. To show that, we recall that $\xi_0'$ can at most differ from $\xi_0$ by a charge, i.e., $\xi_0' =\xi_0 \times q$ for some charge $q$. With this, we use the following  relation from the general Algebraic theory of anyons\cite{kitaev06}
\begin{equation}
R_{\beta\alpha}^\gamma R_{\alpha\beta}^\gamma = e^{i\left(\theta_\gamma - \theta_\alpha - \theta_\beta\right)} {\rm id}_{\mathbb V_{\alpha\beta}^\gamma} \label{spin_formula}
\end{equation}
where $\mathbb V_{\alpha\beta}^\gamma$ is the fusion space of $\alpha$ and $\beta$ in the fusion channel $\gamma$, $R_{\alpha\beta}^\gamma$ is the braid matrix associated with a half-braiding of $\alpha$ and $\beta$ in the fusion channel $\gamma$, and ${\rm id}_{\mathbb V_{\alpha\beta}^\gamma}$ is the identity matrix in $\mathbb V_{\alpha\beta}^\gamma$. Making the substitutions $\alpha \rightarrow \xi_0$, $\beta\rightarrow q$ and $\gamma \rightarrow \xi_0'$, we immediately find that
\begin{equation}
e^{i\pi q_0/m} = e^{i\theta_{\xi_0'} - i\theta_{\xi_0} - i \pi q_0} \label{eq-spin}
\end{equation}
where we have used the facts that the mutual statistics $R_{q \xi_0}^{\xi_0'} R_{\xi_0 q}^{\xi_0'}$ between $q$ and $\xi_0$ is given by the Aharonov-Bohm law (\ref{abphase}), and that the topological spin $\theta_q$ is given by Eq.~(\ref{exchange}). With Eq.~(\ref{eq-spin}), we immediately obtain that $2m\theta_{\xi_0} = 2m\theta_{\xi_0'}$. In particular, if $m$ is odd, we achieve a stronger relation, $m\theta_{\xi_0} = m \theta_{\xi_0'}$. Therefore, we prove point (ii) for $\Theta_0$.


\subsection{Constraints on topological invariants}

The topological invariants $\Theta_\mu, \Theta_{\mu\nu}, \Theta_{\mu\nu\lambda}$ cannot be arbitrarily valued. They satisfy many physical constraints. In this subsection, we study constraints on the topological invariants.

The constraints that the topological invariants satisfy are:
\begin{align}
\Theta_{\mu\nu\lambda} & = {\rm sgn}(\hat p) \Theta_{\hat p(\mu) \hat p(\nu) \hat p(\lambda)} \label{c1}\\
N_{\mu\nu\lambda} \Theta_{\mu\nu\lambda} &=0 \label{c2} \\
\Theta_{\mu\mu\nu} & = \Theta_{\nu\nu\mu}  = m \Theta_{0\mu\nu}  \label{c3}\\
\Theta_{\mu\nu}  & = \Theta_{\nu\mu} \label{c4}\\
N_{\mu\nu} \Theta_{\mu\nu} & =\mathcal F(N^{\mu\nu})\Theta_{\mu\mu\nu} \label{c5}\\
\frac{N_i}{2}\Theta_{ii} & = \frac{N_{0i}}{2}\Theta_{0i} + \left[\frac{N_i}{2}\mathcal F(m)+m\mathcal F\left(\frac{N_i}{2}\right)\right]\Theta_{00i}, \nonumber\\
& \quad\quad\text{(only for $N_i$ being even)}\label{c6}\\
\Theta_{ii} & = \left\{\begin{array}{ll}
2\Theta_i + \mathcal F(N_i) \Theta_{iii}, & \text{if $N_i$ is even} \\
\Theta_i, & \text{if $N_i$ is odd}
\end{array}
\right. \label{c7}\\
\Theta_{00} & = \left\{\begin{array}{ll}
2\Theta_0, & \text{if $m$ is even}\\
4\Theta_0 + \Theta_{000}, & \text{if $m$ is odd}
\end{array}
\right. \label{c8}
\end{align}
where all equations are defined modulo $2\pi$, the Greek indices $\mu,\nu,\lambda$ take values in the range $0, 1,\dots, K$, and the Roman indices $i,j,k$ take values in the range $1,2,\dots, K$. The symbol $\hat p$ denotes a permutation on the indices $\mu,\nu,\lambda$, and its signature ${\rm sgn}(\hat p)=\pm1$. The function $\mathcal F(n) = n(n-1)/2$, where $n$ is an integer. We have used $N_{\mu\dots\lambda}$ to denote the greatest common divisor of $N_\mu, \dots, N_\lambda$, and have used $N^{\mu\dots\lambda}$ to denote the least common multiple of $N_\mu,\dots,N_\lambda$. We keep this notation throughout the paper.

Some of the above constraints follow from simple properties of braiding statistics, e.g. (\ref{c4}) is a direct consequence of the fact that braiding is symmetric, in the sense that braiding $\alpha$ around $\beta$ is topologically equivalent to braiding $\beta$ around $\alpha$. Nevertheless, to prove most of the constraints, we need to use the general algebraic theory of anyons, outlined in Ref.~\onlinecite{kitaev06}. The constraints follow from various consistency conditions on the braiding statistics. Since the proofs are technical, we have moved them to Appendix \ref{app_proof}.

Several comments are in order. First, the constraint (\ref{c6})  is only for even $N_i$. When $N_i$ is odd, there is no corresponding constraint.   Second, many constraints, such as Eqs.~(\ref{c1}), (\ref{c2}), (\ref{c4}) and (\ref{c5}),  are the same as their BSPT counterparts\cite{wangcj15}. However, others such as Eqs.~(\ref{c3}), (\ref{c6})-(\ref{c8}) are different from their BSPT counterparts, or even do not have BSPT counterparts. Third, the index 0 is special. This speciality originates the special role of fermion-parity flux, which is $m$ times of the type-0 unit flux.

With these constraints, one important question is that: Can every solution to the constraints be realized in physical systems? The answer is affirmative. We discuss this question in detail in  Sec.~\ref{sec:model_con} and Sec.~\ref{sec:exception}. An affirmative answer implies that the solutions of the constraints can provide a (minimal) classification for FSPT phases, which we discuss in Sec.~\ref{sec:classification}.




\subsection{Additional constraints from vanishing chiral central charge $c$}

By definition, FSPT phases are always nonchiral, i.e., the chiral central charge $c$ associated with the edge modes of an FSPT phase always vanishes. The constraints (\ref{c1})-(\ref{c8}) do not include the requirement of vanishing chiral central charge. In general, a solution to Eqs.~(\ref{c1})-(\ref{c8}) may correspond to a gapped fermionic system whose edge is chiral (these phases are sometimes called {\it invertible topological phases}\cite{freed14}). In this subsection, we discuss additional constraints on  $\Theta_{\mu}, \Theta_{\mu\nu}, \Theta_{\mu\nu\lambda}$, imposed by the requirement that  $c$ vanishes for FSPT phases.

To establish the additional constraints, we first establish the following relation
\begin{equation}
e^{i2\pi c/8} =  e^{i\theta_{\alpha_0}}, \label{eqc}
\end{equation}
where $\alpha_0$ is a vortex that carries the fermion-parity flux, i.e., $\phi_{\alpha_0} = (\pi,0,0,0,\dots,0)$. To establish this relation, we first notice that it holds for the case that $G_f = \mathbb Z_2^f$. This is a result of Ref.~\onlinecite{kitaev06}. For example, $p_x+ip_y$ superconductors have $c=1/2$ and the vortex carrying fermion-parity flux has a topological spin $\pi/8$. Next, we consider a general symmetry $G_f$. We imagine breaking the symmetry down to $\mathbb Z_2^f$ only by adding a weak perturbation to the FSPT system. We require the perturbation to be weak enough so that the energy gap does not close. Since the energy gap does not close, we obtain that: (1) the chiral central charge $c$ does not change and (2) after gauging the remaining $\mathbb Z_2^f$ symmetry, the topological spin of a vortex that carries the fermion-parity flux does not change. Note that by breaking $G_f$ to $\mathbb Z_2^f$, most gauge flux in the original $G_f$ gauge theory is killed, but the fermion-parity flux survives, which makes it possible to compare the topological spins of vortices carrying the fermion-parity flux before and after breaking $G_f$.  Since Eq.~(\ref{eqc}) holds in the $\mathbb Z_2^f$-only system, it follows immediately that  Eq.~(\ref{eqc}) also holds in the original $G_f$ gauge theory. Hence, we prove the relation (\ref{eqc}). We note that this relation should hold for any finite symmetry $G_f$, beyond Abelian symmetries.


Next, we express $\theta_{\alpha_0}$ in terms of the topological invariants. We show that
\begin{align}
\theta_{\alpha_0} = \left\{
\begin{array}{ll}
\vspace{5pt}
m\Theta_0 +\frac{m^2-1}{8} \Theta_{000}, & \text{if $m$ is odd,} \\
\frac{m}{2} \Theta_0, & \text{if $m$ is even}
\end{array}
\right. \label{eqc2}
\end{align}
Note that $(m^2-1)/8$ is an integer when $m$ is odd, and $m/2$ is an integer when $m$ is even, making the above equation well defined even if $\Theta_{000}$ and $\Theta_0$ are defined modulo $2\pi$. To prove Eq.~(\ref{eqc2}), we again use a ``symmetry-breaking'' approach. We first notice that Eq.~(\ref{eqc2}) holds for $G_f = \mathbb Z_{2m}^f$. This follows the results of Ref.~\onlinecite{wangcj16}, where braiding statistics in $\mathbb Z_{2m}^f$ gauge theories coupled to fermionic matter were studied.  Next, we consider general Abelian symmetry $G_f$ in the form (\ref{group1}). We imagine breaking $G_f$ down to $\mathbb Z_{2m}^f$ by adding a weak perturbation, which does not close the energy gap. Since the $\mathbb Z_{2m}^f$ symmetry remains, the type-$0$ unit flux survives in the symmetry-broken phase. Since the energy gap does not close, the values of the topological invariants $\Theta_{0}$ and $\Theta_{000}$, as well as the topological spin $\theta_{\alpha_0}$, do not change. Combining all together, we find that Eq.~(\ref{eqc2}) holds for general finite Abelian symmetry $G_f$.

Combining Eqs.~(\ref{eqc}) and (\ref{eqc2}), we find that the requirement of $c=0$ for FSPT phases imposes the following constraints on the topological invariants:
\begin{equation}
m\Theta_0 +\frac{m^2-1}{8} \Theta_{000} =0, \quad \text{if $m$ is odd} \label{ac1}
\end{equation}
and
\begin{equation}
\frac{m}{2} \Theta_0 =0, \quad \text{if $m$ is even} \label{ac2}
\end{equation}

One may notice that even if the constraints (\ref{ac1}) and (\ref{ac2}) are satisfied, it only guarantees that $c$ is a multiple of 8. This is because $\theta_{\alpha_0}$ can only determine $c$ modulo 8 from Eq.~(\ref{eqc}). This uncertainty is compensated by the following interesting fact: there exists a state with $c$ being $8$ but all topological invariants $\Theta_{\mu}, \Theta_{\mu\nu}, \Theta_{\mu\nu\lambda}$ vanish, which is usually referred to as the $E_8$ state.\cite{e8,lu12} Therefore, if the topological invariants of an FSPT phase satisfy (\ref{ac1}) and (\ref{ac2}),  it is always possible to turn it to a state with $c=0$, without changing the value of topological invariants, by stacking multiple copies of $E_8$ state or its time reversal (The quantities $c,\Theta_\mu,\Theta_{\mu\nu},\Theta_{\mu\nu\lambda}$ are additive under stacking; see Sec.~\ref{sec:stacking} for stacking of FSPT phases.)

\section{Classification of FSPT phases}

\label{sec:classification}


One way to classify topological phases is that: (i) find a {\it complete set} of topological invariants, such that this set distinguishes {\it every} phase under consideration; (ii) find all possible values that the topological invariants can take. It follows from the completeness that there is a one-to-one correspondence between topological phases and values of the topological invariants. Accordingly, the classification of topological phases can be inferred from the topological invariants.  However, in general, it is hard to judge if a given set of topological invariants is complete or not. For 2D FSPT phases with unitary Abelian symmetry, we have defined a set of topological invariants $\{\Theta_\mu, \Theta_{\mu}, \Theta_{\mu\nu\lambda}\}$ in Sec.~\ref{sec:invariants}, but we are not able to prove that this set is complete or not. 


The main purpose of this section is to obtain a classification of FSPT phases with unitary finite Abelian symmetry $G_f$, using the set  $\{\Theta_\mu, \Theta_{\mu}, \Theta_{\mu\nu\lambda}\}$ under the assumption that it is complete. Our strategy is as follows. We first solve the constraints (\ref{c1})-(\ref{c8}), as well as (\ref{ac1}) and (\ref{ac2}), and find all solutions. The solutions consist of all possible values that the topological invariants can take. We assume that the solutions have a one-to-one correspondence to FSPT phases. Accordingly, we read out a classification from the structure of the solutions.

For this classification scheme to work, we have made two assumptions: (1) the set $\{\Theta_\mu, \Theta_{\mu}, \Theta_{\mu\nu\lambda}\}$ is complete and (2) all solutions to constraints (\ref{c1})-(\ref{c8}), (\ref{ac1}) and (\ref{ac2}) are realizable in physical systems. We cannot prove the first assumption, but we show some evidence for the completeness of our topological invariants in Sec.~{\ref{sec:completeness}}.  The second assumption will be discussed in Sec.~\ref{sec:model_con}, where we construct models to realize solutions to the constraints.







\subsection{Group structure of FSPT phases}
\label{sec:stacking}

By classification of FSPT phases, we mean two pieces of information: the total number of phases for a given symmetry $G_f$, and the group structure of phases under stacking operation. The latter can be observed as follows: (1) ``identity''---there  exists a trivial phase, the conventional atomic insulators; (2) ``group multiplication''---stacking two FSPT phases, we obtain a new phase; and (3) ``inverse''---given an FSPT phase, there exists an inverse phase, such that stacking the two produces the trivial phase. In this paper, we denote the stacking group of FSPT phases as $H_{\rm stack}$. Since stacking is a symmetric operation, $H_{\rm stack}$ is Abelian. It is obvious that the total number of FSPT phases is given by the order of the group, $|H_{\rm stack}|$. For finite Abelian symmetry, we believe that $|H_{\rm stack}|$ is finite; indeed, it is finite in our classification.

In order to obtain classification of FSPT phases from the topological invariants $\Theta_\mu, \Theta_{\mu\nu}, \Theta_{\mu\nu\lambda}$, one question is how to infer the group $H_{\rm stack}$ from their possible values. To answer that, we first notice that the topological invariants of the trivial phases all vanish. This can be easily checked by studying gauge theories coupled to conventional atomic insulators. Second, the topological invariants are {\it additive} under stacking operation. More precisely, given two FSPT phases with the values of the topological invariants being $(\Theta_{\mu}^a, \Theta_{\mu\nu}^a, \Theta_{\mu\nu\lambda}^a)$ and $(\Theta_{\mu}^b, \Theta_{\mu\nu}^b, \Theta_{\mu\nu\lambda}^b)$ respectively, the values of the topological invariants for the new phase obtained by stacking are given by
\begin{equation}
(\Theta_{\mu}^a+\Theta_{\mu}^b, \Theta_{\mu\nu}^a+\Theta_{\mu\nu}^b, \Theta_{\mu\nu\lambda}^a + \Theta_{\mu\nu\lambda}^b)
\end{equation}
To see that, we notice that the topological invariants are Berry phases associated with gauge flux. Intuitively, after stacking, gauge flux should pierce both layers. Hence, the total Berry phase should be the sum of Berry phases from each layer. In addition, one can check that if $\{\Theta_\mu, \Theta_{\mu\nu}, \Theta_{\mu\nu\lambda}\}$ is a solution to the constraints (\ref{c1})-(\ref{c8}), so is  $\{-\Theta_\mu, -\Theta_{\mu\nu}, -\Theta_{\mu\nu\lambda}\}$. Therefore, $H_{\rm stack}$ is also the group formed by all possible values of the topological invariants under the addition modulo $2\pi$.


\subsection{Classification}

We now solve the constraints (\ref{c1})-(\ref{c8}), (\ref{ac1}) and (\ref{ac2}), from which we determine the stacking group $H_{\rm stack}$ of FSPT phases with general finite Abelian symmetry $G_f$ given in (\ref{group1}). We show that the group $H_{\rm stack}$ has the following form
\begin{equation}
H_{\rm stack} = A \times \prod_i  B_i\times \prod_{i<j}  C_{ij}\times \prod_{i<j<k}  D_{ijk} \label{fspt_group}
\end{equation}
where $A$, $B_{i}$, $C_{ij}$ and $D_{ijk}$ are finite Abelian groups listed in Table \ref{tab1}, and the indices $i,j,k$ take values in the range $1,\dots, K$ (see Table \ref{tab2} for several specific examples). The purpose of this subsection is to derive $A$, $B_i$, $C_{ij}$ and $D_{ijk}$.

The constraints (\ref{c1})-(\ref{c8}), (\ref{ac1}) and (\ref{ac2}) are linear equations of the tensors $\Theta_\mu, \Theta_{\mu\nu}, \Theta_{\mu\nu\lambda}$. So, solving them is straightforward, though a bit tedious due to the fact that all equations are defined modulo $2\pi$. We first notice that the constraints only relate those components of the tensors whose indices differ at most by the index 0. Therefore, we can divide the components of the tensors into four categories
\begin{align}
(a): & \ \Theta_{0}, \ \Theta_{00}, \ \Theta_{000} \nonumber \\
(b): & \ \Theta_{i}, \ \Theta_{0i}, \Theta_{ii}, \ \Theta_{00i}, \ \Theta_{0ii}, \Theta_{iii}\nonumber\\
(c): & \ \Theta_{ij}, \ \Theta_{0ij}, \ \Theta_{iij}, \ \Theta_{jji} \nonumber \\
(d): & \ \Theta_{ijk}\nonumber
\end{align}
where $i\neq j$ in category (c), $i\neq j \neq k$ in category (d), and $i,j,k$ take values in the range $1,2,\dots, K$. Since $\Theta_{\mu\nu\lambda}$ is fully antisymmetric due to (\ref{c1}) and $\Theta_{\mu\nu}$ is symmetric due to (\ref{c4}), we do not list other components of the tensors, whose indices are permutations of the ones listed above. The components from different categories are independent, and the components with different values of indices in each category are also independent. This allows us to solve the constraints for {\it fixed values} of indices, and solve them separately for each category.  The groups $A, B_i, C_{ij}, D_{ijk}$ are determined by the components in the four categories respectively.  Below, we solve the constraints for each category.

\subsubsection{Category (d)}

Let us begin with the simplest case, category (d). Due to the antisymmetry of $\Theta_{ijk}$, it is enough to consider $i<j<k$. One can see that the only constraint related to $\Theta_{ijk}$ is (\ref{c2}), i.e., $N_{ijk}\Theta_{ijk}=0$. Hence, $\Theta_{ijk}$ can take $N_{ijk}$ distinct values:
\begin{equation}
0, \ \frac{2\pi}{N_{ijk}}, \ \frac{4\pi}{N_{ijk}},\  \dots, \ \frac{(N_{ijk}-1)2\pi}{N_{ijk}} \nonumber
\end{equation}
Obviously, under addition modulo $2\pi$, these values form a group
\begin{equation}
D_{ijk} = \mathbb Z_{N_{ijk}}
\end{equation}
Correspondingly, there are $N_{ijk}$ distinct FSPT phases and they are characterized by the $N_{ijk}$ distinct values of $\Theta_{ijk}$. In particular, all these phases can be obtained by the one characterized by $\Theta_{ijk} = 2\pi/N_{ijk}$ through stacking operation. In other words, the phase with $\Theta_{ijk} = 2\pi/N_{ijk}$  is the {\it generating phase}.  By varying the values of $i,j,k$, we obtain the part $\prod_{i<j<k} D_{ijk}$ of the group $H_{\rm stack}$ in Eq.~(\ref{fspt_group}).

\subsubsection{Category (c)}
Next, we solve the constraints for category (c).  Due to the symmetry of $\Theta_{\mu\nu}$ and antisymmetry of $\Theta_{\mu\nu\lambda}$, it is enough to consider $i<j$. The constraints that are relevant to this category are:
\begin{align}
2\Theta_{iij} & = 2\Theta_{jji} =0 \label{c99} \\
N_{0ij}\Theta_{0ij} & =0 \\
\Theta_{iij} & = \Theta_{jji} =m \Theta_{0ij} \\
N_{ij} \Theta_{ij} & =\frac{N^{ij}(N^{ij}-1)}{2}\Theta_{iij} \label{c100}
\end{align}
which are special cases of (\ref{c1}), (\ref{c2}), (\ref{c3}) and (\ref{c5}) respectively. It is obvious that $\Theta_{iij}$ and $\Theta_{jji}$ are determined by $\Theta_{0ij}$.  So, we can focus on possible values of $\Theta_{ij}$ and $\Theta_{0ij}$. In fact, we will focus on the values of  $\Theta_{ij}$ and $\Theta_{0ij}$ of the generating phases.

First of all, the right-hand side of (\ref{c100}) does not vanish only if $m$ is odd, and $N_i,N_j$ are both odd multiples of $2$. This can be seen by considering the following cases: (i) if either $N_i$ or $N_j$ are odd, $m$ is a multiple of $N_{0ij}$ and thereby $\Theta_{iij}=m\Theta_{0ij}=0$; (ii) if either $N_i$ or $N_j$ are even multiples of 2,  $N^{ij}(N^{ij}-1)/2$ is an even number and thereby the right-hand side of (\ref{c100}) vanishes because of (\ref{c99}); and (iii) if $N_i, N_j$ are odd multiples of $2$ and $m$ is even, $m$ is a multiple of $N_{0ij}$ and thereby $m\Theta_{0ij}=0$.

Accordingly, when $m$ is odd and $N_i,N_j$ are both odd multiples of $2$, we solve the constraints in (\ref{c100}) and find two generating phases, which are described by
\begin{equation}
(\Theta_{ij}, \Theta_{0ij})  = \left(\frac{\pi}{N_{ij}}, \frac{2\pi}{N_{0ij}}\right), \quad \left(0, \frac{4\pi}{N_{0ij}}\right)
\end{equation}
For all other cases, i.e., when the right-hand side of (\ref{c100}) vanishes, we find two generating phases that are described by
\begin{equation}
(\Theta_{ij}, \Theta_{0ij})  = \left(\frac{2\pi}{N_{ij}},0 \right), \quad \left(0, \frac{2\pi}{N_{0ij}}\right)
\end{equation}
Therefore, by stacking the generating phases, we obtain FSPT phases with a group structure
\begin{equation}
C_{ij} = \left\{
\begin{array}{ll}
\mathbb Z_{2N_{ij}}\times \mathbb Z_{N_{0ij}/2},  & \text{if $m, \frac{N_i}{2}, \frac{N_j}{2}$ are odd integers}\\
\mathbb Z_{N_{ij}}\times \mathbb Z_{N_{0ij}}, & \text{otherwise}
\end{array}
\right.
\end{equation}
By varying the indices $i,j$, the part $\sum_{i<j}C_{ij}$ of the group $H_{\rm stack}$ is obtained.



\subsubsection{Category (b)}

Now, we work on category (b). From Eq.~(\ref{c3}), we know that $\Theta_{iii} = m \Theta_{0ii}$ and $\Theta_{0ii} = m\Theta_{00i}$. Also, from (\ref{c7}), we observe that $\Theta_{ii}$ is determined by $\Theta_i$ and $\Theta_{iii}$. Hence, we only need to consider three independent components, $\Theta_i$, $\Theta_{0i}$ and $\Theta_{00i}$. Below, we find values of $\Theta_i$, $\Theta_{0i}$ and $\Theta_{00i}$ for generating phases, by solving the relevant constraints in four cases.

First, we consider the case that $N_i$ is odd. In this case, $N_{00i}$ is also odd. According to (\ref{c1}) and (\ref{c2}), we have $N_{00i}\Theta_{00i} =2\Theta_{00i} =0$. Accordingly, $\Theta_{00i}=0$. It then leads to $\Theta_{iii} = \Theta_{0ii}=0$. With this result and the constraints (\ref{c5}) and (\ref{c7}), we further obtain that $N_i \Theta_i = N_{0i} \Theta_{0i} =0$. Accordingly, we find two generating phases, which are characterized by
\begin{equation}
(\Theta_{i}, \Theta_{0i}, \Theta_{00i}) = \left(\frac{2\pi}{N_i}, 0, 0\right), \ \left( 0,  \frac{2\pi}{N_{0i}},
0\right)\end{equation}
Other phases can be obtained by stacking the two generating phases.

Second, we consider the case that $m$ and $N_i$ are both even. According to (\ref{c1}) and (\ref{c3}), we have $2\Theta_{00i}=0$ and $\Theta_{00i} = m \Theta_{00i}$. Consequently, $\Theta_{00i}$ must be 0 when $m$ is even. Combining this result with the constraints (\ref{c5}) and (\ref{c7}), we obtain $2N_i\Theta_i = N_{0i} \Theta_{0i}=0$. At the same time, Eqs.~(\ref{c6}) and (\ref{c7}) lead to $N_i\Theta_i = (N_{0i}/2)\Theta_{0i}$. With these, we find two generating phases that are described by
\begin{equation}
(\Theta_{i}, \Theta_{0i}, \Theta_{00i})= \left(\frac{\pi}{N_i}, \frac{2\pi}{N_{0i}},0\right), \ \left(0, \frac{4\pi}{N_{0i}},0\right)
\end{equation}
All other solutions to the constraints can be generated by the above two.

Third, we consider the case that $m$ is odd and $N_{i}$ is an odd multiple of $2$. In this case, after some minor simplifications to the general constraints (\ref{c1})-(\ref{c7}),  we find that $\Theta_{iii}=\Theta_{0ii}=\Theta_{00i}$, $2\Theta_{00i}=0$, $N_{0i}\Theta_{0i} = \Theta_{00i}$, $N_{i}\Theta_{ii} = \Theta_{00i}$, $\Theta_{ii}  = 2\Theta_i + \Theta_{00i}$, and $(N_i/2)\Theta_{ii} = (N_{0i}/2)\Theta_{0i} +[m(m-1)/2 + (N_i/2)(N_i/2-1)/2]\Theta_{00i}$. Then, after some straightforward calculations, we find two generating phases that are characterized by
\begin{equation}
(\Theta_{i}, \Theta_{0i}, \Theta_{00i})= \left(\frac{\pi}{2N_i}, \mp \frac{\pi}{N_{0i}},\pi\right), \ \left(0, \frac{4\pi}{N_{0i}},0\right)
\end{equation}
where the ``$-$'' sign applies when $m = N_i/2 \modulo{4}$, and the ``$+$'' sign applies when $m=N_i/2+2 \modulo{4}$. All other solutions can be generated by the above two.

Finally, we consider the case that $m$ is odd and $N_i$ is a multiple of $4$.  After some minor simplifications to the general constraints, we find that $\Theta_{iii}=\Theta_{0ii}=\Theta_{00i}$, $2\Theta_{00i}  =0$, $N_{0i}\Theta_{0i}  = 0$, $N_{i}\Theta_{ii} = 0 $, $\Theta_{ii} = 2\Theta_i $, and $(N_i/2)\Theta_{ii} = (N_{0i}/2)\Theta_{0i} +[(N_i/2)(N_i/2-1)/2]\Theta_{00i}$. We find that there are two generating phases with
\begin{equation}
(\Theta_{i}, \Theta_{0i}, \Theta_{00i})= \left(\frac{\pi}{N_i}, \frac{2\pi}{N_{0i}},0\right), \ \left(0, \lambda\frac{2\pi}{N_{0i}},\pi\right)
\end{equation}
where $\lambda = 1$ if $N_i = 4 \modulo{8}$ and $\lambda = 2$ if $N_i=0 \modulo{8}$. All other phases can be generated by the above generating phases.

Combining all cases, we conclude that the stacking group $B_{i}$ is given by
\begin{equation}
B_i  = \left\{
\begin{array}{ll}
\mathbb Z_{N_i} \times\mathbb Z_{N_{0i}}, & \text{if $N_i$ is odd}\\
\mathbb Z_{2N_i} \times\mathbb Z_{N_{0i}/2}, & \text{if $m, N_i$ are even}\\
\mathbb Z_{4N_i} \times\mathbb Z_{N_{0i}/2}, & \text{if $m$ is odd, $N_i =2 \ ({\rm mod}\ 4) $}\\
\mathbb Z_{2N_i} \times\mathbb Z_{N_{0i}}, & \text{if $m$ is odd, $N_i =0 \ ({\rm mod}\ 4) $}
\end{array}
\right.
\end{equation}
By varying the index $i$, the part $\sum_i B_i$ of the group $H_{\rm stack}$ is obtained.

\subsubsection{Category (a)}

Finally, we solve the constraints for category (a). In this case, relevant constraints include $2\Theta_{000}=0$, $\Theta_{000}= m \Theta_{000}$, $N_0\Theta_{00} = m \Theta_{000}$, the constraint (\ref{c8}),  and the additional constraints (\ref{ac1}) and (\ref{ac2}) from vanishing chiral central charge. We divide the discussion into two cases.

First, we consider the case that $m$ is even. In this case, $\Theta_{000}=0$. According to (\ref{c8}), the only independent invariant is $\Theta_0$. With (\ref{ac2}), we find that the generating phase is described by
\begin{equation}
\Theta_0 = \frac{4\pi}{m}
\end{equation}
Next, we consider that $m$ is odd. In this case, combining Eqs.~(\ref{c8}), (\ref{ac1}) and $N_0\Theta_{00} = \Theta_{000}$, we find that $\Theta_{000}=0$ too. Therefore, Eq.~(\ref{ac1}) reduces to $m\Theta_{0}=0$, and we find one generating phase described by
\begin{equation}
\Theta_0 = \frac{2\pi}{m}
\end{equation}
Other FSPT phases can be obtained by stacking the generating phases. Combining both cases, we obtain the group $A$:
\begin{align}
A = \left\{
\begin{array}{cl}
\mathbb Z_m  & \text{if $m$ is odd}\\
\mathbb Z_{m/2} & \text{if $m$ is even}
\end{array}
\right.
\end{align}



\subsection{On assumptions of the classification}
\label{sec:completeness}

Let us repeat the two assumptions that we rely on in order to obtain the above classification: (i) The topological invariants are complete, in the sense that they distinguish every FSPT phase with symmetry $G_f$ in (\ref{group}) and (ii) every solution to the constraints (\ref{c1})-(\ref{c8}), (\ref{ac1}) and (\ref{ac2}) can be realized in physical systems. In this subsection, we show some evidence that support the first assumption. The second assumption will be justified in Sec.~\ref{sec:model_con}.

The first evidence supporting assumption (i) is that our classification reproduces several known examples. For example, for $G_f = \mathbb Z_2^f\times\mathbb Z_2$, our classification gives $H_{\rm stack} = \mathbb Z_8$, agreeing with Ref.~\onlinecite{gu14b}. Our classification also agrees with Ref.~\onlinecite{cheng15} for other small groups, such as $\mathbb Z_2^f\times \mathbb Z_{N}$.

The second evidence is that our classification gives the same counting of FSPT phases as the general classification in Ref.~\onlinecite{cheng15}. These works only consider symmetry $G_f$ of the form $\mathbb Z_2^f\times G$, and find that the classification comes in three types, described by cohomology groups $H^1(G, \mathbb Z_2)$, $H^2(G, \mathbb Z_2)$\footnote{There is an obstruction for FSPT phases described by $H^2(G,\mathbb Z_2)$, in the sense that not every phase in $H^2(G, \mathbb Z_2)$ can be realized in 2D. However, one can show that this obstruction always vanishes for finite Abelian group $G$.} and $H^3(G, U(1))$ respectively (note that the three types mix under stacking operation). For Abelian symmetry $G= \prod_{i} \mathbb Z_{N_i}$, we find that cohomology groups are given by
\begin{align}
H^1(G, \mathbb Z_2) & = \prod_i\mathbb Z_{N_{0i}} \nonumber\\
H^2(G, \mathbb Z_2) & = \prod_i\mathbb Z_{N_{0i}} \prod_{i<j}\mathbb Z_{N_{0ij}}\nonumber\\
H^3(G,  U(1)) & = \prod_i \mathbb Z_{N_{i}}\prod_{i<j}\mathbb Z_{N_{ij}} \prod_{i<j<k} \mathbb Z_{N_{ijk}} \nonumber
\end{align}
where we have set $N_0=2$. Then, the total number of FSPT phases is given by $|H^1(G, \mathbb Z_2)|\times |H^2(G, \mathbb Z_2) | \times |H^3(G,  U(1))|$.

In our classification, the counting of FSPT phases is as follows. When $m$ is even, we have
\begin{equation}
|H_{\rm stack}| = \frac{m}{2}\prod_{i}\left(N_i N_{0i}\right) \prod_{i<j} \left(N_{ij} N_{0ij}\right) \prod_{i<j<k} N_{ijk}
\end{equation}
and when $m$ is odd
\begin{equation}
|H_{\rm stack}| = m \prod_{i}\left(N_i N_{0i} M_i\right) \prod_{i<j} N_{ij} N_{0ij} \prod_{i<j<k} N_{ijk}
\end{equation}
where have denoted $M_i = {\rm gcd}(2,N_i)$. Here, ``gcd'' stands for greatest common divisor.

One can easily check that for $m=1$, $|H_{\rm stack}| = |H^1(G, \mathbb Z_2)|\times |H^2(G, \mathbb Z_2) | \times |H^3(G,  U(1))|$. In fact, since we have the isomorphism $\mathbb Z_{2m}^f = \mathbb Z_{2}^f \times \mathbb Z_m$ for odd $m$, the case $m=1$ is general enough to represent all odd-$m$ cases. Hence, for general Abelian symmetry $G_f$ with $m$ being odd, the counting of FSPT phases in our classification agrees with that of Refs.~\onlinecite{cheng15}. Since Ref.~\onlinecite{cheng15} does not work out the stacking group for general symmetries, we are not able to make a comparison.

The case with even $m$ is rarely studied.  The simplest case $G_f=\mathbb Z_4^f$ was considered by Ref.~\onlinecite{kapustin14, wangcj16, lan16}.  Our classification agrees with these results.



\section{Model construction}
\label{sec:model_con}

One of the two assumptions in our classification (see Sec.~\ref{sec:completeness}) is that all solutions to the constraints (\ref{c1})-(\ref{c8}), (\ref{ac1}) and (\ref{ac2}) can be realized in physical systems. In this section, we justify this assumption by constructing models for the phases in our classification. We successfully construct models for almost all FSPT phases, except one class --- case (C-4) in Sec.~\ref{sec:con_cij_even} --- which will be further discussed in Sec.~\ref{sec:exception}.


\subsection{Two types of existing models}
\label{exmod}

The idea behind our model construction is simple: We take two types of existing models, (1) free-fermion models and (2) FSPT models that are adapted from BSPT models, which we call BSPT-embedded models. Then, we make a layer construction out of the two types of models in an appropriate way. We do not introduce any coupling between different layers. In this way, we are able to realize various FSPT phases. (As a comparison, Refs.~\onlinecite{tarantino16, gaiotto16} constructed exactly soluble models to realize 2D FSPT phases with $\mathbb Z_2^f\times G$ onsite unitary symmetry; see also Ref.~\onlinecite{ware16} for a related construction.) Below, we review properties of the free-fermion models and BSPT-embedded models that we will use in our construction. In particular, we list the values of the topological invariants $\Theta_\mu, \Theta_{\mu\nu}, \Theta_{\mu\nu\lambda}$ of these models.

We start with free-fermion models. Two well-known free-fermion states are the $p_x+ip_y$ superconductors\cite{read00,ivanov01} and integer quantum Hall (IQH) states.\cite{wen-book} (These states are not FSPT phases since they carry chiral edge models, however, they will be very useful for our model construction.) The $p_x+ip_y$ superconductors preserve the fermion parity $\mathbb Z_2^f$ only. They are chiral states with the chiral central charge $c=1/2$. If we gauge the $\mathbb Z_2^f$ symmetry, it is known that the resulting system has a non-Abelian Ising topological order.\cite{ivanov01,kitaev06} The IQH states preserve a charge $U(1)$ symmetry. They are also chiral states with the chiral central charge $c=\nu$, where $\nu$ is the integer filling factor. One may gauge a subgroup $\mathbb Z_{2m}^f$ of the charge $U(1)$ symmetry and obtain a gapped gauged model. In contrast to $p_x+ip_y$ superconductors, braiding statistics in these gauged IQH states are always Abelian.

For our purpose, we are more interested in another class of gapped nearly-free-fermion models, namely  {\it charge-$2m$ superconductors}, i.e., fermion systems with $\mathbb Z_{2m}^f$ symmetry. It was shown in Ref.~\onlinecite{wangcj16} that general charge-$2m$ superconductors can be constructed by stacking $p_x+ip_y$ superconductors and IQH states in an appropriate way. Since IQH states respect charge $U(1)$ symmetry, we need to add a weak perturbation, which does not close the energy gap, to break $U(1)$ down to $\mathbb Z_{2m}^f$. Such perturbation is not quadratic, but since it is weak, we still consider these models as free-fermion models.  According to Ref.~\onlinecite{wangcj16}, the topological invariants $\Theta_0, \Theta_{00}, \Theta_{000}$ of charge-$2m$ superconductors are given by
\begin{align}
\Theta_0 & = \left\{
\begin{array}{ll}
\vspace{5pt}
\frac{\pi}{8m}p, & \text{if $m$ is odd} \\
\frac{\pi}{2m}p, & \text{if $m$ is even}
\end{array}
\right. \nonumber\\
\Theta_{000}& = \left\{
\begin{array}{ll}
\vspace{5pt}
\pi, & \text{if $m, p$ are odd} \\
0, & \text{otherwise}
\end{array}
\right.\label{2msc}
\end{align}
where $p$ is an integer. The value of $\Theta_{00}$ can be determined through Eq.~(\ref{c8}). In general, charge-$2m$ superconductors are chiral, and the chiral central charge $c$ is given by
\begin{equation}
c = \left\{
\begin{array}{ll}
p \ \modulo{8}, & \text{if $m$ is even} \\[3pt]
\frac{p}{2} - \frac{m^2-1}{2}\sigma(p)\ \modulo{8}, & \text{if $m$ is odd}
\end{array}
\right.\label{2msc-c}
\end{equation}
where $\sigma(p)=1$ if $p$ is odd, and $\sigma(p)=0$ if $p$ is even. The case that $m=p=1$ describes $p_x+ip_y$ superconductors. It was show in Ref.~\onlinecite{wangcj16} that when $\Theta_{000}=0$, all excitations in gauged charge-$2m$ superconductors are Abelian.



The second type of models are built out of BSPT phases: We first let fermions form strongly bound pairs, then put the pairs (bosons) into a BSPT phases. In other words, we ``embed'' a BSPT phase into the fermionic system. Hence, we call these models {\it BSPT-embedded models}. For a fermionic system with symmetry $G_f = \mathbb Z_{2m}^f\times\prod_{i} \mathbb Z_{N_i}$, the corresponding bosonic system should have a symmetry
\begin{equation}
G_b \equiv G_f/\mathbb Z_2^f =  \mathbb Z_m \times \prod_{i=1}^K \mathbb Z_{N_i}
\end{equation}
For bosonic systems with $G_b$ symmetry, Ref.~\onlinecite{chen13} constructed a large class of exactly soluble models, labeled by the elements of the cohomology group $H^3[G_b, U(1)]$. It is believed that the group cohomology $H^3[G_b, U(1)]$ classifies BSPT phases with $G_b$ symmetry.

These cohomology models of bosons can be similarly characterized by a BSPT version of the topological invariants\cite{wangcj15}.  The relation between the BSPT version and  the FSPT version of topological invariants will be discussed in Sec.~\ref{sec:stability}. Here, we list the values of the FSPT topological invariants for these group cohomology models after they are embedded into fermionic systems. The values of independent topological invariants are\footnote{These values are obtained by combining Eqs.~(\ref{binv}) and (\ref{dis-b3}) from Sec.~\ref{sec:stability}.}
\begin{align}
\Theta_0 & =\left\{ \begin{array}{ll}
\frac{2\pi}{m}p_0, \ \ &\text{if $m$ is odd} \\[5pt]
\frac{4\pi}{m}p_0,\ \  &\text{if $m$ is even}
\end{array}
\right. \nonumber\\[5pt]
\Theta_i &  = 
\frac{2\pi}{N_i} p_i, 
\quad \Theta_{0i}  = \frac{N^{0i}}{\bar N^{0i}}\frac{2\pi}{ \bar N_{0i}} p_{0i}, \quad \Theta_{00i} =0 \nonumber\\[5pt]
\Theta_{ij} & = \frac{2\pi}{N_{ij}} p_{ij}, \quad \Theta_{0ij}  = \frac{2\pi}{\bar N_{0ij}} p_{0ij} \nonumber\\[5pt]
\Theta_{ijk} &  =\frac{2\pi}{N_{ijk}} p_{ijk} \label{bspt}
\end{align}
where $p_0$, $p_i$, $p_{0i}$, $p_{ij}$ with $i<j$, $p_{0ij}$ with $i<j$, and $p_{ijk}$ with $i<j<k$,  are independent integers. Here, $\bar N_{0i} = \gcd(m, N_i)$, $\bar N_{0ij} = \gcd(m, N_i, N_j)$ and $\bar N^{0i}={\rm lcm}(m, N_i)$, where ``gcd'' and ``lcm'' stand for greatest common divisor and least common multiple respectively. Note that $N_{0i} = \gcd(N_0, N_i)$,  $N_{0ij} = \gcd(N_0, N_i, N_j)$, and $ N^{0i}={\rm lcm}(N_0, N_i)$, where $N_0=2m$.  All group cohomology models are nonchiral.

One can see that the BSPT-embedded models realize a large class of FSPT phases, but not all of them. Sometimes, we will call those phases that do not have a BSPT-embedded-model realization {\it intrinsic} FSPT phases. Many free-fermion models are intrinsic FSPT phases.



\subsection{Procedure of the construction}
\label{procedure}

The idea of our construction is to make a multi-layer construction using the free-fermion models and BSPT-embedded models. Below we discuss the general procedure of our construction.

First, since a general FSPT phase can be obtained by stacking the generating phases, it is enough to construct models for the generating phases in $H_{\rm stack}$ (see Table \ref{tab1}).

Second, to construct models for the generating phases, it is enough to consider the following four simpler symmetry groups
\begin{align}
G_f&= \mathbb Z_{2m}^f \nonumber\\
G_f&= \mathbb Z_{2m}^f \times \mathbb Z_{N_i}\nonumber\\
G_f&= \mathbb Z_{2m}^f \times \mathbb Z_{N_i} \times \mathbb Z_{N_j}\nonumber\\
G_f&= \mathbb Z_{2m}^f \times \mathbb Z_{N_i} \times \mathbb Z_{N_j} \times \mathbb Z_{N_k} \label{sgroup}
\end{align}
where $m, N_i, N_j, N_k$ are arbitrary integers (the indices $i,j,k$ are arbitrary but fixed). Let us take an example to illustrate the reason. Suppose that we would like to construct models for the generating phases associated with the $B_i$ component for a fixed index $i$ in $H_{\rm stack}$ for general symmetry group $G_f = \mathbb Z_{2m}^f\times \prod_{l=1}^K\mathbb Z_{N_l}$.  These generating phases are characterized by the topological invariants $\Theta_{i}, \Theta_{0i}, \Theta_{00i}$, etc, which only involve the indices ``0'' and ``$i$''. Physically, it means that these generating phases are fully characterized by properties of the vortices carrying type-0 and type-$i$ unit flux. Therefore, we can ignore the existence of vortices carrying other types of unit flux. In other words, we can view the group $\mathbb Z_{2m}^f\times \prod_{l=1}^K\mathbb Z_{N_l}$ as if it is $\mathbb Z_{2m}^f \times \mathbb Z_{N_i}$ without loosing any generality. One can similarly argue that the model construction of generating phases associated with the components $A$, $C_{ij}$ and $D_{ijk}$ in $H_{\rm stack}$ can be reduced to the rest of the symmetry groups in (\ref{sgroup}).

Third, we build multi-layer models for the groups in (\ref{sgroup}) using free-fermion and BSPT-embedded models in an appropriate way. Note that the indices $i,j,k$ in (\ref{sgroup}) encodes the information about how these simpler groups are mapped back to the general symmetry group $\mathbb Z_{2m}^f\times \prod_{l=1}^K\mathbb Z_{N_l}$. These indices are not relevant for model construction, but we keep them for consistency of our notation.

Hence, if we are able to construct models for the symmetry groups in (\ref{sgroup}), models for general FSPT phases can be easily obtained through the above steps in a reversed order.

By comparing Eq.~(\ref{bspt}) and Table \ref{tab1},  we notice that the generating phases associated with the $A$ and $D_{ijk}$ components are already realized by the BSPT-embedded phases.\footnote{Generating phases associated with the component $A$ can also be realized in free-fermion systems\cite{wangcj16}. } Therefore, we are left with the construction of models for the generating phases associated with the $B_i$ and $C_{ij}$ components, which we do in Sec.~\ref{sec:con_bi} and Sec.~\ref{sec:con_cij} respectively.





\begin{table*}%
\caption{Summary of models for the generating phases associated with the subgroups $B_i$ and $C_{ij}$ of $H_{\rm stack}$. The ``Case'' columns list various cases discussed in the main text, and the ``Generator'' columns list the equations that give the values of the topological invariants in the corresponding generating phase. In the ``Model'' columns, ``BSPT-embedded'' means that the corresponding generating phase can be realized through the BSPT-embedded models, while others are beyond BSPT-embedded models (i.e., intrinsically fermionic). For the latter phases, the layer constructions from the main text are depicted. For the generating phase described by (\ref{c4gen2}) in case (C-4), we do not have models in certain situations; see Sec.~\ref{sec:con_cij_even} and \ref{sec:exception} for detailed discussion.} \label{tab3}
\begin{tabular}
[c]{ccccc|ccccc}\hline\hline
\ \ Case \ \ &    \quad & Generator  &\quad & Model & \ \ Case \ \ & \quad & Generator & \quad & Model\\\hline
(B-1) & & (\ref{b1gen}) & & BSPT-embedded & (C-1) & & (\ref{c1gen})& & BSPT-embedded \\\hline
 (B-2) & & (\ref{b2gen}) & &
\begin{tikzpicture}[scale=1]
\fill[draw=black, gray!20] (0,0)--(3,0)--(4,1)--(1,1)--cycle;
\draw [gray](0,0)--(3,0)--(4,1)--(1,1)--cycle;
\node at (1.5,0.15)[scale=0.8]{charge-2};
\node at (0.3,0.15)[scale=0.8]{$a$};

\begin{scope}[yshift=0.3cm]
\fill[gray!20] (0,0)--(3,0)--(4,1)--(1,1)--cycle;
\draw [gray](0,0)--(3,0)--(4,1)--(1,1)--cycle;
\node at (1.5,0.15)[scale=0.8]{charge-$N_i$};
\node at (0.3,0.15)[scale=0.8]{$b$};
\node at (0,1.05){};
\end{scope}
\end{tikzpicture} & (C-2) &&(\ref{c2gen})&&
\begin{tikzpicture}[scale=1]
\fill[draw=black, gray!20] (0,0)--(3,0)--(4,1)--(1,1)--cycle;
\draw [gray](0,0)--(3,0)--(4,1)--(1,1)--cycle;
\node at (1.5,0.15)[scale=0.8]{charge-2};
\node at (0.3,0.15)[scale=0.8]{$a$};

\begin{scope}[yshift=0.3cm]
\fill[gray!20] (0,0)--(3,0)--(4,1)--(1,1)--cycle;
\draw [gray](0,0)--(3,0)--(4,1)--(1,1)--cycle;
\node at (1.5,0.15)[scale=0.8]{charge-$N_i$};
\node at (0.3,0.15)[scale=0.8]{$b$};
\node at (0,1.05){};
\end{scope}

\begin{scope}[yshift=0.6cm]
\fill[gray!20] (0,0)--(3,0)--(4,1)--(1,1)--cycle;
\draw [gray](0,0)--(3,0)--(4,1)--(1,1)--cycle;
\node at (1.5,0.15)[scale=0.8]{charge-$N_j$};
\node at (0.3,0.15)[scale=0.8]{$c$};
\node at (0,1.05){};
\end{scope}

\begin{scope}[yshift=0.9cm]
\fill[gray!20] (0,0)--(3,0)--(4,1)--(1,1)--cycle;
\draw [gray](0,0)--(3,0)--(4,1)--(1,1)--cycle;
\node at (1.5,0.15)[scale=0.8]{charge-$2$};
\node at (0.3,0.15)[scale=0.8]{$d$};
\node at (0,1.05){};
\end{scope}
\end{tikzpicture} \\ \hline
(B-3) & & (\ref{b3gen1}) & &
\begin{tikzpicture}[scale=1]
\fill[draw=black, gray!20] (0,0)--(3,0)--(4,1)--(1,1)--cycle;
\draw [gray](0,0)--(3,0)--(4,1)--(1,1)--cycle;
\node at (1.5,0.15)[scale=0.8]{ charge-2};
\node at (0.3,0.15)[scale=0.8]{$a$};

\begin{scope}[yshift=0.3cm]
\fill[gray!20] (0,0)--(3,0)--(4,1)--(1,1)--cycle;
\draw [gray](0,0)--(3,0)--(4,1)--(1,1)--cycle;
\node at (1.5,0.15)[scale=0.8]{charge-$N_i$};
\node at (0.3,0.15)[scale=0.8]{$b$};
\node at (0,1.05){};
\end{scope}
\end{tikzpicture} &
(C-3)& & (\ref{c3gen1})&& BSPT-embedded \\\hline
(B-3) & & (\ref{monb3-2}) & & \begin{tikzpicture}[scale=1]
\fill[draw=black, gray!20] (0,0)--(3,0)--(4,1)--(1,1)--cycle;
\draw [gray](0,0)--(3,0)--(4,1)--(1,1)--cycle;
\node at (1.5,0.15)[scale=0.8]{charge-2};
\node at (0.3,0.15)[scale=0.8]{$a$};

\begin{scope}[yshift=0.3cm]
\fill[gray!20] (0,0)--(3,0)--(4,1)--(1,1)--cycle;
\draw [gray](0,0)--(3,0)--(4,1)--(1,1)--cycle;
\node at (1.5,0.15)[scale=0.8]{charge-$N_i$};
\node at (0.3,0.15)[scale=0.8]{$b$};
\end{scope}

\begin{scope}[yshift=0.6cm]
\fill[gray!20] (0,0)--(3,0)--(4,1)--(1,1)--cycle;
\draw [gray](0,0)--(3,0)--(4,1)--(1,1)--cycle;
\node at (1.5,0.15)[scale=0.8]{charge-$2$};
\node at (0.3,0.15)[scale=0.8]{$c$};
\node at (0,1.05){};
\end{scope}
\end{tikzpicture}
&(C-3)&&(\ref{c3gen2})&&
\begin{tikzpicture}[scale=1]
\fill[draw=black, gray!20] (0,0)--(3,0)--(4,1)--(1,1)--cycle;
\draw [gray](0,0)--(3,0)--(4,1)--(1,1)--cycle;
\node at (1.5,0.15)[scale=0.8]{charge-2};
\node at (0.3,0.15)[scale=0.8]{$a$};

\begin{scope}[yshift=0.3cm]
\fill[gray!20] (0,0)--(3,0)--(4,1)--(1,1)--cycle;
\draw [gray](0,0)--(3,0)--(4,1)--(1,1)--cycle;
\node at (1.5,0.15)[scale=0.8]{charge-$N_i$};
\node at (0.3,0.15)[scale=0.8]{$b$};
\node at (0,1.05){};
\end{scope}

\begin{scope}[yshift=0.6cm]
\fill[gray!20] (0,0)--(3,0)--(4,1)--(1,1)--cycle;
\draw [gray](0,0)--(3,0)--(4,1)--(1,1)--cycle;
\node at (1.5,0.15)[scale=0.8]{charge-$N_j$};
\node at (0.3,0.15)[scale=0.8]{$c$};
\node at (0,1.05){};
\end{scope}

\begin{scope}[yshift=0.9cm]
\fill[gray!20] (0,0)--(3,0)--(4,1)--(1,1)--cycle;
\draw [gray](0,0)--(3,0)--(4,1)--(1,1)--cycle;
\node at (1.5,0.15)[scale=0.8]{charge-$2$};
\node at (0.3,0.15)[scale=0.8]{$d$};
\node at (0,1.05){};
\end{scope}
\end{tikzpicture}\\\hline
(B-4) & & (\ref{b4gen}) & & BSPT-embedded &(C-4)&&(\ref{c4gen1})&& BSPT-embedded\\ \hline
(B-5) & & (\ref{b5gen1}) & & \begin{tikzpicture}[scale=1]
\fill[draw=black, gray!20] (0,0)--(3,0)--(4,1)--(1,1)--cycle;
\draw [gray](0,0)--(3,0)--(4,1)--(1,1)--cycle;
\node at (1.5,0.15)[scale=0.8]{charge-$2m$};
\node at (0.3,0.15)[scale=0.8]{$a$};

\begin{scope}[yshift=0.3cm]
\fill[gray!20] (0,0)--(3,0)--(4,1)--(1,1)--cycle;
\draw [gray](0,0)--(3,0)--(4,1)--(1,1)--cycle;
\node at (1.5,0.15)[scale=0.8]{charge-$mN_i$};
\node at (0.3,0.15)[scale=0.8]{$b$};
\node at (0,1.05){};
\end{scope}
\end{tikzpicture}
&(C-4)&&(\ref{c4gen2})&& BSPT-embedded/no model\\\hline
(B-5) & & (\ref{eqsec}) & & BSPT-embedded\\
\hline\hline
\end{tabular}
\end{table*}

\subsection{Generating phases of $B_i$}
\label{sec:con_bi}

In this subsection, we construct models for the generating phases associated with the $B_i$ component in $H_{\rm stack}$ for a fixed index $i$. As discussed in Sec.~\ref{procedure}, it is enough to consider the simpler group $G_f = \mathbb Z_{2m}^f\times \mathbb Z_{N_i}$.

\subsubsection{$m$ being odd}
\label{sec:con_bi_odd}

We first consider odd $m$. In this case, it is enough to consider $m=1$. To see that, we notice that  $\mathbb Z_{2m}^f$ is isomorphic to $\mathbb Z_2^f \times \mathbb Z_m$ for odd $m$. Then, $\mathbb Z_{m}$ can be absorbed into the  $\prod_{l} \mathbb Z_{N_l}$ part in $G_f$, making $G_f$ of the form $\mathbb Z_2^f\times G$. Accordingly, we can set $m=1$ without loosing generality.

According to the classification in Table \ref{tab1}, the component $B_i$ with $m=1$ is given by
\begin{align}
B_i & = \left\{
\begin{array}{ll}
\vspace{3pt}
\mathbb Z_{N_i} & \text{if $N_i$ is odd}\\
\vspace{3pt}
\mathbb Z_{4N_i} & \text{if $N_i=2 \modulo{4}$}\\
\mathbb Z_{2N_i} \times \mathbb Z_2 & \text{if $N_i =0 \modulo{4}$}
\end{array}
\right.
\end{align}
Below we construct models for the generating phases in each case for the group $G_f = \mathbb Z_{2}^f \times \mathbb Z_{N_i}$.

{\bf Case (B-1)}---If $N_i$ is odd, we have $B_i = \mathbb Z_{N_i}$. According to Table \ref{tab1}, the generating phase is described by
\begin{equation}
(\Theta_0,\Theta_{i},\Theta_{0i}, \Theta_{00i}) = (0, 2\pi/N_i, 0, 0) \label{b1gen}
\end{equation}
Here, we require $\Theta_0=0$, so that this phase is a pure generating phase associated with $B_i$, i.e., not a mixture of the generating phases associated with both $A$ and $B_i$. Other components of the topological invariants are determined by $\Theta_0, \Theta_{i},\Theta_{0i} $ and $ \Theta_{00i}$ through the constraints (\ref{c1})-(\ref{c8}). Comparing to Eq.~(\ref{bspt}), one immediately sees that this phase can be realized by a BSPT-embedded model.

{\bf Case (B-2)}---If $N_i =2 \ ({\rm mod} \ 4)$, we have $B_i = \mathbb Z_{4N_i}$. According to Table \ref{tab1}, the generating phase is described by
\begin{equation}
(\Theta_0, \Theta_{i}, \Theta_{0i}, \Theta_{00i})= (0,\pi/2N_i, \pm \pi/2, \pi) \label{b2gen}
\end{equation}
where the ``$-$'' sign applies when $N_i/2=1 \modulo{4}$, and the ``$+$'' sign applies when $N_i/2 = 3 \modulo{4}$. All other components are determined by $\Theta_0, \Theta_{i}, \Theta_{0i}$ and $\Theta_{00i}$. This FSPT phase is beyond the BSPT-embedded models.



To obtain this phase, we consider a two-layer construction. The first layer $a$ is a charge-2 superconductor with the topological invariant $\Theta_0^a$ and  the chiral central charge $c^a$  given by
\begin{equation*}
 \Theta_0^a = \frac{\pi}{8}\left(\frac{N_i^2}{4}-2\right), \quad \Theta_{000}^a=\pi, \quad c^a=\frac{N_i^2}{8}-1
\end{equation*}
That is, we pick the case that $m=1$ and $p=N_i^2/4-2$ in Eqs.~(\ref{2msc}) and (\ref{2msc-c}). The second layer $b$ is a charge-$N_i$ superconductor. Since $N_i$ is an odd multiple of $2$, we can choose it to have
\begin{equation*}
\Theta_{0}^b = \frac{\pi}{4N_i}, \quad \Theta_{000}^b=\pi, \quad c^b=1 - \frac{N_i^2}{8}
\end{equation*}
That is, we pick the case that $m=N_i/2$ and $p=1$ in Eqs.~(\ref{2msc}) and (\ref{2msc-c}). The total chiral central charge of this two-layer system vanish, thereby this model is nonchiral.

Let us check that this system indeed has a $\mathbb Z_2^f\times\mathbb Z_{N_i}$ symmetry. Let $F_a$ be the fermion number operator of layer $a$, and $F_b$ be the fermion number operator of layer $b$. By construction, $(-1)^{F_a}$ and $\exp(i2\pi F_b/N_i)$ are symmetry operators of the system. We observe that the following two operators are also symmetry operators
\begin{align}
P_f & = (-1)^{F_a + F_b} \nonumber\\
g_i & = e^{i2\pi F_b/N_i} \label{monb2-1}
\end{align}
where $P_f$ is  the fermion parity operator by definition.  It is then obvious that the system has a $\mathbb Z_2^f\times\mathbb Z_{N_i}$ symmetry.

We now show that the invariants $\Theta_i, \Theta_{0i}, \Theta_{00i}$ in this two-layer system are indeed given by (\ref{b2gen}). First, according to (\ref{monb2-1}), it is obvious that the type-$i$ unit flux (we still call it type-$i$ flux, even though there are two components in the symmetry group $\mathbb Z_2^f\times\mathbb Z_{N_i}$) is the same as the unit flux in layer $b$. Therefore,
\begin{align}
\Theta_i = 2\Theta_0^b =  \frac{\pi}{2N_i}
\end{align}
where the factor $2$ comes from the difference in the definitions of $\Theta_0$ and $\Theta_i$ (see Sec.~\ref{inv_def}). The calculations of $\Theta_{0i}$ and $\Theta_{00i}$ require some extra effort. Note that according to (\ref{monb2-1}), after we gauge the symmetry, a vortex carrying the type-0 unit flux is composed out of a unit-flux vortex in layer $a$ and a vortex in layer $b$ which carries $N_i/2$ times of the unit flux. After some algebras, we find that
\begin{align}
\Theta_{0i} & = k\Theta_{00}^b + \frac{k(k-1)}{2}\frac{N_i(N_i-1)}{2}\Theta_{000}^b \nonumber\\
\Theta_{00i} & = k^2 \Theta_{000}^b
\end{align}
where $k=N_i/2$ for abbreviation. With these relations, the values of $\Theta_0^b$ and $\Theta_{000}^b$ given above, and the relation $\Theta_{00}^b=4\Theta_0^b + \Theta_{000}^b$ [following (\ref{c8})], it is straightforward to see that $\Theta_{0i}$ and $\Theta_{00i}$ are indeed given by (\ref{b2gen}). One may explicitly check that $\Theta_0=0$, however, this is guaranteed by the fact that $m=1$ and the fact that the total chiral central charge vanishes.  Hence, this two-layer construction indeed realizes the demanded generating phase.

{\bf Case (B-3)}---If $N_i =0\ ({\rm mod} \ 4)$, we have $B_i = \mathbb Z_{2N_i} \times \mathbb Z_2$. There are two generating phases. According to Table \ref{tab1}, the first generating phases is described by
\begin{equation}
(\Theta_0, \Theta_{i}, \Theta_{0i}, \Theta_{00i})= (0, \pi/N_i, \pi, 0) \label{b3gen1}
\end{equation}
and the second is described by
\begin{equation}
(\Theta_0, \Theta_{i}, \Theta_{0i}, \Theta_{00i})= (0, 0,\lambda \pi, \pi) \label{monb3-2}
\end{equation}
where $\lambda=1$ when $N_i =4 \modulo{8}$ and $\lambda=2$ when $N_i= 0 \modulo{8}$. As before, all other components are determined by $\Theta_0, \Theta_{i}, \Theta_{0i}, \Theta_{00i}$ through the constraints on topological invariants. Both phases are beyond the BSPT-embedded phases.

The model for the first generating phase can be constructed in a similar way as in case (B-2).  We consider a two-layer construction. Layer $a$ is a charge-2 superconductor with $\Theta_0^a=-\pi/4$, $\Theta_{000}^a =0$, and $c^a=-1$. Since $N_i=0\modulo{4}$, we choose layer $b$ to be a charge-$N_i$ superconductor, such that $\Theta_0^b = \pi/N_i$, $\Theta_{000}^b=0$ and $c^b=1$. Following the same argument as in the case (B-2), we obtain that
\begin{align}
\Theta_i & = \Theta_0^b =  \frac{\pi}{N_i} \nonumber\\
\Theta_{0i}& = N_i\Theta^b_{0}  = \pi \nonumber\\
\Theta_{00i} & = \Theta_{000}^b=0
\end{align}
In addition, one can show that $\Theta_0=0$. Accordingly, this model indeed realizes the first generating phase characterized by (\ref{b3gen1}).

To construct models for the second generating phase, we consider a three-layer construction. The three layers are charge-$2$, charge-$N_i$ and charge-$2$ superconductors respectively. They are characterized by
\begin{align*}
\Theta_0^a & =\frac{\pi}{8}\left(\frac{N_i^2}{4}-1\right), \quad \Theta_{000}^a =\pi, \quad c^a =\frac{N_i^2}{8} -\frac{1}{2} \\
\Theta_{0}^b &= -\frac{N_i\pi}{8}, \quad \Theta_{000}^b=0, \quad c^b=-\frac{N_i^2}{8} \\
\Theta_{0}^c & = \frac{\pi}{8}, \quad \Theta_{000}^c=\pi, \quad c^c=\frac{1}{2}
\end{align*}
One can check with Eqs.~(\ref{2msc}) and (\ref{2msc-c}) that these values of topological invariants are legitimate.

Let us check the symmetry of this three-layer model. Let $F_a, F_b, F_c$ be the fermion number operators of each layer. One can see that the whole system has a symmetry $\mathbb Z_2^f\times\mathbb Z_{N_i} \times \mathbb Z_2$, generated respectively by the operators
\begin{align}
P_f &= (-1)^{ F_a + F_b+ F_c} \nonumber\\
g_i &= e^{i2\pi F_b/N_i} (-1)^{ F_c} \nonumber\\
\tilde g &=  (-1)^{ F_c}  \label{monb3-1}
\end{align}
We have chosen the generators in such a way that the symmetry group is of the form (\ref{group1}). This symmetry is larger than the demanded $\mathbb Z_2\times\mathbb Z_{N_i}$ symmetry. One can just ignore the additional $\mathbb Z_2$ symmetry or break it if one wishes.

Now we would like to compute the topological invariants $\Theta_0, \Theta_{i}, \Theta_{0i}$ and $\Theta_{00i}$, to see if they are given by (\ref{monb3-2}). According to (\ref{monb3-1}) and the correspondence between group elements and gauge flux, we see that after gauge the symmetry, the type-$i$ unit flux is composed out of a unit flux from layer $b$ and  a unit flux from layer $c$. Also, the type-0 unit flux is composed out of a unit flux from layer $a$, $N_i/2$ times of unit flux from layer $b$, and a unit flux from layer $c$.  With this picture in mind, we find that
\begin{align}
\Theta_0 & =0 \nonumber\\
\Theta_{i}  &= \Theta_{0}^b + N_i \Theta_0^c = 0, \nonumber\\
\Theta_{0i} &= N_i\Theta_{0}^b + \frac{N_i}{2}(4\Theta_0^c + \Theta_{000}^c) = \frac{3\pi}{4}N_i  \nonumber\\
\Theta_{00i}&= \Theta_{000}^c = \pi
\end{align}
It is straightforward to check that the above values of topological invariants agree with Eq.~(\ref{monb3-2}). Hence, this three-layer construction realizes the second generating phase of this case.

\subsubsection{$m$ being even}

\label{sec:con_bi_even}

Next, we consider even $m$. According to the classification in Table \ref{tab1}, we have
\begin{align}
B_i & = \left\{
\begin{array}{ll}
\vspace{3pt}
\mathbb Z_{N_i} \times\mathbb Z_{N_{0i}}, & \text{if $N_i$ is odd}\\
\mathbb Z_{2N_i} \times\mathbb Z_{N_{0i}/2}, & \text{if $N_i$ is even}
\end{array}
\right.
\end{align}
Below we construct models for the generating phases associated with the above $B_i$ groups, fro the reduced symmetry group $G_f = \mathbb Z_{2m}^f\times \mathbb Z_{N_i}$.

{\bf Case (B-4)}---When $N_i$ is odd, we have $B_i = \mathbb Z_{N_i} \times\mathbb Z_{N_{0i}}$. According to Table \ref{tab1}, the two generating phases are described by
\begin{equation}
(\Theta_0, \Theta_{i}, \Theta_{0i}, \Theta_{00i})= (0, 2\pi/N_i, 0, 0), \ (0, 0, 2\pi/N_{0i}, 0) \label{b4gen}
\end{equation}
All other components of the topological invariants are determined by the ones listed above. It is obvious that the first generating phase can be realized by the BSPT-embedded models. Moreover, the second generating phase can also be realized by BSPT-embedded models. To see this, we notice that $N_{0i} = \gcd (2m, N_i) = \gcd (m, N_i) = \bar N_{0i}$, and $N^{0i} = {\rm lcm}(2m, N_i) =2\ {\rm lcm}(m, N_i)= 2\bar N^{0i}$. Therefore, according to Eq.~(\ref{bspt}), BSPT-embedded models can realize phases characterized by
\begin{equation*}
\Theta_{0i} = \frac{2\pi}{N_{0i}} 2p_{0i}
\end{equation*}
where $p_{0i}$ is some integer. Since  $N_{0i}$ is odd, it is possible to find an integer $p_{0i}$ such that $2p_{0i} =1 ({\rm mod } \  N_{0i})$. Hence, the second generating phase can also be realized by BSPT-embedded models.

{\bf Case (B-5)}---When $N_i$ is even, we have $B_i = \mathbb Z_{2N_i} \times\mathbb Z_{N_{0i}/2}$. According to Table \ref{tab1}, the two generating phases are described respectively by
\begin{equation}
(\Theta_0, \Theta_{i}, \Theta_{0i}, \Theta_{00i})= (0, \pi/N_i, 2\pi/N_{0i}, 0) \label{b5gen1}
\end{equation}
and
\begin{equation}
(\Theta_0, \Theta_{i}, \Theta_{0i}, \Theta_{00i})= (0, 0, 4\pi/N_{0i}, 0) \label{eqsec}
\end{equation}
As before, $\Theta_0=0$ is enforced so that these phases are pure generating phases associated with $B_i$ (i.e., not a mixture of generating phases of $A$ and $B_i$). All remaining components of the topological invariants are determined by the ones listed above.

First, we show that the second generating phase can be realized by the BSPT-embedded models. Let $m = 2^ar$ with $r$ being odd, and $N_i = 2^b t$ with $t$ also being odd. If $b > a$, we have
\begin{align}
N_{0i} &= {\rm gcd}(2m, N_i) = 2^{a+1}{\rm gcd}(r,t) = 2 \bar N_{0i} \nonumber\\
N^{0i} &= \lcm(2m, N_i) = 2^b\lcm (r,t) = \bar N^{0i} \nonumber
\end{align}
On the other hand, if $b\le a$, we have
\begin{align}
N_{0i} & =  2^{b}\ {\rm gcd}(r,t) = \bar N_{0i} \nonumber\\
N^{0i} & = 2^{a+1}\lcm (r,t) = 2\bar N^{0i} \nonumber
\end{align}
With these relations, we can now directly compare (\ref{eqsec}) and (\ref{bspt}). It is straightforward to see that for both $b> a$ and $b\le a $, the generating phase characterized by (\ref{eqsec}) can be realized by the BSPT-embedded models.

In contrast, the first generating phase is beyond BSPT-embedded models. We now construct a model for this phase. It is a two-layer construction. The first layer $a$ is a charge-$2m$ superconductor, and the second layer $b$ is a charge-$mN_i$ superconductor. The two layers are characterized by
\begin{align*}
\Theta_{0}^a& = -\frac{\pi}{2m}, \quad \Theta_{000}^a=0, \quad c^a=-1 \\
\Theta_0^b & = \frac{\pi}{mN_i}, \quad \Theta_{000}^b=0, \quad c^b=1
\end{align*}
One can check with Eqs.~(\ref{2msc}) and (\ref{2msc-c}) that the above choices are legitimate.

Let us check the symmetry of this two-layer model. Let $F_a,F_b$ be the fermion number operators of each layer respectively. This two-layer system has a $\mathbb Z_{2m}^f\times \mathbb Z_{mN_i}$ symmetry, generated by
\begin{align}
g_0 & = e^{i\pi(F_a+F_b)/m} \nonumber\\
g_i & = e^{i2\pi F_b/(mN_i)}  \label{b5basis}
\end{align}
The fermion parity $P_f$ is equal to $g_0^m$. Note that the total symmetry is larger that $\mathbb Z_{2m}^f\times\mathbb Z_{N_i}$. One may choose to break  $\mathbb Z_{2m}^f\times \mathbb Z_{mN_i}$ down to $\mathbb Z_{2m}^f\times\mathbb Z_{N_i}$ by adding a weak perturbation that does not close the energy gap. More conveniently, one can just ignore the enlarged part of the symmetry.

Now we would like to compute $\Theta_0, \Theta_{i}, \Theta_{0i}, \Theta_{00i}$, and show that they are indeed given by (\ref{b5gen1}). First of all, we notice that since $m$ and $N_i$ are even, all vortices are Abelian if we gauge the full $\mathbb Z_{2m}\times\mathbb Z_{mN_i}$ symmetry. With this in mind, we notice that the topological spin of a vortex that carries the unit flux from layer $b$ is given by $\frac{\Theta_0^b}{mN_i}+\frac{2\pi}{mN_i}\times\text{integer}$. According to (\ref{b5basis}), the type-$i$ unit flux associated with the $\mathbb Z_{2m}^f\times\mathbb Z_{N_i}$ symmetry is $m$ times of the unit flux of layer $b$. Accordingly, we the topological spin of a vortex $\xi_i$ that carries the type-$i$ unit flux is given by
\begin{align}
\theta_{\xi_i} &  = m^2 \left(\frac{\pi}{(mN_i)^2}+\frac{2\pi}{mN_i}\times\text{integer}\right) \nonumber\\
 & = \frac{\pi}{N_i^2} + \frac{2\pi m}{N_i}\times\text{integer}
\end{align}
Then, the topological invariant $\Theta_i$ is given by
\begin{align}
\Theta_i   = N_i \theta_{\xi_i} = \frac{\pi}{N_i}
\end{align}
which agrees with (\ref{b5gen1}). At the same time, it is not hard to show that $\Theta_{0i} = 2\pi/N_{0i}$, $\Theta_{00i}=0 $, and $\Theta_0 = 0$. Hence, we have constructed a model for the first generating phase.

\subsection{Generating phases of $C_{ij}$}
\label{sec:con_cij}

In this subsection, we construct models for the generating phases associated with the $C_{ij}$ component in $H_{\rm stack}$ for fixed indices $i,j$ with $i\neq j$. As discussed in Sec.~\ref{procedure}, it is enough to consider the reduced group $G_f = \mathbb Z_{2m}^f\times \mathbb Z_{N_i}\times \mathbb Z_{N_{j}}$.

\subsubsection{$m$ being odd}
\label{sec:con_cij_odd}

We first consider odd $m$. As argued before, it is enough to consider $m=1$. That is, we consider group $G_f = \mathbb Z_2^f \times \mathbb Z_{N_i} \times \mathbb Z_{N_j}$.  Accordingly to Table \ref{tab1}, the component $C_{ij}$ is given by
\begin{align}
C_{ij} & = \left\{
\begin{array}{ll}
\vspace{3pt}
\mathbb Z_{N_{ij}}, & \text{if either $N_i$  or $N_j$ is odd} \\
\vspace{3pt}
\mathbb Z_{2N_{ij}},  & \text{if both $N_i, N_j = 2 \modulo{4}$}\\
\mathbb Z_{N_{ij}}\times \mathbb Z_2, & \text{otherwise}
\end{array}
\right.
\end{align}
Below we construct models for the generating phases in each case.

{\bf Case (C-1)}---When  either $N_i$ or $N_j$ is odd, we have $C_{ij} = \mathbb Z_{N_{ij}}$. According to the classification in Table \ref{tab1}, the generating phase is described by
\begin{equation}
(\Theta_{ij}, \Theta_{0ij})= ( 2\pi/N_{ij}, 0)\label{c1gen}
\end{equation}
In addition, we require that $\Theta_0$, $\Theta_i$, $\Theta_{0i}$, $\Theta_{00i}$, $\Theta_j$, $\Theta_{0j}$, $\Theta_{00j}$ all vanish, so that no phases associated with $B_i$ and $A$ in $H_{\rm stack}$ are mixed in. Checking with Eq.~(\ref{bspt}), one can see that this phase can be realized by the BSPT-embedded models.

{\bf Case (C-2)}---When both $N_i$ and $N_j$ are odd multiples of 2, we have $C_{ij} = \mathbb Z_{2N_{ij}}$. The generating phase is described by
\begin{equation}
(\Theta_{ij}, \Theta_{0ij}) = (\pi/N_{ij}, \pi) \label{c2gen}
\end{equation}
As before, we require that $\Theta_0, \Theta_i, \Theta_{0i}, \Theta_{00i}, \Theta_j, \Theta_{0j}, \Theta_{00j}$ all vanish. This phase is beyond BSPT-embedded models.

We consider a four-layer construction to realize this generating phase. The four layers are charge-2, charge-$N_i$, charge-$N_j$ and charge-$2$ superconductors respectively. We denote the four layers as $a,b,c,d$ respectively. The four layers are chosen to have the following values of topological invariants and chiral central charge:
\begin{align}
\Theta_0^a & = -\frac{\pi}{8}, \ \Theta_{000}^a = \pi, \ c^a = -\frac{1}{2} \nonumber\\
\Theta_0^b & = \frac{N_i\pi}{16}, \ \Theta_{000}^b = \pi, \ c^b = \frac{1}{2} \nonumber\\
\Theta_0^c & = \frac{N_j\pi}{16}, \ \Theta_{000}^c = \pi, \ c^c = \frac{1}{2} \nonumber\\
\Theta_0^d & = -\frac{\pi}{8}, \ \Theta_{000}^d = \pi, \ c^d = -\frac{1}{2}
\end{align}
The total chiral central charge is 0, thereby the system is nonchiral.

This system has a total symmetry $\mathbb Z_2^f \times \mathbb Z_{N_i} \times \mathbb Z_{N_j}\times \mathbb Z_2$. If we let $F_a, F_b, F_c, F_d$ be the fermion number operators of each layer, the symmetry is generated by the following operators
\begin{align}
P_f &= (-1)^{F_a+F_b+F_c+ F_d} \nonumber\\
g_i & =  e^{i 2\pi F_b/N_i}(-1)^{F_d} \nonumber\\
g_j & = e^{i2\pi F_c/N_j}(-1)^{F_d} \nonumber\\
\tilde g & = (-1)^{F_d}\label{c2generator}
\end{align}
Again, we have used the same trick as before by enlarging the symmetry $G_f$ to include an auxiliary $\mathbb Z_2$. One may break the auxiliary $\mathbb Z_2$ symmetry or just ignore it.

We now calculate the values of the topological invariants $\Theta_{ij}$ and $\Theta_{0ij}$ associated with the type-0, type-$i$ and type-$j$ unit flux. With some algebra, it is not hard to check that
\begin{align}
\Theta_{ij} & = \frac{N^{ij}}{2}(4\Theta_{0}^d + \Theta_{000}^d) =\frac{\pi}{4}N^{ij}\nonumber\\
\Theta_{0ij} &  = \Theta_{000}^d = \pi \label{case-c2}
\end{align}
At the same time, one can show that $\Theta_0$, $\Theta_i$, $\Theta_{0i}$, $\Theta_{00i}$, $\Theta_j$, $\Theta_{0j}$, $\Theta_{00j}$ all vanish.

We have not achieved our goal yet. Next, we stack $x$ layers of this four-layer system with $y$ layers of a BSPT-embedded model with the same symmetry, where $x, y$ are two integers to be determined. We choose the BSPT-embedded model to have $\Theta_{ij} = 2\pi/N_{ij}$, and all other invariants vanish. After stacking, the topological invariants of the stacked system  are
\begin{equation}
\Theta_{ij} = \frac{\pi}{4} N^{ij} x + \frac{2\pi}{N_{ij}} y, \quad \Theta_{0ij} = \pi x
\end{equation}
Let us write $N_i=2k_i$ and $N_j=2k_j$, where $k_i, k_j$ are odd integers. Then, we have
\begin{equation}
\Theta_{ij} = \frac{\pi}{2k_{ij}}(k_ik_j x + 2 y)
\end{equation}
where $k_{ij} = \gcd(k_i,k_j)$. Since $k_ik_j$ is odd, there always exist non-negative $x,y$ such that $k_ik_j x + 2 y =1$.  For such $x$ and $y$, we obtain that $\Theta_{ij}=\pi/2k_{ij} = \pi/N_{ij}$. Obviously, to satisfy $k_ik_j x + 2 y =1$,  $x$ must be odd. Hence, $\Theta_{0ij} = x\pi = \pi$. Therefore, we have constructed a model that realizes the  generating phase characterized by (\ref{c2gen}).

{\bf Case (C-3)}---If either $N_i$ or $N_j$ is a multiple of 4, we have $C_{ij} = \mathbb Z_{N_{ij}}\times\mathbb Z_2$. According to the classification in Table \ref{tab1}, the two generating phases are described by
\begin{equation}
(\Theta_{ij}, \Theta_{0ij}) = (2\pi/N_{ij}, 0) \label{c3gen1}
\end{equation}
and
\begin{equation}
(\Theta_{ij}, \Theta_{0ij}) =  (0, \pi) \label{c3gen2}
\end{equation}
respectively. As before, we require that $\Theta_0$, $\Theta_i$, $\Theta_{0i}$, $\Theta_{00i}$, $\Theta_j$, $\Theta_{0j}$, $\Theta_{00j}$ all vanish. Other components are determined by the ones listed out through the constraints on topological invariants. Checking with Eq.~(\ref{bspt}), one can see that the first generating phase can be realized by the BSPT-embedded models.

The second generating phase is beyond BSPT-embedded models. It can be constructed using a similar four-layer model as in Case (C-2). The four layers are charge-$2$, charge-$N_i$, charge-$N_j$ and charge-$2$ superconductors respectively, with the topological invariants and chiral central charges given by
\begin{align}
\Theta_0^a & = \frac{\pi}{8}, \ \Theta_{000}^a = \pi, \ c^a = \frac{1}{2} \nonumber\\
\Theta_0^b & = 0, \ \Theta_{000}^b = 0, \ c^b = 0 \nonumber\\
\Theta_0^c & = 0, \ \Theta_{000}^c = 0, \ c^c = 0 \nonumber\\
\Theta_0^d & = -\frac{\pi}{8}, \ \Theta_{000}^d = \pi, \ c^d = -\frac{1}{2}
\end{align}
Again, this four-layer model has an enlarged $\mathbb Z_2^f\times\mathbb Z_{N_i}\times\mathbb Z_{N_j}\times\mathbb Z_2$ symmetry, with the generators given by (\ref{c2generator}). We find that in this four-layer construction, the topological invariants are given by
\begin{align}
\Theta_{ij} &= \frac{\pi}{4}N^{ij}, \quad \Theta_{0ij} = \pi \nonumber\\
\Theta_{00i}& = \Theta_{00j} = \pi, \nonumber \\
\Theta_{0i} &= \frac{\pi}{4}N_i, \quad \Theta_{0j} = \frac{\pi}{4}N_j\nonumber \\
\Theta_{i} & = -\frac{\pi}{8}N_i,\quad \Theta_j = -\frac{\pi}{8}N_j\nonumber\\
\Theta_0& =0 \label{4layer}
\end{align}

Next, we stack other FSPT phases onto this four-layer model. We observe that the following three FSPT phases exist: (i) from the results of Sec.~\ref{sec:con_bi_odd}, one can show that that as long as $N_i$ is even, there always exist FSPT phases with
\begin{align}
(\Theta_0, \Theta_{i}, \Theta_{0i}, \Theta_{00i})& = (0, N_i\pi/8, -\pi N_i/4, \pi) \label{xx1}
\end{align}
and with vanishing $\Theta_{j}$, $\Theta_{0j}$, $\Theta_{00j}$, $\Theta_{ij}$ and $\Theta_{0ij}$; (ii) Similarly, one can show that there exist FSPT phases with
\begin{align}
(\Theta_0, \Theta_{j}, \Theta_{0j}, \Theta_{00j}) &=  (0, N_j\pi/8, -\pi N_j/4, \pi) \label{xx2}
\end{align}
and with vanishing $\Theta_{i}$, $\Theta_{0i}$, $\Theta_{00i}$, $\Theta_{ij}$ and $\Theta_{0ij}$; and (iii) There exist BSPT models characterized by
\begin{equation}
\Theta_{ij}=\pi, \quad \Theta_{0ij}=0 \label{xx3}
\end{equation}
and with vanishing $\Theta_0$, $\Theta_i$, $\Theta_{0i}$, $\Theta_{00i}$, $\Theta_j$, $\Theta_{0j}$, $\Theta_{00j}$. Stacking the phases characterized by (\ref{xx1}) and (\ref{xx2}) to the above four-layer system (\ref{4layer}),  we realize a phase with $\Theta_{ij} = \pi N^{ij}/4 $, $\Theta_{0ij}=\pi$, and all other invariants vanish. If $N^{ij}$ is an even multiple of $4$, this phase is already the second generating phase (\ref{c3gen2}). If $N^{ij}$ is an odd multiple of 4, we further stack the system with the phase characterized by (\ref{xx3}), and then we obtain the second generating phase.

\subsubsection{$m$ being even}
\label{sec:con_cij_even}

Finally, we consider the case that $m$ is even for $C_{ij}$. According to Table \ref{tab1}, the classification in this case is given by
\begin{align}
C_{ij} & = \mathbb Z_{N_{ij}}\times \mathbb Z_{N_{0ij}}
\end{align}
Below we construct models for this case. We consider the reduced symmetry group $G_f = \mathbb Z_{2m}^f\times\mathbb Z_{N_i}\times \mathbb Z_{N_j}$.

{\bf Case (C-4)}---According to the classification in Table \ref{tab1}, the two generating phases are described by
\begin{equation}
(\Theta_{ij}, \Theta_{0ij}) = (2\pi/N_{ij}, 0) \label{c4gen1}
\end{equation}
and
\begin{equation}
(\Theta_{ij}, \Theta_{0ij}) = (0, 2\pi/N_{0ij}) \label{c4gen2}
\end{equation}
respectively. The components $\Theta_0$, $\Theta_i$, $\Theta_{0i}$, $\Theta_{00i}$, $\Theta_j$, $\Theta_{0j}$, $\Theta_{00j}$ are enforced to be 0. Other components are determined by the ones listed out. Checking with Eq.~(\ref{bspt}), we find that the first generating phase can be realized by a BSPT-embedded model.

The second generating phase may or may not be realized by BSPT-embedded models, depending on whether $N_{0ij}$ and $\bar N_{0ij}$ are equal. Remind that $N_{0ij}=\gcd(N_0, N_i, N_j)$ and $\bar N_{0ij}=\gcd(m, N_i, N_j)$. If $N_{0ij}= \bar N_{0ij}$, BSPT-embedded models can realize the second generating phase. Otherwise, they cannot. To find when $N_{0ij}$ and $\bar N_{0ij}$ are not equal, let us denote $m = 2^a r$ and $N_{ij} = 2^b s$, where $r, s$ are odd integers. Since $m$ is even, $a\ge 1$. It is easy to check that if $b\ge a+1$, we find $N_{0ij} = 2 \bar N_{0ij}$; otherwise, $N_{0ij} = \bar N_{0ij}$.

In the case that $N_{0ij}$ is not equal to $\bar N_{0ij}$, we are not able to construct models through layer construction based on free-fermion and BSPT-embedded models. We will discuss these phases in detail in Sec.~\ref{sec:exception}. The simplest symmetry to support these FSPT phases is $\mathbb Z_4^f\times\mathbb Z_4\times\mathbb Z_4$ symmetry. We will argue in Sec.~\ref{sec:exception} that these FSPT phases actually belong to the third kind of FSPT phases discussed in the introduction.





\section{Examples}

\label{sec:examples}

In the above section, we have focused on the topological invariants of the models that we construct. It is worth to analyzing some examples in more detail. In this section, we discuss the full excitation spectrum and their braiding statistics for some simple symmetry groups. We are interested in examples that are beyond the BSPT-embedded models\footnote{The topological order of gauged BSPT-embedded models with $G_f$ in the form $\mathbb Z_2^f\times G$ is simply a stack of the toric code and the topological order of gauged BSPT phase with symmetry $G$. However, the topological order of gauged BSPT-embedded models with symmetries beyond the form $\mathbb Z_2^f\times G$ is more complicated.}, i.e., examples that can be thought of as {\it intrinsically fermionic}. All examples that we discuss below can be realized by free fermions.

The simplest example with no realization through BSPT-embedding is associated with $\mathbb Z_2^f\times\mathbb Z_2$ symmetry [case (B-2)]. A detailed analysis for this example was given in Ref.~\onlinecite{gu14b}, so we do not discuss it here.


\subsection{$G_f=\mathbb Z_4^f\times \mathbb Z_2$}

Our first example is $\mathbb Z_4^f\times \mathbb Z_2$ symmetry, which is the simplest symmetry for case (B-5). According to Eq.~(\ref{stackgroup}) and Table \ref{tab1}, the stacking group of FSPT phases is $H_{\rm stack} = \mathbb Z_4$, and the generating phase is characterized by
\begin{equation}
(\Theta_0,\Theta_{1}, \Theta_{01}, \Theta_{001}) = (0, \pi/2, \pi, 0) \label{ex-a1}
\end{equation}
According to Sec.~\ref{sec:con_bi_even}, the generating phase can be realized in a two-layer construction. The first layer is a charge-4 superconductor with chiral central charge $c=-1$, and the second layer is a charge-4 superconductor with $c=1$. The system has a $\mathbb Z_4^f\times\mathbb Z_4$ symmetry, so we eventually break it down to $\mathbb Z_4^f\times\mathbb Z_2$ symmetry. Below, we study the excitation spectrum and braiding statistics in the generating phase. 

To do that, we first argue that the braiding statistics is Abelian, i.e., all charge and vortex excitations are Abelian anyons. We notice that charge-4 superconductors only admit Abelian anyons after gauging the symmetry.\cite{wangcj16} So, the two-layer system also admits Abelian anyons only, if we gauge the full $\mathbb Z_4^f\times\mathbb Z_4$ symmetry. Next, we understand that breaking $\mathbb Z_4^f\times\mathbb Z_4$ to $\mathbb Z_4^f\times\mathbb Z_2$ {\it before} we gauge the symmetry is equivalent to driving a Higgs transition {\it after} we gauge the symmetry. Driving a Higgs transition from  $\mathbb Z_4^f\times\mathbb Z_4$ gauge theory to $\mathbb Z_4^f\times\mathbb Z_2$ gauge theory  is done by condensing the $(0,2)$ bosonic charge  in $\mathbb Z_4^f\times\mathbb Z_4$ gauge theory. This boson condensation does not change the fact that braiding statistics is Abelian.

Having understood that braiding statistics is Abelian, the full excitation spectrum becomes clear. First of all, there are 8 distinct charge excitations, labeled by $q=(q_0,q_1)$, with $q_0=0,1,2,3$ and $q_1 = 0, 1$.  Second, for each gauge flux $\phi=(\pi k_0/2, \pi k_1)$ with $k_0=0,1,2,3$ and $k_1=0,1$, we can obtain 8 distinct vortex excitations. The vortices with the same gauge flux differ by charge attachment. Since braiding statistics is Abelian, attaching different charges to a vortex always produces different vortices. Therefore, there are 64 excitations in total.

The full braiding statistics data can be deduced from the values of topological invariants in (\ref{ex-a1}), together with the exchange statistics (\ref{exchange}) of charges and the Aharonov-Bohm law (\ref{abphase}). First, consider vortices  $\xi_0, \xi_1$, which carry unit flux $(\pi/2, 0)$ and $(0,\pi)$ respectively. Since braiding is Abelian and according to the definitions of topological invariants, we have that the exchange statistics $\theta_{\xi_0} = \Theta_0/4 + \pi p_0/2$, $\theta_{\xi_1} = \Theta_1/2 +\pi p_1$, and mutual statistics $\theta_{\xi_0, \xi_1} = \Theta_{01}/4 + \pi p_{01}/2$, where $p_0, p_1, p_{01}$ are some integers. Following Eq.~(\ref{ex-a1}), it is easy to show that, through appropriate charge attachments to $\xi_0$ and $\xi_1$, one can find two reference vortices $\hat \xi_0$ and $\hat \xi_1$ such that \begin{equation}
\theta_{\hat \xi_0} = 0,  \quad \theta_{\hat \xi_1} = \frac{\pi}{4}, \quad \theta_{\hat\xi_0,\hat \xi_1} = \frac{\pi}{4} \label{ex-a2}
\end{equation}
With the two reference vortices, a general excitation can be obtained by fusing $k_0$ copies of $\hat \xi_0$, $k_1$ copies of $\hat\xi_1$, and a charge $q$. We denote the excitation as $(k,q)$, where $k= (k_0, k_1)$ and $q=(q_0,q_1)$. The full braiding statistics can be obtained through (\ref{abphase}), (\ref{exchange}) and (\ref{ex-a2}) using the linearity of Abelian statistics:
\begin{align}
\theta_{x, y+y'} & = \theta_{x,y} + \theta_{x,y'} \nonumber\\
\theta_{x +y } &= \theta_{x}+\theta_y + \theta_{x,y}
\end{align}
where $x, y, y'$ are any Abelian anyons, and $\theta_x$ is the exchange statistics (topological spin) of $x$, and $\theta_{x,y}$ is the mutual statistics between $x$ and $y$. In addition, the mutual statistics $\theta_{x,x}=2\theta_x$. Using these relations, we find that the exchange statistics of $(k,q)$ is given by
\begin{equation}
\theta_{(k,q)} = \frac{\pi}{4}k_1^2 + \frac{\pi}{4} k_0k_1 + \frac{\pi}{2}k_0q_0 + \pi k_1 q_1 + \pi q_0
\end{equation}
and the mutual statistics between $(k,q)$ and $(k',q')$ is given by
\begin{align}
\theta_{(k,q),(k',q')} = & \frac{\pi}{4}(k_0k'_1 + k_1 k'_0 +2k_1k_1') \nonumber\\
 &+ \frac{\pi}{2}(k_0q_0'+k_0' q_0) + \pi (k_1 q_1'+ k_1' q_1)
\end{align}


\subsection{$G_f= \mathbb Z_2^f\times \mathbb Z_4$}
\label{sec:example2}
Next, we consider the symmetry $\mathbb Z_2^f\times \mathbb Z_4$, which is the simplest example for case (B-3). According to our classification, the stacking group for FSPT phases with this symmetry is $H_{\rm stack} = \mathbb Z_{8}\times \mathbb Z_2$.

The generating phase for the $\mathbb Z_8$ component is described by the topological invariants
\begin{equation}
(\Theta_0,\Theta_{1}, \Theta_{01}, \Theta_{001})= (0,\pi/4, \pi, 0) \nonumber
\end{equation}
According to Sec.~\ref{sec:con_bi_odd}, this phase can be realized in a two-layer construction: the first layer is a regular charge-2 superconductor with chiral central charge $c=-1$ and the second layer is a charge-4 superconductor with $c=1$. Like the $\mathbb Z_4^f\times\mathbb Z_2$ example,  this phase supports only Abelian anyons after gauging the symmetries. Hence, one can go through a similar argument to obtain the full excitation spectrum and full set of braiding statistics data. We do not repeat the argument here.

Unlike the above phase, the generating phase for the $\mathbb Z_2$ component in $H_{\rm stack}$ supports non-Abelian statistics. This phase is characterized by the topological invariants
\begin{equation}
(\Theta_0,\Theta_{1}, \Theta_{01}, \Theta_{001})= (0,0, \pi, \pi) \nonumber
\end{equation}
The braiding statistics must be non-Abelian because $\Theta_{001}\neq 0$. The fact that nonvanishing $\Theta_{001}$ implies non-Abelian statistics follows from the definition of $\Theta_{001}$. According to Sec.~\ref{sec:con_bi_odd}, this phase is realized in a three-layer construction: layer $a$ is a charge-2 superconductor with chiral central charge $c^a=3/2$, layer $b$ is a charge-4 superconductor with $c^b=-2$, and layer $c$ is a $p_x+ip_y$ superconductor with $c^c=1/2$. The three-layer system has a total symmetry $\mathbb Z_2^f\times\mathbb Z_4\times\mathbb Z_2$. One can break it down to $\mathbb Z_2^f\times\mathbb Z_4$, or just ignore the additional $\mathbb Z_2$.

To obtain the excitation spectrum and braiding statistics in the FSPT system after gauging the $\mathbb Z_2^f\times\mathbb Z_4$ symmetry, we play the following trick. We first gauge the full $\mathbb Z_2^f\times\mathbb Z_4\times\mathbb Z_2$ symmetry in the three-layer model. The excitation spectrum and braiding statistics of the $\mathbb Z_2^f\times\mathbb Z_4\times\mathbb Z_2$  gauge theory is just a simple stacking of the anyons from each layer, and the excitations in each layer are known. Then, we drive a Higgs transition in the $\mathbb Z_2^f\times\mathbb Z_4\times\mathbb Z_2$ gauge theory by condensing the unit charge associated with the $\mathbb Z_2$ gauge symmetry. In this way, we can eventually obtain excitation spectrum and braiding statistics of the gauged $\mathbb Z_2^f \times\mathbb Z_4$ FSPT phase.

Let us first look at the excitation spectrum and braiding statistics in each layer. The full braiding statistics data of gauged charge-$2m$ superconductors can be found in Refs.~\onlinecite{kitaev06} and \onlinecite{wangcj16}. According to these works, layer $a$ supports Ising-like excitation after gauging the symmetry. There are three excitations, $1, \psi_a, \sigma_a$. They satisfy the usual non-Abelian Ising fusion rules
\begin{equation}
\psi_a\times\psi_a=1, \ \psi_a\times\sigma_a = \sigma_a, \ \sigma_a\times\sigma_a=1+ \psi_a
\end{equation}
and the fusion between 1 and any $x$ gives rise to $x$, where $x=1,\psi_a,\sigma_a$. The anyon $\psi_a$ is the charge excitation and $\sigma_a$ is the vortex excitation of charge-2 superconductors. The anyon $\psi_a$ is a fermion, and $\sigma_a$ is a non-Abelian anyon with topological spin $\theta_{\sigma_a} = 3\pi/8$ and quantum dimension $d_{\sigma_a} = \sqrt 2$.  Layer $b$ supports Abelian statistics only. Let us denote the unit charge as $\psi_b$, and the unit vortex as $m_b$. The charge $\psi_b$ is a fermion, and the vortex $m_b$ is chosen to have a topological spin $\theta_{m_b} = -\pi/8$. A general excitation can be labeled as $m_b^x\psi_b^y$, with $x,y=0,1,2,3$. Since they are Abelian anyons, the full braiding statistics can be easily obtained.  Layer $c$ also supports Ising-like anyons, $1, \psi_c, \sigma_c$. However, $\sigma_c$ has a different topological spin from $\sigma_a$, with $\theta_{\sigma_c} = \pi/8$. Without any confusion, we do not distinguish the vacuum 1 from different layers.

With the above information, we now consider the $\mathbb Z_2^f\times\mathbb Z_4\times\mathbb Z_2$ gauge theory. The excitations are just a simple stacking of those from each layer. The total number of excitations is $3\times 16\times 3 = 144$. According to our general discussion of excitations in gauge theories in Sec.~\ref{sec:gauge}, there are 16 charges, which can be labeled as $\psi_a^x\psi_b^y\psi_c^z$ with $x,z=0,1$ and $y=0,1,2,3$. There are 15 sectors of gauge flux. Representative vortices of each flux sector are $m_b^x, \sigma_am_b^y,\sigma_cm_b^z,\sigma_a\sigma_cm_b^w$, where $x=1,2,3$ and $y,z,w=0,1,2,3$. Other vortices can be obtained by fusing (attaching) charges into the representatives.

It is worth establishing a translation between the above notation and the general notation used throughout this paper where we use $\xi_0, \xi_1,\xi_2$ to denote vortices that carry unit flux and use $(1,0,0), (0,1,0), (0,0,1)$  to denote the unit charges. According to the correspondence between group elements and gauge flux and the expression (\ref{monb3-1}) of the generators of the group $\mathbb Z_2^f\times\mathbb Z_4\times\mathbb Z_2$, we find that
\begin{equation}
\xi_0 = \sigma_a\sigma_cm_b^2, \quad \xi_1 = m_b \sigma_c, \quad \xi_2 = \sigma_c
\end{equation}
Then, by matching the Aharonov-Bohm phases between unit charges and the vortices $\xi_0,\xi_1,\xi_2$, we find that unit charges are given by
\begin{align}
(1,0,0) &= \psi_a\nonumber\\
(0,1,0) &= \psi_a\psi_b \nonumber \\
(0,0,1) &= \psi_a\psi_b^2\psi_c
\end{align}
Clearly, $(1,0,0)$ is a fermion and the other unit charges are bosons, in agreement with the expectation.

With the above properties of anyons in the $\mathbb Z_2^f\times\mathbb Z_4\times\mathbb Z_2$ gauge theory, we now derive properties of anyons in the $\mathbb Z_2^f\times\mathbb Z_4$ theory. To do that, we drive a Higgs transition by condensing the $(0,0,1)=\psi_a\psi_b^2\psi_c$ bosonic charge. Two physical consequences of the condensation are that (1) two anyons will be identified if they differ by the condensed anyon $\psi_a\psi_b^2\psi_c$ and (2) those anyons with nontrivial mutual braiding around  $\psi_a\psi_b^2\psi_c$ will be confined. (For sophisticated theory of anyon condensation in topological orders, we refer readers to Ref.~\onlinecite{bais09}.)  Accordingly, we find that there remain 8 charges,  $\psi_a^x\psi_b^y$ with $x=0,1$ and $y=0,1,2,3$. In addition, there are 7 deconfined flux sectors, which is expected for $\mathbb Z_2^f\times\mathbb Z_4$ gauge theory.  Representative vortices of each flux sector are listed as follows, where we have grouped them into three kinds:
\begin{align}
(i)\! : &\ m_b^2 \nonumber\\
(ii)\! : &\ \sigma_a m_b, \ \sigma_a m_b^3, \ \sigma_c m_b, \ \sigma_cm_b^3 \nonumber\\
(iii)\! : &\ \sigma_a\sigma_c,\ \sigma_a\sigma_cm_b^2
\end{align}
Other vortices can be obtained by attaching charges to the representative vortices. In group (i), one can find 8 vortices by attaching charges to $m_b^2$, all of which are Abelian anyons. Vortices in group (ii) are non-Abelian, and they have quantum dimension $\sqrt 2$.  One can find 4 distinct vortices in each flux sector in this group. For example, in the sector of $\sigma_am_b$, we have four vortices: $\sigma_am_b$, $\sigma_am_b\psi_b$, $\sigma_am_b\psi_b^2$, and $\sigma_am_b\psi_b^3$.  One may wonder that $\sigma_am_b\psi_c$ is a distinct vortex. However, because of the condensation of $\psi_a\psi_b^2\psi_c$, we have the following identifications
\begin{equation}
\sigma_am_b\psi_c \sim\sigma_am_b\psi_c\times\psi_a\psi_b^2\psi_c \sim \sigma_am_b\psi_b^2
\end{equation}
where we have used the fusion rules $\psi_c\times\psi_c = 1$ and $\sigma_a\psi_a=\sigma_a$. Vortices in group (iii) have quantum dimension 2, and for each flux sector, one can find 2 distinct vortices. Hence, we find 8 charges, $8+4\times 4 + 2\times 2 = 28$ vortices, and in total $36$ anyons. The braiding statistics and fusion rules of the anyons follow those before the condensation. One may explicitly check that the topological invariants $\Theta_0,\Theta_1, \Theta_{01}$
and $\Theta_{001}$ acquire the demanded values.


\subsection{$G_f=\mathbb Z_2^f\times \mathbb Z_2\times \mathbb Z_2$}

The third example is $G_f=\mathbb Z_2^f \times \mathbb Z_2\times \mathbb Z_2$ symmetry. The stacking group of FSPT phases with this symmetry is given by $H_{\rm stack} = \mathbb Z_8 \times\mathbb Z_8\times \mathbb Z_4$. The two $\mathbb Z_8$ components correspond to case (B-2) in Sec.~\ref{sec:con_bi_odd}. These FSPT phases are protected by the two ($\mathbb Z_2^f$, $\mathbb Z_2$) pairs in $G_f$ respectively. The physics there are discussed in Ref.~\onlinecite{gu14b}, so we do not repeat the discussion here. The $\mathbb Z_4$ component in $H_{\rm stack}$ requires protection from the whole $\mathbb Z_2^f \times \mathbb Z_2\times \mathbb Z_2$ symmetry. Below we study properties of the generating phase of the $\mathbb Z_4$ component. This phase is the simplest example of case (C-2).

According to Table \ref{tab1}, the generating phase of the $\mathbb Z_4$ component in $H_{\rm stack}$ is characterized by the topological invariants
\begin{align}
(\Theta_{12}, \Theta_{012})=(\pi/2, \pi)\nonumber
\end{align}
All other independent topological invariants vanish,
\begin{equation}
\Theta_0=\Theta_1=\Theta_{01}=\Theta_{001}=\Theta_2=\Theta_{02}=\Theta_{002}=0 \nonumber
\end{equation}
Since $\Theta_{012}$ does not vanish, the gauged FSPT system must support non-Abelian statistics. According to Sec.~\ref{sec:con_cij_odd}, this phase is realized by stacking four layers of regular charge-2 superconductors, with chiral central charges being $-\frac{1}{2}, \frac{1}{2}, \frac{1}{2}, -\frac{1}{2}$ respectively. The four-layer system has an enlarged $\mathbb Z_2^f\times\mathbb Z_2\times\mathbb Z_2\times\mathbb Z_2$ symmetry, so we eventually break it down to demanded $\mathbb Z_2^f\times\mathbb Z_2\times\mathbb Z_2$ symmetry, or just ignore the last $\mathbb Z_2$.

To analyze the excitation spectrum and the braiding statistics for the gauged system, we play the same trick as in Sec.~\ref{sec:example2}. We first consider the excitation spectrum in the gauged system with the full $\mathbb Z_2^f\times\mathbb Z_2\times\mathbb Z_2\times\mathbb Z_2$ gauge symmetry, which is easy to obtain. Then, we drive a Higgs transition by condensing the charge excitation corresponding to the last $\mathbb Z_2$. In this way, we obtain the excitation spectrum and their braiding statistics for the gauged $\mathbb Z_2^f\times\mathbb Z_2\times\mathbb Z_2$ phase associated with the $\mathbb Z_4$ component in $H_{stack}$.

The excitations in the $\mathbb Z_2^f\times\mathbb Z_2\times\mathbb Z_2\times\mathbb Z_2$  gauge theory are just compositions of the excitations from each layer. Each layer supports Ising-like anyons. We denote the anyons by $1, \psi_t, \sigma_t$, with $t=a,b,c,d$. Here, $\psi_t$ is the charge excitation in each layer, and $\sigma_t$ is the vortex in each layer. They satisfy the Ising fusion rules
\begin{equation}
\psi_t\times\psi_t =1, \ \psi_t\times\sigma_t = \sigma_t, \ \sigma_t\times\sigma_t = 1+\psi_t
\end{equation}
The topological spins of the vortices are
\begin{equation}
\theta_{\sigma_a} =\theta_{\sigma_d}= -\frac{\pi}{8}, \quad \theta_{\sigma_b} =\theta_{\sigma_c}  = \frac{\pi}{8}
\end{equation}
and the charges $\psi_a,\psi_b,\psi_c,\psi_d$ are all fermions. Accordingly, there are 81 anyons in total, with 16 charges and 15 flux sectors. Representative vortices of each flux sector are
\begin{align}
&\sigma_a,\ \sigma_b,\ \sigma_c,\ \sigma_d \nonumber\\
&\sigma_a\sigma_b,\ \sigma_a\sigma_c,\ \sigma_a\sigma_d,\ \sigma_b\sigma_c,\ \sigma_b\sigma_d,\ \sigma_c\sigma_d \nonumber\\
& \sigma_a\sigma_b\sigma_c,\ \sigma_a\sigma_b\sigma_d, \ \sigma_a\sigma_c\sigma_d, \ \sigma_b\sigma_c\sigma_d \nonumber\\
& \sigma_a\sigma_b\sigma_c\sigma_d \label{vort}
\end{align}
The topological spins of these vortices can be obtained by summing over the topological spins of their components, e.g., $\theta_{\sigma_a\sigma_b}=\theta_{\sigma_a}+\theta_{\sigma_b} =-\frac{\pi}{8}+\frac{\pi}{8}=0$. Other vortices can be obtained by fusing charges to the representative vortices.

It is worth establishing a translation between the above notation and the general notation used throughout this paper where we use $\xi_0, \xi_1,\xi_2,\xi_3$ to denote vortices that carry unit flux and use $(1,0,0,0), (0,1,0,0), (0,0,1,0),(0,0,0,1)$  to denote the unit charges. According to the correspondence between group elements and gauge flux and the expressions (\ref{c2generator}) of the generators of the group $\mathbb Z_2^f\times\mathbb Z_2\times\mathbb Z_2\times\mathbb Z_2$, we find that the vortices $\xi_0,\xi_1,\xi_2,\xi_3$ correspond to
\begin{align}
\xi_0 &= \sigma_a\sigma_b\sigma_c\sigma_d \nonumber\\
\xi_1 &= \sigma_b\sigma_d \nonumber\\
\xi_2 &= \sigma_c\sigma_d \nonumber\\
\xi_3 &= \sigma_d
\end{align}
and the unit charges are
\begin{align}
(1,0,0,0) &= \psi_a \nonumber\\
(0,1,0,0) &= \psi_a\psi_b \nonumber\\
(0,0,1,0) &= \psi_a\psi_c \nonumber\\
(0,0,0,1) &= \psi_a\psi_b\psi_c\psi_d
\end{align}

Next, we drive a Higgs condensation by condensing the charge $(0,0,0,1)=\psi_a\psi_b\psi_c\psi_d$, so that we achieve a $\mathbb Z_2^f\times\mathbb Z_2\times\mathbb Z_2$ gauge theory. (We refer readers to Ref.~\onlinecite{bais09} for general theory of boson condensation in topological orders.) We show that after the Higgs transition, there are 22 excitations in the theory, with 8 Abelian charges and 14 non-Abelian vortices with quantum dimension 2. The charges are $\psi_a^x\psi_b^y\psi_c^z$, with $x,y,z=0,1$. Condensing $\psi_a\psi_b\psi_c\psi_d$ will lead to confinement of many vortices in the original $\mathbb Z_2^f\times\mathbb Z_2\times\mathbb Z_2\times\mathbb Z_2$ gauge theory---those with nontrivial mutual braiding around $\psi_a\psi_b\psi_c\psi_d$ will be confined.  Those left deconfined fall into 7 flux sectors, as expected for $\mathbb Z_2^f\times\mathbb Z_2\times\mathbb Z_2$ gauge theory. More specifically, the vortices in (\ref{vort}) that contain even number of $\sigma$'s are left deconfined. For those vortices that contain two $\sigma$'s,  we find 12 vortices in total through charge attachments to the representatives in (\ref{vort}), with their topological spins given by
\begin{align}
\theta_{\sigma_a\sigma_d} &=-\frac{\pi}{4},\quad \theta_{\sigma_b\sigma_c} = \frac{\pi}{4}\nonumber\\
\theta_{\sigma_a\sigma_b} &=\theta_{\sigma_a\sigma_c}=\theta_{\sigma_b\sigma_d} =\theta_{\sigma_c\sigma_d}=0\nonumber\\
\theta_{\sigma_a\sigma_d\psi_b} &=\frac{3\pi}{4},\quad \theta_{\sigma_b\sigma_c\psi_a} = -\frac{3\pi}{4}\nonumber\\
\theta_{\sigma_a\sigma_b\psi_c} &=\theta_{\sigma_a\sigma_c\psi_b}=\theta_{\sigma_b\sigma_d\psi_a} =\theta_{\sigma_c\sigma_d\psi_a}=\pi\end{align}
All these vortices have quantum dimension 2. Other vortices with two $\sigma$'s are identified with the above ones after condensing $\psi_a\psi_b\psi_c\psi_d$. The vortex $\sigma_a\sigma_b\sigma_c\sigma_d$ will split into vortices, $(\sigma_a\sigma_b\sigma_c\sigma_d)_1$ and $(\sigma_a\sigma_b\sigma_c\sigma_d)_2$, after condensing $\psi_a\psi_b\psi_c\psi_d$. To understand the occurrence of splitting, one needs to go to general anyon condensation theory\cite{bais09}, which is beyond the scope of the current work. According to the general anyon condensation theory, the topological spins of the new vortices are the same as that before splitting. Hence, we have
\begin{equation}
\theta_{(\sigma_a\sigma_b\sigma_c\sigma_d)_1}=\theta_{(\sigma_a\sigma_b\sigma_c\sigma_d)_2}=0
\end{equation}
Both vortices have quantum dimension 2. There are no other anyons in the theory. Hence, we obtain $8+14=22$ excitations in total. 

Finally, we comment that there exists a bosonic analog of this example. That theory can be obtained by gauging a particular BSPT phase with $\mathbb Z_2\times\mathbb Z_2\times\mathbb Z_2$ symmetry\cite{propitius95, chen13}. The latter also have 22 anyons with 8 being Abelian and 14 being non-Abelian with quantum dimension 2. However, the topological spins are different from the current example. In particular, in the current example there are 4 fermionic charges and 4 bosonic charges, but the bosonic counterpart has all 8 charges being bosonic.

\section{$G_f = \mathbb Z_4^f \times \mathbb Z_4\times\mathbb Z_4$: FSPT phases of the third kind}
\label{sec:exception}

In this section, we discuss the ``exceptional'' FSPT phases --- those mentioned in case (C-4) in Sec.~\ref{sec:con_cij_even}.  We are not able to construct models for these exceptional phases based on free-fermion and BSPT-embedded models. These phases are generated by the generating phases of case (C-4), or generated by a combination of the generating phases of case (C-4) and other generating phases.

Generally speaking, there are three possible fates for these ``exceptional'' phases:
\begin{enumerate}
\item They do not exist in physical systems. They are unphysical solutions of the constraints (\ref{c1}-\ref{c8}), (\ref{ac1}) and (\ref{ac2}), implying that these constraints are incomplete.

\item They do exist, and can be realized through free-fermion models, or BSPT-embedded models, or a combination of them, in an appropriate way that we do not know yet.

\item They do exist, and can only be realized in interacting fermionic systems. They are {\it intrinsically fermionic} and {\it intrinsically interacting}, i.e., they belong to the third kind of FSPT phases discussed in the Introduction.

\end{enumerate}
The main purpose of this section is to argue that it is the third possibility. 

To simplify the discussion,  we focus on $G_f=\mathbb Z_{4}^f\times \mathbb Z_{4} \times \mathbb Z_{4}$, which is the simplest symmetry to support these exceptional FSPT phases. Our analysis can be straightforwardly extended to more general symmetries. According to Eq.~\ref{stackgroup} and Table \ref{tab1}, for $G_f= \mathbb Z_4^f\times\mathbb Z_4\times \mathbb Z_4$, the stacking group $H_{\rm stack} = \mathbb Z_8^2\times\mathbb Z_4^2\times\mathbb Z_2^2$. There are $4096$ distinct FSPT phases. Accordingly to the discussion in Sec.~\ref{sec:con_cij_even}, half of the phases are ``exceptional'', which are characterized by
\begin{equation}
\Theta_{012} = \frac{\pi}{2} \ \text{or} \ \frac{3\pi}{2} \label{dis-a1}
\end{equation}
For the other half that are characterized by $\Theta_{012}=0$ or $\pi$, we do have models to realize them as discussed in Sec.~\ref{sec:model_con}. The inability to construct models for the exceptional half can be traced back to the inability to construct models for the second generating phase in case (C-4) in Sec.~\ref{sec:con_cij_even}.

Below, we argue that the $\mathbb Z_4^f\times\mathbb Z_4 \times \mathbb Z_4$ FSPT phases characterized by Eq.~(\ref{dis-a1}) can not be realized by free-fermion models, BSPT-embedded models, or a combination of them. In addition, we show evidence for their existence. Accordingly, it is the third possibility mentioned above. Finally, we also argue that similar FSPT phases exist in 1D systems with a $\mathbb Z_4^f\times\mathbb Z_4$ symmetry.

\subsection{No BSPT-embedding realization}
\label{nobspt}

To begin, we argue that BSPT-embedded models cannot realize the phases characterized by Eq.~(\ref{dis-a1}). The key point in our argument is that the fermion parity flux plays a nontrivial role in FSPT phases characterized by (\ref{dis-a1}), while in BSPT-embedded models it always plays a trivial role.

Consider a gauged FSPT phase with $\mathbb Z_4^f\times\mathbb Z_4 \times \mathbb Z_4$ symmetry. Take $\Pi_0$, $\xi_1$, and $\xi_2$ to be three vortices, where $\Pi_0$ carries the fermion-parity flux, $\xi_1$ carries type-1 unit flux, and $\xi_2$ carries type-2 unit flux. Note that the fermion-parity flux is twice of the type-0 unit flux. We then imagine the following braiding process: $\Pi_0$ is first braided around $\xi_1$, then around $\xi_2$, and then around $\xi_1$ in the opposite direction, and finally around $\xi_2$ in the opposite direction. This braiding process is identical to the one in the definition of $\Theta_{012}$. Similarly to $\Theta_{012}$, one can show that this braiding process leads to an Abelian Berry phase $\Omega$. Moreover, one can show that $\Omega=2\Theta_{012}$, where the factor 2 follows from the fact that $\Pi_0$ carries twice of the type-0 unit flux.

Now we compute the values of $\Omega$, both in the exceptional FSPT phases and in BSPT-embedded phases. According to (\ref{dis-a1}), we have $\Omega=\pi$ for the exceptional FSPT phases. In BSPT-embedded phases, the fermion parity operator acts like the identity operator on the bosons. Accordingly, the vortex $\Pi_0$ either has trivial braiding statistics with respect to $\xi_1$ and $\xi_2$, or has braiding statistics with respect to $\xi_1$ and $\xi_2$  resulting from charge attachment to the vortices. In either case, $\Omega=0$. Considering distinct values of $\Omega$,  we conclude that BSPT-embedded models cannot realize FSPT phases characterized by (\ref{dis-a1}).

\subsection{No free-fermion realization}

In this subsection, we argue that $\mathbb Z_4^f\times\mathbb Z_4\times\mathbb Z_4$ FSPT phases from free-fermion realization can only lead to Abelian statistics after gauging the symmetry. Then, $\Theta_{012}=0$ by its definition. Accordingly, free-fermion models cannot realize FSPT phases characterized by (\ref{dis-a1}). Combining this result with that of Sec.~\ref{nobspt} and using the additivity of $\Theta_{012}$ under stacking, we see that a simple stack of BSPT-embedded models and free-fermion models can only realize FSPT phases with $\Theta_{012}=0$ or $\pi$. Hence, we prove the claim that BSPT-embedded models, free-fermion models, and a simple stack of them cannot realize FSPT phases characterized by (\ref{dis-a1}).

We now show that free-fermion models can only lead to Abelian statistics after gauging the symmetry. Our argument takes several steps. First, we notice that any free-fermion Hamiltonian $ H_{\rm free}$ that respects $\mathbb Z_4^f$ symmetry must also respect the charge $U_c(1)$ symmetry. To see this,  note that $ H_{\rm free}$ is a sum of fermion bilinear terms of the form $f^\dag f$, $ff^\dag$, $ff$ and $f^\dag f^\dag$ (lattice and flavor indices are omitted for simplicity). The presence of $\mathbb Z_4^f$ symmetry rules out  terms of the form $ff$ and $f^\dag f^\dag$, leaving terms of the form $f^\dag f$ and $ff^\dag$ only. Accordingly,  $ H_{\rm free}$ is symmetric under $U_c(1)$ symmetry, with the symmetry transformation given by $f\rightarrow e^{-i\alpha} f$ and $f^\dag \rightarrow e^{i\alpha} f^\dag$ where $\alpha$ is the $U_c(1)$ angle ($\alpha=\pi/2$ corresponds to the generator of $\mathbb Z_4^f$ symmetry). With this observation, it is straightforward to see that $\mathbb Z_4^f\times\mathbb Z_4\times\mathbb Z_4$ symmetric free-fermion systems are also symmetric under $U_c(1) \times \mathbb Z_4\times\mathbb Z_4$ symmetry.

Next, we obtain the free-fermion classification of $\mathbb Z_4^f\times\mathbb Z_4\times\mathbb Z_4$ symmetric gapped systems. To do this, we first block diagonalize  $H_{\rm free}$ into the following form:
\begin{equation}
H_{\rm free} = \left(\begin{array}{cccc}
H_1 & \\
& H_2 \\
&& \ddots \\
&&& H_{16}
\end{array}\right)
\end{equation}
where the 16 blocks correspond to the 16 irreducible representations of $\mathbb Z_4\times\mathbb Z_4$ symmetry.  Physically, it can be viewed as 16 layers of free fermions, where each layer is composed of fermions that carry a distinct $\mathbb Z_4\times\mathbb Z_4$ charge. Then, the classification of FSPT phases of $H_{\rm free}$ is decomposed into the classification of each $H_i$, where $i=1,2,\dots, 16$. As argued above, every $H_i$ must respect $U_c(1)$ symmetry. In fact, this ``16-layer'' system has a $[U_c(1)]^{16}$ symmetry, since the fermion number is conserved individually in each layer.  According Refs.~\onlinecite{ff1,ff2}, each $H_i$ describes a fermionic system in the A class and has a $\mathbb Z$ classification in two dimensions. The $\mathbb Z$ classification corresponds to the IQHEs at various integer filling factors.  Hence, the overall classification of $H_{\rm free}$ is given by $(\mathbb Z)^{16}$. Physically, every phase in the classification can be thought as a stack of 16 layers of IQHEs, associated with the $[U_c(1)]^{16}$ symmetry. (Note that the total chiral central charge may not be zero; but this is irrelevant to our discussion below.)

With the above understanding, we now argue that free-fermion models only support Abelian statistics after gauging the $\mathbb Z_4^f\times\mathbb Z_4\times\mathbb Z_4$ symmetry. Our key observation is the fact that gauging any finite subgroup of $U_c(1)$ symmetry in IQHE systems, we only obtain Abelian statistics. This can be obtained through the standard Chern-Simon description of IQHEs\cite{wen-book}. Similarly, gauging any finite subgroup of the $[U_c(1)]^{16}$ symmetry of a stack of 16 layers of IQHEs, we also only obtain Abelian statistics. Note that $\mathbb Z_4^f\times\mathbb Z_4\times\mathbb Z_4$ is a subgroup of $[U_c(1)]^{16}$. More specifically, the $\mathbb Z_4^f\times\mathbb Z_4\times\mathbb Z_4$ symmetry is generated by the following three operators
\begin{align}
g_0 &= \prod_{m,n=0}^3 e^{i\pi F_{(m,n)}/2} \nonumber\\
g_1&= \prod_{m,n=0}^3 e^{im\pi F_{(m,n)}/2} \nonumber\\
g_2&= \prod_{m,n=0}^3 e^{in\pi F_{(m,n)}/2}
\end{align}
where $(m,n)$ labels the 16 irreducible representations of $\mathbb Z_4\times\mathbb Z_4$, and $F_{(m,n)}$ is the fermion number operator in the corresponding layer. The $U_c(1)^{16}$ symmetry is generated by $\exp[i\alpha_{(m,n)} F_{(m,n)}]$, where $m,n=0,1,2,3$ and $\alpha_{(m,n)}$ is an angle. It is obvious that $\mathbb Z_4^f\times\mathbb Z_4\times\mathbb Z_4$ is a subgroup of $[U_c(1)]^{16}$. Hence, we prove our claim.

\subsection{Evidence of existence}

We now argue that FSPT phases characterized by (\ref{dis-a1}) do exist. Our argument takes two steps: We first argue that the corresponding {\it gauged} FSPT phases exist, then we argue that the actual FSPT phases can be obtained from the gauged phases by {\it ungauging} the symmetry.

The gauged FSPT phases do not exist if the values (\ref{dis-a1}) of topological invariants cannot be consistently extended to a full set of braiding statistics, including a set of anyon labels, fusion rules, braiding data, etc. By ``consistently extended'', we mean that the full braiding statistics should satisfy unitarity, the pentagon equation, the hexagon equation, etc \cite{kitaev06}. We show that topological invariants with the values in (\ref{dis-a1}) can be consistently extended to a set of full braiding statistics. To do that, we first gauge $\mathbb Z_4^f$ only. Following Ref.~\onlinecite{wangcj16}, we find that the resulting topological order has 16 anyons, $e^i m^j$, with  $i,j=0,1,2,3$ and $e^4=m^4=1$. Here, $e$ is the fermionic unit charge, and $m$ is a bosonic vortex carrying the unit flux. This topological order has a remaining $\mathbb Z_4\times\mathbb Z_4$ symmetry. That is, it is a {\it symmetry enriched topological} (SET) phases.\cite{essin13, mesaros13, tarantino16, barkeshli14} The property (\ref{dis-a1}) can be translated into the following property of the SET phase: the vortex $m$ carries a four dimensional projective representation $V_m$ of the $\mathbb Z_4\times\mathbb Z_4$ symmetry, whose generators $g_1$ and $g_2$ satisfies
\begin{equation}
V_m(g_1)V_m(g_2) = e^{i\Theta_{012}} V_m(g_2)V_m(g_1) \label{proj}
\end{equation}
On the other hand, $e$ does not carry any projective representation of $\mathbb Z_4\times\mathbb Z_4$, since $e$ corresponds to the local fermion in the original FSPT phases.  With this manipulation, the existence of gauged $\mathbb Z_4^f\times\mathbb Z_4\times\mathbb Z_4$ FSPT phases characterized by (\ref{dis-a1}) is translated to the question whether the $\mathbb Z_4\times\mathbb Z_4$ SET phase characterized by (\ref{proj}) can be gauged in a consistent way, such that the resulting theory is  strictly 2D, i.e., {\it anomaly-free}. The anomaly-detection problem in topological orders has been widely studied\cite{chen14,barkeshli14,bbc}. Applying the formulas from Ref.~\onlinecite{bbc} (see also Ref.~\onlinecite{chen14}) with the property (\ref{proj}) and the fact the $e$ does not carry projective representation, we find that the above SET is indeed anomaly-free. This proves that the guaged $\mathbb Z_4^f\times\mathbb Z_4\times\mathbb Z_4$ FSPT phases characterized by (\ref{dis-a1}) exist. (One may consult Refs.~\onlinecite{heinrich16,cheng16} for a general scheme for realizing such anomaly-free SETs using exactly soluble string-net models).

With the gauged FSPT phases, in principle we can obtain the actual FSPT phase by ungauging the symmetry. In fact, it is more convenient to start with the $\mathbb Z_4\times\mathbb Z_4$ SET mentioned above. To ``ungauge'' the symmetry, one way is to formally ``condense'' the fermionic charge $e$. Or more physically, one can stack the SET with a trivial fermionic system, where there is a local fermion $f$. Then, we condense the bosonic pair $ef$. Since $ef$ does not carry any quantum number of $\mathbb Z_4\times\mathbb Z_4$ symmetry, condensing it does not break the symmetry. After condensation, $m$ will be confined and $e$ is identified with the local fermion $f$. Hence, there is only a local fermion $f$ in the condensed phase. Since we condense unit charge $e$, the resulting theory has an emergent $\mathbb Z_4^f$ symmetry, making the whole symmetry being $\mathbb Z_4^f\times\mathbb Z_4\times\mathbb Z_4$. (The symmetry defect of $\mathbb Z_4^f$ corresponds to the strings pulled out by the confined $m$ particle; see Ref.~\onlinecite{hung14} for a connection between SET/SPT and anyon condensation).

The argument is abstract, but nevertheless shows the existence of FSPT phases characterized by (\ref{dis-a1}). Of course, it is desirable to construct explicit (exactly soluble) models to realize these FSPT phases. We leave such model construction for future work.


\subsection{1D FSPT phases of the third kind }
\label{sec:1dfspt}

In passing, we point out that there exist analogous 1D FSPT phases of the third kind. The simplest symmetry to support these 1D FSPT phases that we find is $\mathbb Z_4^f\times\mathbb Z_4$ symmetry. According to Fidkowski and Kitaev \cite{fidkowski11}, 1D FSPT phases with $\mathbb Z_4^f\times\mathbb Z_4$ symmetry is classified by $H^2(\mathbb Z_4^f\times\mathbb Z_4,U(1)) = \mathbb Z_4$. We find that the generating phase and three copies of it are the third-kind of FSPT phases. In fact, they are related to the above 2D $\mathbb Z_4^f\times\mathbb Z_4\times\mathbb Z_4$ phases through an appropriate {\it dimensional reduction} procedure.


Let us describe the dimensional reduction procedure. Consider a $\mathbb Z_4^f\times\mathbb Z_4\times\mathbb Z_4$ 2D FSPT phase defined on a cylinder geometry. We assume that the FSPT phase satisfies the property (\ref{dis-a1}). Imagine that we insert a type-2 {\it external} unit flux into the cylinder. Here, we view the gauge field as non-dynamical external field. Then, the two ends of the cylinder can be viewed as two defects that carry type-2 unit flux. Each defect carries a projective representation of the symmetry. In this external gauge field setting, the property (\ref{dis-a1}) leads to the following relation
\begin{equation}
U(g_1)U (g_0) = e^{i\Theta_{012}}U(g_0)U(g_1) \label{dis-a4}
\end{equation}
where $U$ is the projective representation carried by one of the defects, $g_0$, $g_1$ are the first and second generators of $\mathbb Z_4^f\times\mathbb Z_4\times\mathbb Z_4$. Next, we ignore the last $\mathbb Z_4$, take the thin-cylinder limit, and view the system as a 1D system. In that limit, (\ref{dis-a4}) does not change. With this procedure, we obtain a 1D gapped fermionic system with $\mathbb Z_4^f\times\mathbb Z_4$ symmetry, characterized by the projective representation (\ref{dis-a4}). Comparing to Ref.~\onlinecite{fidkowski11}, we find that the case $\Theta_{012}=\pi/2$ corresponds to the generating phase in the $H^2(\mathbb Z_4^f\times\mathbb Z_4,U(1)) = \mathbb Z_4$ classification.

Similarly to the 2D case, one can argue that the $\mathbb Z_4^f\times\mathbb Z_4$ 1D FSPT phases characterized by (\ref{dis-a4}) with $\Theta_{012}=\pm \pi/2$ cannot be realized by free-fermion and BSPT-embedeed models. There is no BSPT-embedding realization because the fermion parity plays a nontrivial role in the projective representation (\ref{dis-a4}). In addition, there is no free-fermion realization, because the classification of free-fermion $\mathbb Z_4^f\times\mathbb Z_4$ FSPT phases can be reduced to the classification of free fermions in the A class, and the latter has no nontrivial FSPT phases in 1D\cite{ff1,ff2}. Hence, these 1D FSPT phases are of the third kinds, i.e., intrinsically fermionic and intrinsically interacting.


\section{Stability of BSPT phases}
\label{sec:stability}

In Sec.~\ref{sec:model_con}, we use the BSPT-embedded fermionic models to construct models for general FSPT phases. These BSPT-embedded fermionic models are obtained by following embedding procedure: first let the fermions form strongly bound pairs, and then put the pairs in a BSPT phase. The BSPT phase should have a symmetry
\begin{equation}
G_b = G_f/\mathbb Z_2^f = \mathbb Z_m \times \prod_{i=1}^K \mathbb  Z_{N_i} \label{gb}
\end{equation}
That is, the fermion-parity element is treated as the identity element. 

One of the interesting phenomena that occur is that: in certain cases, two inequivalent BSPT phases are identified as the same FSPT phase through the above embedding procedure. In particular, a nontrivial BSPT phase may be identified with the trivial phase after embedding. In other words, BSPT phases may be unstable under embedding. Below we discuss the stability/instability issue for BSPT phases with symmetry $G_b$ in (\ref{gb}), when they are embedded into fermionic systems with symmetry $G_f$ in (\ref{group}).


First, we discuss the classification and characterization of BSPT phases with symmetry $G_b$ in (\ref{gb}). According to the group cohomology classification,\cite{chen13} they are classified by the cohomology group $H^3[G_b, U(1)]$.  Each element in $H^3[G_b, U(1)]$ corresponds to one BSPT phase. These BSPT phases can be characterized by a bosonic version of topological invariants, which we denote as $\tilde\Theta_\mu, \tilde \Theta_{\mu\nu}, \tilde\Theta_{\mu\nu\lambda}$. It was shown in Ref.~\onlinecite{wangcj15} that the topological invariants are able to distinguish every BSPT phase in the group cohomology classification. More explicitly, they take values in the following form
\begin{align}
\tilde \Theta_0 & = \frac{2\pi}{m}p_0 \nonumber \\
\tilde \Theta_i &  = \frac{2\pi}{N_i} p_i, \quad \tilde\Theta_{0i} = \frac{2\pi}{\bar N_{0i} }p_{0i}, \quad \tilde\Theta_{00i}=0 \nonumber  \\
\tilde \Theta_{ij} & = \frac{2\pi}{N_{ij}} p_{ij}, \quad \tilde \Theta_{0ij} = \frac{2\pi}{\bar N_{0ij}} p_{0ij} \nonumber\\
\tilde \Theta_{ijk} & =\frac{2\pi}{N_{ijk}} p_{ijk} \label{binv}
\end{align}
where the integers $p_0, p_i, p_{0i}$, $p_{ij}$ and $p_{0ij}$ with $i<j$, and  $p_{ijk}$ with $i<j<k$ are independent. The numbers $\bar N_{0i} ={\rm gcd}(m, N_i)$ and $\bar N_{0ij} = {\rm gcd}(m, N_i, N_j)$. (The index ``0'' is not special for BSPT phases; however, we separate it out for a better comparison to their fermionic counterparts.) Every assignment of the independent integers describes one physically realizable BSPT phase.  The values of other components of topological invariants (e.g. $\Theta_{iii}$ and $\tilde\Theta_{ij}$ with $i>j$) are either determined by the ones listed above or constrained to be 0. If all topological invariants vanish, it corresponds to the trivial phase.

Next, we find the relation between the fermionic topological invariants $\Theta_\mu, \Theta_{\mu\nu}, \Theta_{\mu\nu\lambda}$ and the bosonic topological invariants  $\tilde\Theta_\mu, \tilde \Theta_{\mu\nu}, \tilde\Theta_{\mu\nu\lambda}$ in the BSPT-embedded fermionic models. From the definitions of topological invariants given in Ref.~\onlinecite{wangcj15} and Sec.~\ref{inv_def}, we find that they satisfy the following relation:
\begin{align}
\Theta_0 & =\left\{\begin{array}{ll}
\tilde \Theta_0, & \text{if $m$ is odd} \\[3pt]
2\tilde \Theta_0, & \text{if $m$ is even}
\end{array}
\right. \nonumber\\
\Theta_i & =\left\{\begin{array}{ll}
2\tilde \Theta_i, & \text{if $N_i$ is odd} \\[3pt]
\tilde \Theta_i, & \text{if $N_i$ is even}
\end{array}
\right. \nonumber\\
\Theta_{0i} & = \frac{N^{0i}}{\bar N^{0i}} \tilde \Theta_{0i}, \quad
\Theta_{ij}  = \tilde\Theta_{ij} \nonumber\\[3pt]
\Theta_{00i} & = \tilde\Theta_{00i}, \quad
\Theta_{0ij}  = \tilde\Theta_{0ij}, \quad \Theta_{ijk}  = \tilde\Theta_{ijk} \label{dis-b3}
\end{align}
where $\bar N^{0i}$ is the least common multiple of $m$ and $N_i$.

Then, unstable BSPT phases can be found as follows: For a nontrivial BSPT phase described by non-vanishing $\tilde\Theta_\mu, \tilde \Theta_{\mu\nu}, \tilde\Theta_{\mu\nu\lambda}$, we calculate $\Theta_\mu, \Theta_{\mu\nu}, \Theta_{\mu\nu\lambda}$ according to Eq.~(\ref{dis-b3}); if the fermionic topological invariants all vanish,  the BSPT phase is unstable.  There are two cases that such instability can occur:
\begin{enumerate}
\item When $m$ is even, there is a BSPT phase characterized by $\tilde\Theta_0 = \pi$ and all other invariants vanish.  According to (\ref{dis-b3}), $\Theta_0=0$. Hence, it is embedded into the trivial FSPT phase.  The simplest example of this case is that $G_b=\mathbb Z_2$ and $G_f = \mathbb Z_4^f$.

\item When $m, N_i$ are even, there is a BSPT phase characterized by $\tilde \Theta_{0i} = \pi$ and all other invariants vanish. If $N^{0i}/\bar N^{0i} = 2$, we find that $\Theta_{0i} = 0$. Hence, the BSPT phase is embedded into the trivial FSPT phase.  The simplest example in this case is that $G_b = \mathbb Z_2\times\mathbb Z_2$ and $G_f = \mathbb Z_4^f\times\mathbb Z_2$.
\end{enumerate}
Other unstable BSPT phases can be reduced to the above two cases. More generally, if two BSPT phases are found to collapse to the same FSPT phase through embedding, they may be thought of as differing by an unstable BSPT phase in the sense of stacking.

Finally, two comments are in order. First, a BSPT phase with symmetry $G_b$ can be embedded into FSPT systems with different $G_f$'s. The instability of the BSPT phase depends on the embedding. For example, $\mathbb Z_2$ symmetric BSPT phase can be embedded into $G_f=\mathbb Z_4^f$ or $G_f=\mathbb Z_2^f\times\mathbb Z_2$ fermionic systems.  The nontrivial $\mathbb Z_2$ BSPT phase is unstable when embedded into a $\mathbb Z_4^f$ system, but it is stable when embedded into a $\mathbb Z_2^f\times\mathbb Z_2$ system.  Second, our instability analysis above assumes that the topological invariants $\Theta_{\mu}, \Theta_{\mu\nu}, \Theta_{\mu\nu\lambda}$ are complete in the sense that they distinguish every FSPT phase. Accordingly, we have interpreted the vanishing of fermionic topological invariants as the trivial FSPT phase.



\section{Conclusion}

\label{sec:conclusion}
In summary, we study various aspects of 2D FSPT phases with general finite unitary Abelian symmetry. By gauging the symmetry and using braiding statistics in the resulting gauge theory, we define a set of topological invariants, denoted as $\Theta_\mu, \Theta_{\mu\nu}, \Theta_{\mu\nu\lambda}$, to characterize the FSPT phases. Under the assumption that these topological invariants form a complete set, in the sense that they distinguish every FSPT phase, we obtain a classification of 2D FSPT phases with Abelian symmetry. To further support the classification, we construct models to realize the phases in our classification. Most of the phases in our classification can be realized through free-fermion models, or models obtained through embedding of BSPT phases, or a simple stack of them.

Nevertheless, there is an exceptional class of FSPT phases which we are not able to construct models. We argue that these exceptional FSPT phases can only be realized in interacting fermionic systems. The simplest symmetry to support these 2D FSPT phases is $\mathbb Z_4^f\times\mathbb Z_4\times\mathbb Z_4$ symmetry. We also find that 1D fermionic systems with $\mathbb Z_4^f\times\mathbb Z_4$ can also support similar FSPT phases. Since we do not have models for these FSPT phases, it is desirable to build explicit models to realize them.

It remains an open question to classify 2D interacting FSPT phases with general symmetry. For onsite unitary symmetry $G_f$, where $G_f$ is a nontrivial $\mathbb Z_2^f$ group extension and $G_f$ can be non-Abelian, a general classification scheme is still missing. It is also interesting to include antiunitary symmetry such as time-reversal symmetry.

Another open question is to classify 3D interacting FSPT phases. For Abelian unitary symmetries, we expect a straightforward generalization of the topological invariants defined in this work to three dimensions, using the idea of three-loop braiding statistics.\cite{threeloop,ran14} Nevertheless, it might not be easy to obtain a reasonably complete set of constraints on the topological invariants\cite{wangcj15}.  Hence a reasonably complete classification might not be easily obtained. Besides classification, another interesting question that one can ask is that: are there any 3D FSPT phases beyond those that can be obtained through BSPT embedding? Note that in 3D, there are no free-fermion FSPT phases protected by unitary symmetry. A preliminary calculation shows that such FSPT phases do exist, and can be found in the supercohomology models\cite{cheng-wang}.

\begin{acknowledgments}
CW thanks M. Cheng, A. Furusaki, M. Levin, Y. Wan, and in particular M. Metlitski, for helpful discussions.  This research was supported in part by Perimeter Institute for Theoretical Physics. Research at Perimeter Institute is supported by the Government of Canada through the Department of Innovation, Science and Economic Development Canada and by the Province of Ontario through the Ministry of Research, Innovation and Science. CHL acknowledges the funding from the Canada Research Chair (CRC) program and the University of Alberta.
\end{acknowledgments}

\appendix

\section{Canonical form of $G_f$}
\label{appd_canonical_form}

In this appendix, we show that any finite Abelian unitary symmetry $G_f$ of a fermionic system can be written in the canonical form, Eq.~(\ref{group1}), with the fermion-parity element being $(m,0,\dots,0)$.

To begin, we understand that any finite Abelian group can be written as
\begin{equation}
G_f = \prod_{\mu=0}^K \mathbb Z_{\bar N_\mu} \label{gf}
\end{equation}
where $\bar N_\mu, K$ are positive integers. Group elements can be labeled by integer vectors $\bar a=(\bar a_0,\dots, \bar a_K)$, where the component $\bar a_\mu$ takes values in the range $0,1,\dots, (\bar N_\mu-1)$. Alternatively, we can label the group elements by the equivalence classes of all $(K+1)$-component integer vectors, under the following equivalence relation:
\begin{equation}
\bar a \equiv  \bar b, \quad \text{if } \bar a_\mu = \bar b_\mu \ ({\rm mod \ } \bar N_\mu) \label{equiv}
\end{equation}
The latter labeling scheme will be more convenient for our purpose. As discussed in Sec.~\ref{sec:symmetry}, to fully specify the symmetry $G_f$, we need to pick an order-of-2 element that corresponds to the fermion parity. (This is possible if and only if $|G_f|$ is even.) In general, the fermion-parity element can be any integer vector
\begin{equation}
g_f = (g_0, g_1, \dots, g_K) \label{fp}
\end{equation}
where $2g_\mu=0 \ ({\rm mod \ }  \bar N_\mu )$. Here, we choose $0\le g_\mu <\bar N_\mu$. Our goal is to show that the general $G_f$ in (\ref{gf}) with the fermion-parity element given in (\ref{fp}) is isomorphic to a $G_f$ in (\ref{group1}) with the fermion-parity being $(m,0,\dots,0)$.

To do that, we make use of the so-called Smith normal form of integer matrices. We view the fermion-parity element $g_f$ as a one-row integer matrix. According to the Smith normal form, $g_f$ can be written as
\begin{equation}
g_f = (m, 0, \dots, 0) S
\end{equation}
where $S$ is a $(K+1)\times (K+1)$ integer matrix with $\det(S) =\pm 1$, and $m= \gcd(g_0,g_1,\dots,g_K)$. The first row of $S$ is actually $g_f/m$.

Next, we define new integer vectors
\begin{equation}
 a = \bar a S^{-1} \label{mapping}
\end{equation}
where $\bar a$ is any $(K+1)$-component integer vector.  Since $\det(S)=\pm 1$, $S^{-1}$ is also an integer matrix. Accordingly, $a$ is an integer vector. Moreover, one can see that (\ref{mapping}) actually establishes a one-to-one correspondence between integer vectors $\bar a$ and $a$. In particular, $g_f$ is mapped to the vector $(m,0,\dots,0)$.

With the one-to-one mapping (\ref{mapping}),  we can now label the group elements in $G_f$ by integer vectors $a$ under the equivalence relation
\begin{equation}
 a \equiv   b, \quad \text{if }  a_\mu =  b_\mu \ ({\rm mod \ }  N_\mu) \label{equiv2}
\end{equation}
where $\{N_\mu\}$ are integers related to $\{\bar N_\mu\}$ through the matrix $S$. More specifically, for a fixed index $\mu$, $N_\mu$ is the smallest positive integer such that the vector $N_\mu( S_{\mu0}, S_{\mu1},\dots, S_{\mu K})\equiv 0$ under the equivalence relation (\ref{equiv}). In particular, since the first row of $S$ is $g_f/m$, it is not hard to see that $N_0 = 2m$.

With the labeling scheme (\ref{equiv2}) for elements in $G_f$, we see that $G_f$ is indeed given by the canonical form (\ref{group1}) with the fermion parity labeled by $(m,0,\dots,0)$.

\section{Proofs of constraints (\ref{c1})-(\ref{c8})}
\label{app_proof}

\begin{figure*}
\centering
\includegraphics{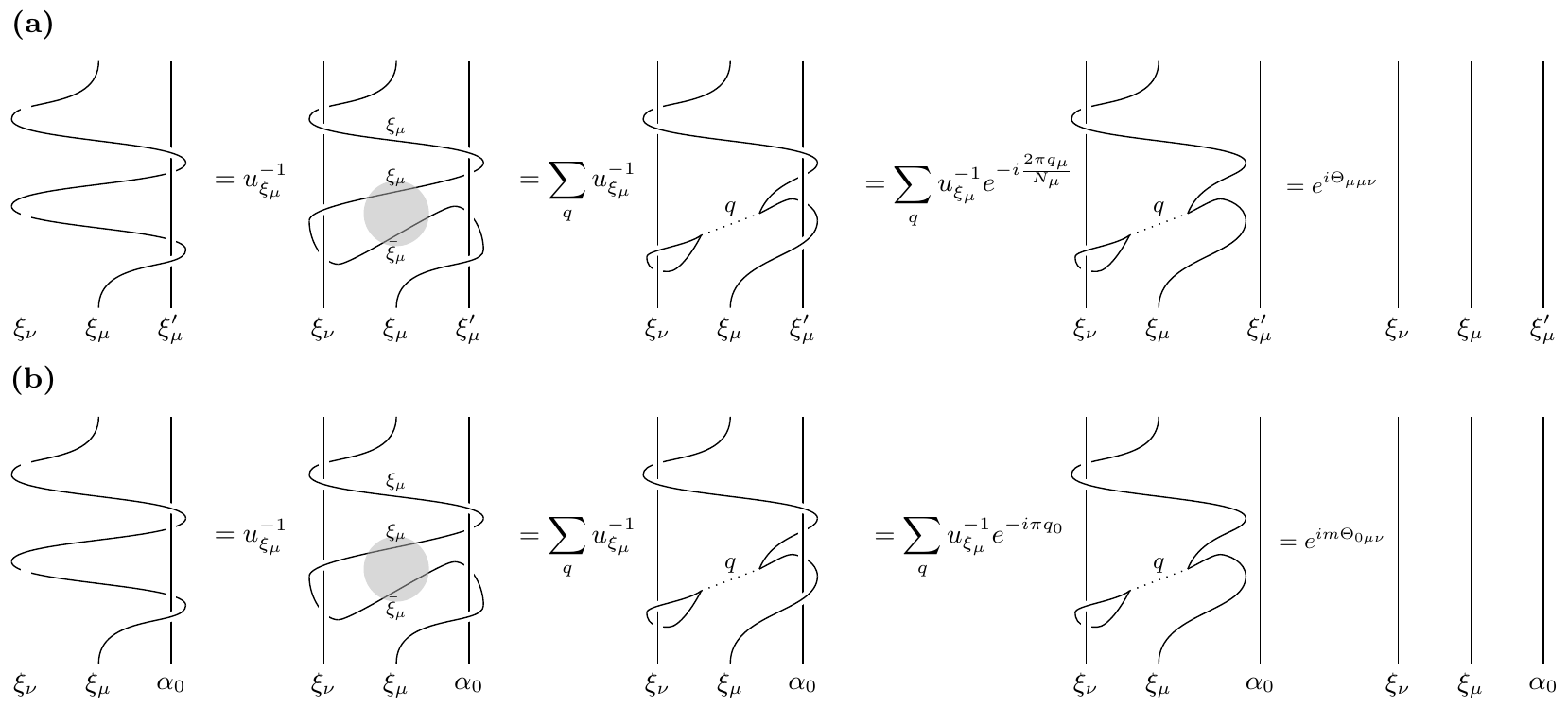}
\caption{Diagrammatic calculations of the Berry phases $\Theta_{\mu\mu\nu}$ (a) and $m\Theta_{0\mu\nu}$ (b). }\label{fig_diagram1}
\end{figure*}

In this section, we prove the constraints (\ref{c1})-(\ref{c8}) of topological invariants $\Theta_\mu,\Theta_{\mu\nu}$ and $\Theta_{\mu\nu\lambda}$. Some of the proofs  involve diagrammatic calculations of braiding statistics. We refer readers to Ref.~\onlinecite{kitaev06} for an introduction of the diagrammatics of braiding statistics.

The constraints (\ref{c1}) and (\ref{c2}) are the same as their counterparts for BSPT phases. The derivations can be carried over from there with no modification. Hence, we do not repeat the proofs here, and instead refer the readers to Ref.~\onlinecite{wangcj15}. The constraint (\ref{c4}) follows immediately from the property that braiding statistics is symmetric, in the sense the braiding anyon $\alpha$ around $\beta$ is topologically equivalent to braiding $\beta$ around $\alpha$.

\subsection{Proof of Eq.~(\ref{c3})}

To prove Eq.~(\ref{c3}), it is enough to show the part $\Theta_{\mu\mu\nu} = m \Theta_{0\mu\nu}$. The other part $\Theta_{\nu\nu\mu} = m \Theta_{0\mu\nu}$ immediately follows from the former and the constraint (\ref{c1}).  To show $\Theta_{\mu\mu\nu} = m \Theta_{0\mu\nu}$, we use diagrammatic calculations on braiding statistics of anyons.\cite{kitaev06} Our strategy is to calculate $\Theta_{\mu\mu\nu}$ and $m\Theta_{0\mu\nu}$ using diagrammics respectively, then compare the two diagrammatic calculations and show that they are equal.

Fig.~\ref{fig_diagram1}a shows the diagrammatic calculation of $\Theta_{\mu\mu\nu}$.  The first diagram can be thought of as the space-time trajectories (with the time direction being upward) of three vortices $\xi_\mu, \xi_\mu', \xi_\nu$ associated  with the braiding process that defines $\Theta_{\mu\mu\nu}$, where $\xi_\mu, \xi_\mu', \xi_\nu$ are vortices carrying type-$\mu$, type-$\mu$ and type-$\nu$ unit flux respectively. The last diagram can be thought of as the trajectories of $\xi_\mu, \xi_\mu', \xi_\nu$ with no braiding happening. By definition, the ratio of the first and last diagrams gives the phase factor $e^{i\Theta_{\mu\mu\nu}}$. The three diagrams in between are intermediate steps of the diagrammatic calculation. The first equation is obtained by applying the following rule of diagrammatics of braiding statistics:\cite{kitaev06}
\begin{equation}
\begin{tikzpicture}[baseline={([yshift=-.5ex]current  bounding  box.center)}]
\draw (0,-0.25)..controls(0, 0.7) and(-0.3, 1)..(-0.5,1)..controls(-0.7,1) and(-0.8,0.5) ..(-1,0.5)..controls(-1.2,0.5) and(-1.5,0.8)..(-1.5,1.75);
\draw (1.2,-0.25)--(1.2,1.75);
\node at (0.6,0.75)[scale=1]{$\displaystyle{=u_\alpha}$};
\node at (0,-0.4)[scale=0.8]{$\alpha$};
\node at (-1.5,1.9)[scale=0.8]{$\alpha$};
\node at (-0.85,0.85)[scale=0.8]{$\bar \alpha$};
\node at (1.2,-0.4)[scale=0.8]{$\alpha$};
\node at (1.2,1.9)[scale=0.8]{$\alpha$};
\end{tikzpicture}\label{eq_app_a1}
\end{equation}
where $\bar \alpha$ is the anti-particle of anyon $\alpha$, and $u_\alpha$ is a complex number that is not important for our purpose. The second equation is obtained by applying the following rule to the shaded region in the second diagram:\cite{kitaev06}
\begin{equation}
\begin{tikzpicture}
\draw (0,0)--(0,2);
\draw (0.7,0)--(0.7,2);
\node at (1.6,0.9)[scale=1]{$=\displaystyle{\sum_{\gamma,n}}$};
\node at (0,-0.15)[scale=0.8]{$\alpha$};
\node at (0.7,-0.15)[scale=0.8]{$\beta$};
\node at (0,2.15)[scale=0.8]{$\alpha$};
\node at (0.7,2.2)[scale=0.8]{$\beta$};
\begin{scope}[xshift=0.5cm]
\draw (2,0)--(2.35,0.7)--(2.35,1.3)--(2,2);
\draw (2.7,0)--(2.35,0.7)--(2.35,1.3)--(2.7,2);
\node at (2,-0.15)[scale=0.8]{$\alpha$};
\node at (2.7,-0.15)[scale=0.8]{$\beta$};
\node at (2,2.15)[scale=0.8]{$\alpha$};
\node at (2.7,2.2)[scale=0.8]{$\beta$};
\node at (2.2,1)[scale=0.8]{$\gamma$};
\node at (2.5,0.7)[scale=0.8]{$n$};
\node at (2.5,1.3)[scale=0.8]{$n$};
\end{scope}
\end{tikzpicture}\label{eq_app_a2}
\end{equation}
where $n$ labels the states in the fusion space $\mathbb V_{\alpha\beta}^\gamma$, i.e., the different ways to fuse $\alpha$ and $\beta$ into $\gamma$. For our calculation, we have used the fusion rule between $\xi_\mu$ and $\bar\xi_\mu$:
\begin{equation}
\xi_\mu\times\bar\xi_\mu =  0 + q+ \dots \label{eqa1-1}
\end{equation}
Since $\xi_\mu$ and $\bar\xi_\mu$ carry opposite gauge flux, only charges appear on the right-hand side. Note that we have set the fusion multiplicity $N_{\alpha\bar\alpha}^q =1$. This is proved in Ref.~\onlinecite{wangcj15} for general Abelian gauge theories.  To pass from the third diagram to the fourth diagram, we decouple $\xi_\mu'$ from the rest of the diagram at the expense of an Aharonov-Bohm phase $e^{-i\frac{2\pi}{N_\mu}q_\mu}$, a consequence of the braiding between $q$ and $\xi_\mu'$.  The minus sign follows that $q$ is braided around $\xi_\mu'$ in a clockwise way. Therefore, we obtain the last equation, which is the main result of this diagrammatic calculation.


The quantity $m\Theta_{0\mu\nu}$ is the Berry phase associated with the following process: a vortex $\xi_\mu$ is first braided around $\alpha_0$, then around $\xi_\nu$, then around $\alpha_0$ in the opposite direction, and finally around $\xi_\nu$ in the opposite direction. Here, $\alpha_0$ is a vortex carrying the fermion-parity flux $\Pi=(\pi, 0,\dots, 0)$. To see that this braiding process indeed leads to the Abelian phase $m\Theta_{0\mu\nu}$, we split $\alpha_0$ into $m$ vortices, $\xi_0^1, \xi_0^2, \dots, \xi_0^m$, each carrying type-0 unit flux. Then, we can write the braiding process in term of the following operator product:
\begin{equation}
B= B_{\xi_\mu\xi_\nu}^{-1}(B_{\xi_\mu\xi_0^m}\dots B_{\xi_\mu\xi_0^1} )^{-1}B_{\xi_\mu\xi_\nu}(B_{\xi_\mu\xi_0^m}\dots B_{\xi_\mu\xi_0^1} ) \label{eqa1-2}
\end{equation}
where $B_{\xi_\mu\xi_0^t}$ is the braiding operator associated with braiding $\xi_\mu$ around $\xi_0^t$ once. The product  $B$ can be simplified by the commutation relation
\begin{equation}
B_{\xi_\mu\xi_\nu}^{-1}B_{\xi_\mu\xi_0^t}^{-1}B_{\xi_\mu\xi_\nu}B_{\xi_\mu\xi_0^t} = e^{i\Theta_{\mu0\nu}}I \label{eqa1-3}
\end{equation}
where $I$ is the identity operator. This commutation relation follows from the definition of $\Theta_{\mu0\nu}$. Combining Eqs.~(\ref{eqa1-2}) and (\ref{eqa1-3}), we have
\begin{equation}
B = e^{im\Theta_{\mu0\nu}} I
\end{equation}
According to the constraint (\ref{c2}), $m\Theta_{\mu 0\nu}$ can only be 0 or $\pi$. Further using Eq. (\ref{c1}), we have $m\Theta_{\mu 0\nu} = m\Theta_{0\mu\nu}$. Hence, $m\Theta_{0\mu\nu}$ is indeed the Berry phase associated with the described braiding process.

With the above physical interpretation of the phase $m\Theta_{0\mu\nu}$, we perform a similar diagrammatic calculation for $m\Theta_{0\mu\nu}$, in parallel with that for $\Theta_{\mu\mu\nu}$. The diagrammatic calculation is shown in Fig.~\ref{fig_diagram1}b. The calculation is very similar to Fig.~\ref{fig_diagram1}a, with the only difference being that: the factor $e^{-i2\pi q_\mu/N_\mu}$ in the fourth diagram of Fig.~\ref{fig_diagram1}a is replaced by  $e^{-i\pi q_0}$ in Fig.~\ref{fig_diagram1}b.  The latter is the Aharonov-Bohm phase between $q$ and $\alpha_0$.

Now we compare the last equations in Fig.~\ref{fig_diagram1}a and Fig.~\ref{fig_diagram1}b. Since $\xi_\mu'$ is decoupled from the rest of the fourth diagram in Fig.~\ref{fig_diagram1}a and $\alpha_0$ is decoupled in the corresponding diagram, the constraint $\Theta_{\mu\mu\nu} = m\Theta_{0\mu\nu}$ can be established if we can show the following relation
\begin{equation}
e^{i\frac{2\pi}{N_\mu} q_\mu} = e^{i\pi q_0} \label{eqa1-4}
\end{equation}
where $q$ is any charge appearing in the fusion product $\xi_\mu\times\bar\xi_\mu$. We show this relation indeed holds.
A simplified version of (\ref{eqa1-4}) was proved in Ref.~\onlinecite{wangcj16} for symmetry $G_f = \mathbb Z_{2m}^f$. For general Abelian group $G_f$, the proof is very similar. First, according to Eq.~(\ref{spin_formula}), we have
\begin{equation}
R_{\bar\xi_\mu\xi_\mu}^qR_{\xi_\mu\bar\xi_\mu}^q = e^{i(\theta_q - \theta_{\xi_\mu} - \theta_{\bar\xi_\mu})} \label{eqa1-5}
\end{equation}
In addition, the mutual statistics $R_{\bar\xi_\mu\xi_\mu}^qR_{\xi_\mu\bar\xi_\mu}^q$ between $\xi_\mu$ and $\bar\xi_\mu$ in the fusion channel $q$ satisfies the following relation
\begin{equation}
R_{\bar\xi_\mu\xi_\mu}^qR_{\xi_\mu\bar\xi_\mu}^q = e^{i\frac{2\pi q_\mu}{N_\mu}} R_{\bar\xi_\mu\xi_\mu}^0R_{\xi_\mu\bar\xi_\mu}^0  \label{eqa1-6}
\end{equation}
where $R_{\bar\xi_\mu\xi_\mu}^0R_{\xi_\mu\bar\xi_\mu}^0$ is the mutual statistics between $\xi_\mu$ and $\bar{\xi}_\mu$ in the fusion channel 0. That is, the mutual statistics between $\xi_\mu$ and $\bar{\xi}_\mu$ in the fusion channels $q$ and $0$ differ by the Aharonov-Bohm phase $q\cdot \phi_{\xi_\mu}$. Equation (\ref{eqa1-6}) can be proved using the same thought experiment as in Ref.~\onlinecite{wangcj16}, so we do not repeat it here. Combining Eqs.~(\ref{eqa1-5}) and (\ref{eqa1-6}) and using the fact $\theta_q =\pi q_0$,  we immediately obtain the relation (\ref{eqa1-4}). Accordingly, we establish the constraint (\ref{c3}).

\subsection{Proof of Eq.~(\ref{c5})}
To prove Eq.~(\ref{c5}), we consider a vortex $\xi_\mu$ carrying unit flux $ \frac{2\pi}{N_\mu}e_\mu$, and $N_\nu$ vortices $\xi_{\nu}^{1}, \dots, \xi_{\nu}^{N_\nu}$, all of which carry the unit flux $ \frac{2\pi}{N_\nu}e_\nu$. We imagine braiding $\xi_\mu$ around $\xi_{\nu}^{1}$ for $N^{\mu\nu}$ times, then around $\xi_{\nu}^{2}$ for $N^{\mu\nu}$ times, and so on. The result is a total phase of $N_\nu\Theta_{\mu\nu}$. This sequence of braiding processes can be described by a product of operators
\begin{equation}
e^{iN_\nu\Theta_{\mu\nu}}  I=\left(B_{\xi_\mu \xi_{\nu}^{N_\nu}}\right)^{N^{\mu\nu}}\cdots \left(B_{\xi_\mu\xi_\nu^1}\right)^{N^{\mu\nu}} \label{eqa2-1}
\end{equation}
where $B_{\xi_\mu\xi_\nu^t}$ represents the operator associated with braiding $\xi_\mu$ around $\xi_\nu^{t}$ once, for $t=1,\dots, N_\nu$, and $I$ is the identity operator.

Next, we make use of the commutation relation
\begin{equation}
\left(B_{\xi_\mu\xi_\nu^t}\right)^{-1}\left(B_{\xi_\mu\xi_\nu^s}\right)^{-1}B_{\xi_\mu\xi_\nu^t}B_{\xi_\mu\xi_\nu^s} = e^{i \Theta_{\mu\nu\nu}} I
\end{equation}
which follows from the definition of the invariant $\Theta_{\mu\nu\nu}$. Inserting the commutation relation into Eq.~(\ref{eqa2-1}), we find that
\begin{align}
e^{iN_\nu\Theta_{\mu\nu}} I  = & (B_{\xi_\mu \xi_{\nu}^{N_\nu}}\dots B_{\xi_\mu \xi_{\nu}^{1}})^{N^{\mu\nu}} e^{i\zeta_1}
\label{eqa2-2}
\end{align}
where
\begin{equation}
\zeta_1 = \frac{N_\nu (N_\nu-1)}{2}  \frac{N^{\mu\nu} (N^{\mu\nu}-1)}{2} \Theta_{\mu\nu\nu}
\label{eq_zeta1}
\end{equation}
From the constraint (\ref{c2}), we see that $\zeta_1$ is either 0 or $\pi$.

We are left with the evaluation of the product $B_{\xi_\mu \xi_{\nu}^{N_\nu}}\dots B_{\xi_\mu \xi_{\nu}^{1}}$ in Eq.~(\ref{eqa2-2}).  Physically, this product means braiding $\xi_\mu$ around $\xi_\nu^1, \dots, \xi_\nu^{N_\nu}$ as a whole. We notice that the vortices $\xi_\nu^1, \dots, \xi_\nu^{N_\nu}$ all together fuse to some charge. Then, the product $B_{\xi_\mu \xi_{\nu}^{N_\nu}}\dots B_{\xi_\mu \xi_{\nu}^{1}}$ should be proportional to an Aharonov-Bohm phase factor, when  $\xi_\nu^1, \dots, \xi_\nu^{N_\nu}$ stay in a definite fusion channel. Raising to $N^{\mu\nu}$th power,  any Aharonov-Bohm phase factor is equal to 1, regardless the fusion channel of   $\xi_\nu^1, \dots, \xi_\nu^{N_\nu}$. Hence, $(B_{\xi_\mu \xi_{\nu}^{N_\nu}}\dots B_{\xi_\mu \xi_{\nu}^{1}})^{N^{\mu\nu}}=I$ holds in general.

Therefore, we finally have the equation
\begin{equation}
N_\nu\Theta_{\mu\nu} =\zeta_1 \label{eqa2_3}
\end{equation}
where $\zeta_1$ is given in Eq.~(\ref{eq_zeta1}). Similarly, one can show that
\begin{equation}
N_\mu\Theta_{\mu\nu} =\zeta_2 \label{eqa2_4}
\end{equation}
where
\begin{equation}
\zeta_2 = \frac{N_\mu (N_\mu-1)}{2}  \frac{N^{\mu\nu} (N^{\mu\nu}-1)}{2} \Theta_{\nu\mu\mu} \label{eq_zeta2}
\end{equation}
Finally, combining the constraint (\ref{c3}) with Eqs.~(\ref{eq_zeta1})-(\ref{eq_zeta2}) and after some straightforward algebras,  the constraint (\ref{c5}) can be obtained.

\subsection{Proof of Eq.~(\ref{c6})}

\begin{figure*}
\centering
\includegraphics{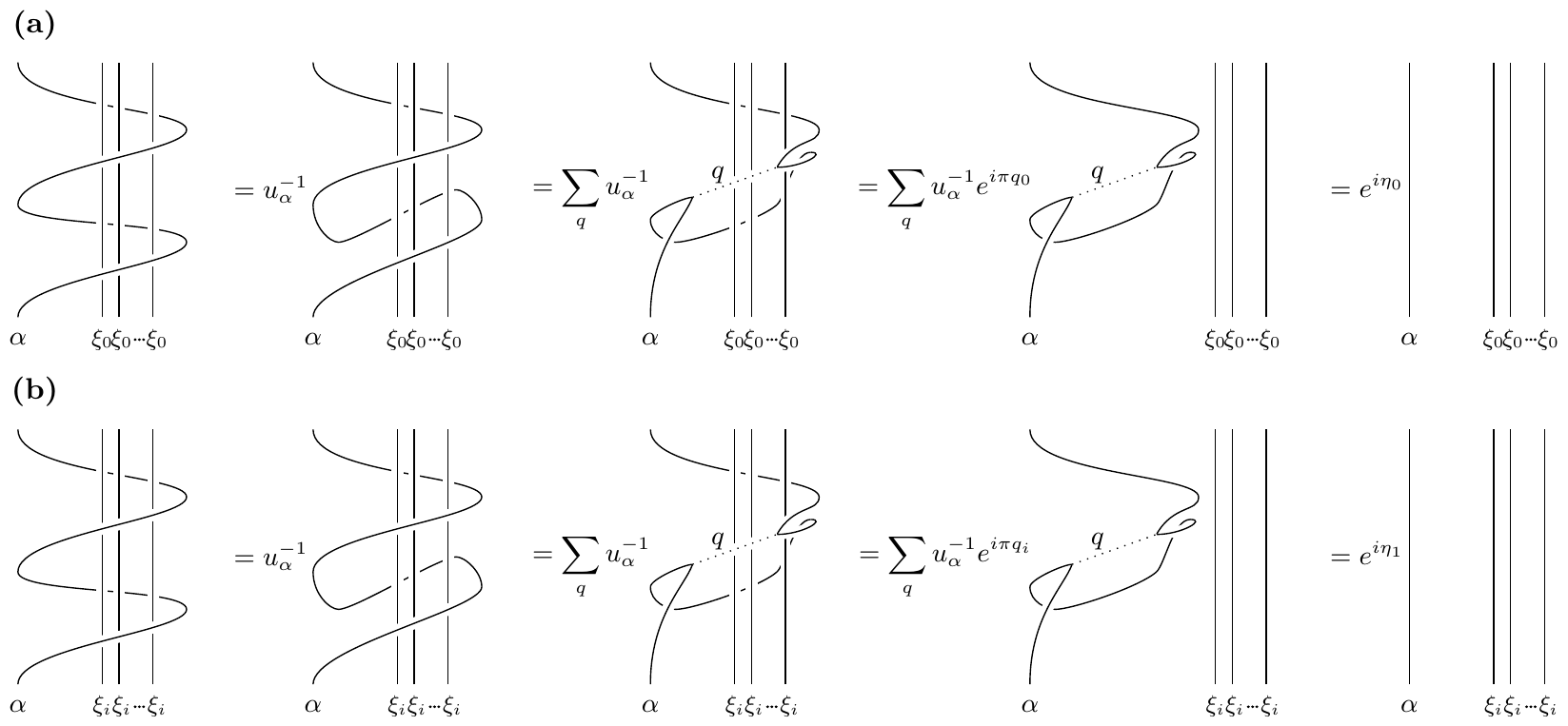}
\caption{Diagrammatic calculations of Berry phases $\eta_0$ (a) and $\eta_1$ (b). } \label{fig_diagram2}
\end{figure*}

We now prove the constraint (\ref{c6}). This is a constraint only for even $N_i$, with $i=1,2,\dots, K$.

To show (\ref{c6}), we first define two Berry phases, $\eta_0$ and $\eta_1$. To define $\eta_0$, we consider a vortex $\alpha$ that carries gauge flux $(0,\dots, \pi, \dots,0)$, with the $i$th entry being $\pi$ and all others being 0, and consider $m$ identical vortices $\xi_0$, each carrying type-$0$ unit flux. Imagine braiding $\alpha$ {\it twice} around the $m$ vortices $\xi_0$'s as a whole. We will show that this braiding process leads to an Abelian Berry phase $\eta_0$. At the same time, we imagine braiding $\alpha$ {\it twice} around $N_i/2$ identical vortices $\xi_i$ as a whole, each carrying type-$i$ unit flux. We will show that this braiding process leads to another Abelian Berry phase $\eta_1$. Below, we show that
\begin{equation}
\eta_0 = \eta_1 \label{eqa3-1}
\end{equation}
At the same time, we show that
\begin{equation}
\eta_1= \frac{N_i}{2}\Theta_{ii} \label{eqa3-2}
\end{equation}
and
\begin{equation}
\eta_0= \left[m\frac{k(k-1)}{2}+k\frac{m(m-1)}{2}\right]\Theta_{00i} + \frac{N_{0i}}{2} \Theta_{0i} \label{eqa3-3}
\end{equation}
where we have set $k=N_i/2$ for abbreviation. Combining the three equations, we prove the constraint (\ref{c6}). Below we prove Eqs.~(\ref{eqa3-1})-(\ref{eqa3-3}) one by one.

To show Eq.~(\ref{eqa3-1}) as well as that the braiding processes associated with $\eta_0$ and $\eta_1$ indeed lead to Abelian Berry phases, we perform diagrammatic calculations for $\eta_0$ and $\eta_1$, shown in Fig.~\ref{fig_diagram2}. Consider Fig.~\ref{fig_diagram2}a, the first diagram shows the space-time trajectories of $\alpha$ and $m$ $\xi_0$'s in the braiding process associated with  $\eta_0$. By using Eq.~(\ref{eq_app_a1}), we obtain the second diagram, and by using Eq.~(\ref{eq_app_a2}), we further obtain the third diagram. Charges $q$ in the third diagram are those appearing in the fusion rule
\begin{equation}
\alpha\times\alpha = q + \dots \label{eqa3-4}
\end{equation}
All fusion channels on the right-hand side are charges, because the total flux carried by two $\alpha$'s is 0. In the fourth diagram, we decouple the ``world lines'' of $\xi_0$'s from that of $\alpha$, at the expense of introducing an Aharonov-Bohm phase $e^{i\pi q_0}$. Note that the braiding of $q$ around each $\xi_0$ gives a factor $e^{i\pi q_0/m}$ and there are $m$ copies of $\xi_0$, hence the total phase factor is $e^{i\pi q_0}$. In the fourth diagram, we see that all $\xi_0$'s are decoupled from $\alpha$, hence the Berry phase in this braiding process does not depend on the fusion channels between $\alpha$ and $\xi_0$'s. Accordingly,   the braiding process gives an Abelian phase. This leads to the last equation in Fig.~\ref{fig_diagram2}a.

Similarly, the diagrammatic calculation for $\eta_1$ is shown in Fig.~\ref{fig_diagram2}b. Everything is the same, except that the Aharonov-Bohm phase in the fourth diagram is $e^{i\pi q_i}$, which is the consequence of braiding $q$ around $N_i/2$ copies of $\xi_i$'s.

Comparing the last equations in Fig.~\ref{fig_diagram2}a and Fig.~\ref{fig_diagram2}b, we see that Eq.~(\ref{eqa3-1}) can be established if we can show that
\begin{equation}
e^{i\pi q_0} = e^{i\pi q_i} \label{eqa3-5}
\end{equation}
where $q$ is a charge appearing in the fusion rule (\ref{eqa3-4}). Below we show that Eq.~(\ref{eqa3-5}) indeed holds. To do that, we note that $N_{\alpha\alpha}^q = N_{\alpha \bar q}^{\bar\alpha}$. Therefore, we have the fusion rule $\alpha \times \bar q = \bar \alpha$. According to Eq.~(\ref{spin_formula}), we have
\begin{equation}
e^{-i\pi q_i} = R_{\alpha \bar q}^{\bar \alpha}  R_{ \bar q \alpha}^{\bar \alpha} = e^{i(\theta_{\bar\alpha} - \theta_{\alpha} - \theta_q)}
\end{equation}
Using the facts that $\theta_{\bar\alpha} = \theta_{\alpha}$ and $\theta_q = \pi q_0$, we immediately obtain Eq.~(\ref{eqa3-5}). Hence, we prove Eq.~(\ref{eqa3-1}).

Next, we prove Eq.~(\ref{eqa3-2}). Let us express the braiding process associated with $\eta_1$ in terms of operators:
\begin{equation}
e^{i\eta_1}I = (B_{\alpha\xi_i^k}\dots B_{\alpha\xi_i^1})^2 \label{eqa3-6}
\end{equation}
where we have used $B_{\alpha\xi_i^s}$ to denote the operator that represents braiding $\alpha$ around the $s$th $\xi_i$ once, for $s=1,\dots, k$ and $k=N_i/2$. Since $\alpha$ carries $k$ times of the type-$i$ unit flux, it is not hard to show that
\begin{equation}
(B_{\alpha\xi_i^s})^{-1}(B_{\alpha\xi_i^t})^{-1}B_{\alpha\xi_i^s}B_{\alpha\xi_i^t} = e^{ik\Theta_{iii}}I\label{eqa3-7}
\end{equation}
In addition, braiding $\alpha$ around any $\xi_i$ twice should leads to an Abelian phase $\Lambda$. This can be shown using a similar diagrammatic calculation as in Fig.~\ref{fig_diagram2}. In terms of operators, we have
\begin{equation}
(B_{\alpha\xi_i^s})^2  = e^{i\Lambda} I \label{eqa3-8}
\end{equation}
Combining Eqs.~(\ref{eqa3-6})-(\ref{eqa3-8}), we obtain
\begin{equation}
\eta_1 = \frac{k(k-1)}{2} k \Theta_{iii} + k\Lambda \label{eqa3-9}
\end{equation}
To further evaluate the phase $k\Lambda$, we understand that it can be thought of as a phase associated with braiding $\alpha$ around a single $\xi_i$ for $N_i$ times. With this, we split $\alpha$ into $k$ vortices, $\xi_i^1, \xi_i^2, \dots, \xi_i^{k}$, each carrying a unit flux.  After the splitting, we can write down the braiding process in terms of operators, like Eq.~(\ref{eqa3-6}), and use similar commutation relations as Eq.~(\ref{eqa3-7}) to make simplification. Eventually, we find that
\begin{equation}
k\Lambda = \frac{k(k-1)}{2} \frac{N_i(N_i-1)}{2}\Theta_{iii} + k \Theta_{ii}  \label{eqa3-10}
\end{equation}
Combining Eqs.~(\ref{eqa3-9}) and (\ref{eqa3-10}) and using the fact that $\Theta_{iii} =0$ or $\pi$,  Eq.~(\ref{eqa3-2}) can be established.

Finally, we prove Eq.~(\ref{eqa3-3}). The proof is similar to Eq.~(\ref{eqa3-2}), so we only briefly sketch it. Following a similar argument as for Eq.~(\ref{eqa3-9}), we find,
\begin{equation}
\eta_0 = \frac{m(m-1)}{2} k\Theta_{00i} + m\Lambda'
\end{equation}
where $\Lambda'$ is the phase associated with braiding $\alpha$ around $\xi_0$ twice. The phase $m\Lambda'$ can be understood as braiding $\alpha$ around a single $\xi_0$ for $N_0$ times. Following a similar argument as for Eq.~(\ref{eqa3-10}), we find that
\begin{equation}
m\Lambda' = \frac{k(k-1)}{2}\frac{N_0(N_0-1)}{2} \Theta_{0ii} + k \Omega
\end{equation}
where $\Omega$ is the phase associated with braiding $\xi_0$ with any $\xi_i$ for $N_0$ times. The phase $\Omega$ depends only on the flux of $\xi_i$, but on the particular choice of $\xi_0$. The phase $k\Omega$ can be understood as the Berry associated with braiding $\xi_0$ around $\xi_i$ for $kN_0$ times. With the definition of $\Theta_{0i}$, one can show that
\begin{equation}
k\Omega = \frac{N_{0i}}{2}\Theta_{0i}
\end{equation}
Combining all equations with the fact $\Theta_{00i}=\Theta_{0ii}=0$ or $\pi$, we prove Eq.~(\ref{eqa3-3}).

\subsection{Proofs of Eqs.~(\ref{c7}) and (\ref{c8})}

We finally prove Eqs.~(\ref{c7}) and (\ref{c8}).

To begin, we prove the following relation
\begin{equation}
\Theta_{\mu\mu} + \Theta_{\mu\bar\mu} = \frac{N_\mu(N_\mu-1)}{2} \Theta_{\mu\mu\mu} \label{eqa4-1}
\end{equation}
where $\Theta_{\mu\bar\mu}$ is defined as braiding $\xi_\mu$ around its antivortex $\bar\xi_\mu$ for $N_\mu$ times. Here, $\xi_\mu$ is again a vortex carrying the type-$\mu$ unit flux. Similar to $\Theta_{\mu\mu}$, the braiding process associated with $\Theta_{\mu\bar\mu}$ indeed leads to an Abelian phase. To show Eq.~(\ref{eqa4-1}), we consider a process that we first braid $\xi_\mu$ around $\bar\xi_\mu$ for $N_\mu$ times and then around $\xi_\mu'$ for $N_\mu$ times, where $\xi_\mu'$ is another vortex that carries type-$\mu$ unit flux. This braiding process gives a Berry phase $\Theta_{\mu\mu} + \Theta_{\mu\bar\mu}$. In terms of braiding operators, we can write the process as
\begin{equation}
e^{i(\Theta_{\mu\mu} + \Theta_{\mu\bar\mu})} I = (B_{\xi_\mu\xi_\mu'} )^{N_\mu} (B_{\xi_\mu\bar\xi_\mu} )^{N_\mu} \label{eqa4-2}
\end{equation}
where $B_{\xi_\mu\bar\xi_\mu}$ and $B_{\xi_\mu\xi_\mu'} $ are operators describing braiding $\xi_\mu$ around $\bar\xi_\mu$ and $\xi_\mu'$ once, respectively. The two operators satisfy the  commutation relation
\begin{equation}
B_{\xi_\mu\bar\xi_\mu}B_{\xi_\mu\xi_\mu'} = e^{-i\Theta_{\mu\mu\mu}} B_{\xi_\mu\xi_\mu'}B_{\xi_\mu\bar\xi_\mu} \label{eqa4-3}
\end{equation}
which follows from the definition of $\Theta_{\mu\mu\mu}$ and the minus sign is due to the fact that $\bar\xi_\mu$ has an opposite flux as $\xi_\mu$. However, the minus sign is irrelevant because $\Theta_{\mu\mu\mu} = -\Theta_{\mu\mu\mu}$ according to Eq.~(\ref{c1}). Inserting Eq.~(\ref{eqa4-3}) into Eq.~(\ref{eqa4-2}), we have that
\begin{equation}
e^{i(\Theta_{\mu\mu} + \Theta_{\mu\bar\mu})} I =  e^{ i[N_\mu(N_\mu-1)/2]\Theta_{\mu\mu\mu}}(B_{\xi_\mu\xi_\mu'} B_{\xi_\mu\bar\xi_\mu} )^{N_\mu} \label{eqa4-4}
\end{equation}
What remains is to evaluate the product $B_{\xi_\mu\xi_\mu'} B_{\xi_\mu\bar\xi_\mu} $. Physically, this product means braiding $\xi_\mu$ around $\bar\xi_{\mu}$ and $\xi_\mu'$ as a whole. Since $\bar\xi_\mu$ and $\xi_\mu'$ fuse to charges only, $B_{\xi_\mu\xi_\mu'} B_{\xi_\mu\bar\xi_\mu} $ should be an Aharonov-Bohm phase when $\bar\xi_\mu$ and $\xi_\mu'$ are in a definite fusion channel. Once raised to $N_\mu$th power, any Aharonov-Bohm phase factor is equal to 1, independent of the fusion channel of $\bar\xi_\mu$ and $\xi_\mu'$. Hence, we obtain $(B_{\xi_\mu\xi_\mu'} B_{\xi_\mu\bar\xi_\mu} )^{N_\mu}  = I$. Accordingly, the relation (\ref{eqa4-1}) holds.

Next, we find constraints between $\Theta_{\mu\bar\mu}$ and $\Theta_{\mu}$. Consider the fusion rule Eq.~(\ref{eqa1-1}) between $\xi_\mu$ and $\bar\xi_\mu$. According to Eq.~(\ref{spin_formula}), we have
\begin{equation}
R_{\bar\xi_\mu \xi_\mu}^qR_{\xi_\mu \bar\xi_\mu}^q = e^{i(\theta_q -\theta_{\xi_\mu} - \theta_{\bar\xi_\mu})}
\end{equation}
Then, we have
\begin{equation}
e^{i\Theta_{\mu\bar\mu}} = \left(R_{\bar\xi_\mu \xi_\mu}^qR_{\xi_\mu \bar\xi_\mu}^q \right)^{N_\mu} =e^{iN_\mu\theta_q - i 2N_\mu\theta_{\xi_\mu}}
\end{equation}
where we have used the fact that $\theta_{\xi_\mu} = \theta_{\bar\xi_\mu}$. Interestingly, for any $q$ appearing in the fusion product $\xi_\mu\times\bar\xi_\mu$, we have that
\begin{equation}
N_\mu \theta_q = \pi q_0 N_\mu =0
\end{equation}
This follows from the relation (\ref{eqa1-4}), which implies that $\pi q_0N_\mu = 2\pi q_\mu =0 \ ({\rm mod} \ 2\pi )$.

Therefore, we obtain $\Theta_{\mu\bar\mu} = - 2N_\mu\theta_{\xi_\mu}$. According to the definition of $\Theta_\mu$, we  have
\begin{equation}
\Theta_{i\bar i} = \left\{
\begin{array}{ll}
-2\Theta_i, & \text{if $N_i$ is even} \\[3pt]
-\Theta_i, & \text{if $N_i$ is odd}
\end{array}
\right.\label{eqa4-10}
\end{equation}
and
\begin{equation}
\Theta_{0\bar 0} = \left\{
\begin{array}{ll}
-2\Theta_0, & \text{if $m$ is even} \\[3pt]
-4\Theta_0, & \text{if $m$ is odd}
\end{array}
\right.\label{eqa4-11}
\end{equation}
Combining Eqs.~(\ref{eqa4-1}), (\ref{eqa4-10}) and (\ref{eqa4-11}), we are led to the constraints (\ref{c7}) and (\ref{c8}).

\bibliography{fspt}

\begin{thebibliography}{58}%
\makeatletter
\providecommand \@ifxundefined [1]{%
 \@ifx{#1\undefined}
}%
\providecommand \@ifnum [1]{%
 \ifnum #1\expandafter \@firstoftwo
 \else \expandafter \@secondoftwo
 \fi
}%
\providecommand \@ifx [1]{%
 \ifx #1\expandafter \@firstoftwo
 \else \expandafter \@secondoftwo
 \fi
}%
\providecommand \natexlab [1]{#1}%
\providecommand \enquote  [1]{``#1''}%
\providecommand \bibnamefont  [1]{#1}%
\providecommand \bibfnamefont [1]{#1}%
\providecommand \citenamefont [1]{#1}%
\providecommand \href@noop [0]{\@secondoftwo}%
\providecommand \href [0]{\begingroup \@sanitize@url \@href}%
\providecommand \@href[1]{\@@startlink{#1}\@@href}%
\providecommand \@@href[1]{\endgroup#1\@@endlink}%
\providecommand \@sanitize@url [0]{\catcode `\\12\catcode `\$12\catcode
  `\&12\catcode `\#12\catcode `\^12\catcode `\_12\catcode `\%12\relax}%
\providecommand \@@startlink[1]{}%
\providecommand \@@endlink[0]{}%
\providecommand \url  [0]{\begingroup\@sanitize@url \@url }%
\providecommand \@url [1]{\endgroup\@href {#1}{\urlprefix }}%
\providecommand \urlprefix  [0]{URL }%
\providecommand \Eprint [0]{\href }%
\providecommand \doibase [0]{http://dx.doi.org/}%
\providecommand \selectlanguage [0]{\@gobble}%
\providecommand \bibinfo  [0]{\@secondoftwo}%
\providecommand \bibfield  [0]{\@secondoftwo}%
\providecommand \translation [1]{[#1]}%
\providecommand \BibitemOpen [0]{}%
\providecommand \bibitemStop [0]{}%
\providecommand \bibitemNoStop [0]{.\EOS\space}%
\providecommand \EOS [0]{\spacefactor3000\relax}%
\providecommand \BibitemShut  [1]{\csname bibitem#1\endcsname}%
\let\auto@bib@innerbib\@empty
\bibitem [{\citenamefont {Gu}\ and\ \citenamefont {Wen}(2009)}]{gu09}%
  \BibitemOpen
  \bibfield  {author} {\bibinfo {author} {\bibfnamefont {Z.-C.}\ \bibnamefont
  {Gu}}\ and\ \bibinfo {author} {\bibfnamefont {X.-G.}\ \bibnamefont {Wen}},\
  }\bibfield  {title} {\enquote {\bibinfo {title}
  {Tensor-entanglement-filtering renormalization approach and
  symmetry-protected topological order},}\ }\href {\doibase
  10.1103/PhysRevB.80.155131} {\bibfield  {journal} {\bibinfo  {journal} {Phys.
  Rev. B}\ }\textbf {\bibinfo {volume} {80}},\ \bibinfo {pages} {155131}
  (\bibinfo {year} {2009})}\BibitemShut {NoStop}%
\bibitem [{\citenamefont {Pollmann}\ \emph {et~al.}(2010)\citenamefont
  {Pollmann}, \citenamefont {Turner}, \citenamefont {Berg},\ and\ \citenamefont
  {Oshikawa}}]{pollmann10}%
  \BibitemOpen
  \bibfield  {author} {\bibinfo {author} {\bibfnamefont {F.}~\bibnamefont
  {Pollmann}}, \bibinfo {author} {\bibfnamefont {A.~M.}\ \bibnamefont
  {Turner}}, \bibinfo {author} {\bibfnamefont {E.}~\bibnamefont {Berg}}, \ and\
  \bibinfo {author} {\bibfnamefont {M.}~\bibnamefont {Oshikawa}},\ }\bibfield
  {title} {\enquote {\bibinfo {title} {Entanglement spectrum of a topological
  phase in one dimension},}\ }\href {\doibase 10.1103/PhysRevB.81.064439}
  {\bibfield  {journal} {\bibinfo  {journal} {Phys. Rev. B}\ }\textbf {\bibinfo
  {volume} {81}},\ \bibinfo {pages} {064439} (\bibinfo {year}
  {2010})}\BibitemShut {NoStop}%
\bibitem [{\citenamefont {Fidkowski}\ and\ \citenamefont
  {Kitaev}(2011)}]{fidkowski11}%
  \BibitemOpen
  \bibfield  {author} {\bibinfo {author} {\bibfnamefont {L.}~\bibnamefont
  {Fidkowski}}\ and\ \bibinfo {author} {\bibfnamefont {A.}~\bibnamefont
  {Kitaev}},\ }\bibfield  {title} {\enquote {\bibinfo {title} {Topological
  phases of fermions in one dimension},}\ }\href {\doibase
  10.1103/PhysRevB.83.075103} {\bibfield  {journal} {\bibinfo  {journal} {Phys.
  Rev. B}\ }\textbf {\bibinfo {volume} {83}},\ \bibinfo {pages} {075103}
  (\bibinfo {year} {2011})}\BibitemShut {NoStop}%
\bibitem [{\citenamefont {Chen}\ \emph
  {et~al.}(2011{\natexlab{a}})\citenamefont {Chen}, \citenamefont {Gu},\ and\
  \citenamefont {Wen}}]{chen11a}%
  \BibitemOpen
  \bibfield  {author} {\bibinfo {author} {\bibfnamefont {X.}~\bibnamefont
  {Chen}}, \bibinfo {author} {\bibfnamefont {Z.-C.}\ \bibnamefont {Gu}}, \ and\
  \bibinfo {author} {\bibfnamefont {X.-G.}\ \bibnamefont {Wen}},\ }\bibfield
  {title} {\enquote {\bibinfo {title} {Complete classification of
  one-dimensional gapped quantum phases in interacting spin systems},}\ }\href
  {\doibase 10.1103/PhysRevB.84.235128} {\bibfield  {journal} {\bibinfo
  {journal} {Phys. Rev. B}\ }\textbf {\bibinfo {volume} {84}},\ \bibinfo
  {pages} {235128} (\bibinfo {year} {2011}{\natexlab{a}})}\BibitemShut
  {NoStop}%
\bibitem [{\citenamefont {Chen}\ \emph
  {et~al.}(2011{\natexlab{b}})\citenamefont {Chen}, \citenamefont {Gu},\ and\
  \citenamefont {Wen}}]{chen11b}%
  \BibitemOpen
  \bibfield  {author} {\bibinfo {author} {\bibfnamefont {X.}~\bibnamefont
  {Chen}}, \bibinfo {author} {\bibfnamefont {Z.-C.}\ \bibnamefont {Gu}}, \ and\
  \bibinfo {author} {\bibfnamefont {X.-G.}\ \bibnamefont {Wen}},\ }\bibfield
  {title} {\enquote {\bibinfo {title} {Classification of gapped symmetric
  phases in one-dimensional spin systems},}\ }\href {\doibase
  10.1103/PhysRevB.83.035107} {\bibfield  {journal} {\bibinfo  {journal} {Phys.
  Rev. B}\ }\textbf {\bibinfo {volume} {83}},\ \bibinfo {pages} {035107}
  (\bibinfo {year} {2011}{\natexlab{b}})}\BibitemShut {NoStop}%
\bibitem [{\citenamefont {Schuch}\ \emph {et~al.}(2011)\citenamefont {Schuch},
  \citenamefont {P\'erez-Garc\'{\i}a},\ and\ \citenamefont {Cirac}}]{schuch11}%
  \BibitemOpen
  \bibfield  {author} {\bibinfo {author} {\bibfnamefont {N.}~\bibnamefont
  {Schuch}}, \bibinfo {author} {\bibfnamefont {D.}~\bibnamefont
  {P\'erez-Garc\'{\i}a}}, \ and\ \bibinfo {author} {\bibfnamefont
  {I.}~\bibnamefont {Cirac}},\ }\bibfield  {title} {\enquote {\bibinfo {title}
  {Classifying quantum phases using matrix product states and projected
  entangled pair states},}\ }\href {\doibase 10.1103/PhysRevB.84.165139}
  {\bibfield  {journal} {\bibinfo  {journal} {Phys. Rev. B}\ }\textbf {\bibinfo
  {volume} {84}},\ \bibinfo {pages} {165139} (\bibinfo {year}
  {2011})}\BibitemShut {NoStop}%
\bibitem [{\citenamefont {Chen}\ \emph {et~al.}(2013)\citenamefont {Chen},
  \citenamefont {Gu}, \citenamefont {Liu},\ and\ \citenamefont {Wen}}]{chen13}%
  \BibitemOpen
  \bibfield  {author} {\bibinfo {author} {\bibfnamefont {X.}~\bibnamefont
  {Chen}}, \bibinfo {author} {\bibfnamefont {Z.-C.}\ \bibnamefont {Gu}},
  \bibinfo {author} {\bibfnamefont {Z.-X.}\ \bibnamefont {Liu}}, \ and\
  \bibinfo {author} {\bibfnamefont {X.-G.}\ \bibnamefont {Wen}},\ }\bibfield
  {title} {\enquote {\bibinfo {title} {Symmetry protected topological orders
  and the group cohomology of their symmetry group},}\ }\href {\doibase
  10.1103/PhysRevB.87.155114} {\bibfield  {journal} {\bibinfo  {journal} {Phys.
  Rev. B}\ }\textbf {\bibinfo {volume} {87}},\ \bibinfo {pages} {155114}
  (\bibinfo {year} {2013})}\BibitemShut {NoStop}%
\bibitem [{\citenamefont {Hasan}\ and\ \citenamefont {Kane}(2010)}]{hasan10}%
  \BibitemOpen
  \bibfield  {author} {\bibinfo {author} {\bibfnamefont {M.~Z.}\ \bibnamefont
  {Hasan}}\ and\ \bibinfo {author} {\bibfnamefont {C.~L.}\ \bibnamefont
  {Kane}},\ }\bibfield  {title} {\enquote {\bibinfo {title}
  {\textit{Colloquium} : Topological insulators},}\ }\href {\doibase
  10.1103/RevModPhys.82.3045} {\bibfield  {journal} {\bibinfo  {journal} {Rev.
  Mod. Phys.}\ }\textbf {\bibinfo {volume} {82}},\ \bibinfo {pages}
  {3045--3067} (\bibinfo {year} {2010})}\BibitemShut {NoStop}%
\bibitem [{\citenamefont {Qi}\ and\ \citenamefont {Zhang}(2011)}]{qi11}%
  \BibitemOpen
  \bibfield  {author} {\bibinfo {author} {\bibfnamefont {X.-L.}\ \bibnamefont
  {Qi}}\ and\ \bibinfo {author} {\bibfnamefont {S.-C.}\ \bibnamefont {Zhang}},\
  }\bibfield  {title} {\enquote {\bibinfo {title} {Topological insulators and
  superconductors},}\ }\href {\doibase 10.1103/RevModPhys.83.1057} {\bibfield
  {journal} {\bibinfo  {journal} {Rev. Mod. Phys.}\ }\textbf {\bibinfo {volume}
  {83}},\ \bibinfo {pages} {1057--1110} (\bibinfo {year} {2011})}\BibitemShut
  {NoStop}%
\bibitem [{\citenamefont {Schnyder}\ \emph {et~al.}(2008)\citenamefont
  {Schnyder}, \citenamefont {Ryu}, \citenamefont {Furusaki},\ and\
  \citenamefont {Ludwig}}]{ff1}%
  \BibitemOpen
  \bibfield  {author} {\bibinfo {author} {\bibfnamefont {A.~P.}\ \bibnamefont
  {Schnyder}}, \bibinfo {author} {\bibfnamefont {S.}~\bibnamefont {Ryu}},
  \bibinfo {author} {\bibfnamefont {A.}~\bibnamefont {Furusaki}}, \ and\
  \bibinfo {author} {\bibfnamefont {A.~W.~W.}\ \bibnamefont {Ludwig}},\
  }\bibfield  {title} {\enquote {\bibinfo {title} {Classification of
  topological insulators and superconductors in three spatial dimensions},}\
  }\href {\doibase 10.1103/PhysRevB.78.195125} {\bibfield  {journal} {\bibinfo
  {journal} {Phys. Rev. B}\ }\textbf {\bibinfo {volume} {78}},\ \bibinfo
  {pages} {195125} (\bibinfo {year} {2008})}\BibitemShut {NoStop}%
\bibitem [{\citenamefont {Kitaev}(2009)}]{ff2}%
  \BibitemOpen
  \bibfield  {author} {\bibinfo {author} {\bibfnamefont {A.}~\bibnamefont
  {Kitaev}},\ }\bibfield  {title} {\enquote {\bibinfo {title} {Periodic table
  for topological insulators and superconductors},}\ }\href {\doibase
  http://dx.doi.org/10.1063/1.3149495} {\bibfield  {journal} {\bibinfo
  {journal} {AIP Conference Proceedings}\ }\textbf {\bibinfo {volume} {1134}},\
  \bibinfo {pages} {22} (\bibinfo {year} {2009})}\BibitemShut {NoStop}%
\bibitem [{\citenamefont {{Kapustin}}(2014)}]{kapustin14a}%
  \BibitemOpen
  \bibfield  {author} {\bibinfo {author} {\bibfnamefont {A.}~\bibnamefont
  {{Kapustin}}},\ }\bibfield  {title} {\enquote {\bibinfo {title} {{Symmetry
  Protected Topological Phases, Anomalies, and Cobordisms: Beyond Group
  Cohomology}},}\ }\href@noop {} {\bibfield  {journal} {\bibinfo  {journal}
  {ArXiv e-prints}\ } (\bibinfo {year} {2014})},\ \Eprint
  {http://arxiv.org/abs/1403.1467} {arXiv:1403.1467} \BibitemShut {NoStop}%
\bibitem [{\citenamefont {{Freed}}(2014)}]{freed14}%
  \BibitemOpen
  \bibfield  {author} {\bibinfo {author} {\bibfnamefont {D.~S.}\ \bibnamefont
  {{Freed}}},\ }\bibfield  {title} {\enquote {\bibinfo {title} {{Short-range
  entanglement and invertible field theories}},}\ }\href@noop {} {\bibfield
  {journal} {\bibinfo  {journal} {arXiv e-prints}\ } (\bibinfo {year}
  {2014})},\ \Eprint {http://arxiv.org/abs/1406.7278} {arXiv:1406.7278}
  \BibitemShut {NoStop}%
\bibitem [{\citenamefont {Kitaev}({\natexlab{a}})}]{kitaevsre}%
  \BibitemOpen
  \bibfield  {author} {\bibinfo {author} {\bibfnamefont {A.}~\bibnamefont
  {Kitaev}},\ }\href
  {http://www.ipam.ucla.edu/abstract/?tid=12389&pcode=STQ2015} {}
  ({\natexlab{a}}),\ \bibinfo {note}
  {http://www.ipam.ucla.edu/abstract/?tid=123 89\&pcode=STQ2015}\BibitemShut
  {NoStop}%
\bibitem [{\citenamefont {Vishwanath}\ and\ \citenamefont
  {Senthil}(2013)}]{vishwanath13}%
  \BibitemOpen
  \bibfield  {author} {\bibinfo {author} {\bibfnamefont {A.}~\bibnamefont
  {Vishwanath}}\ and\ \bibinfo {author} {\bibfnamefont {T.}~\bibnamefont
  {Senthil}},\ }\bibfield  {title} {\enquote {\bibinfo {title} {Physics of
  three-dimensional bosonic topological insulators: Surface-deconfined
  criticality and quantized magnetoelectric effect},}\ }\href {\doibase
  10.1103/PhysRevX.3.011016} {\bibfield  {journal} {\bibinfo  {journal} {Phys.
  Rev. X}\ }\textbf {\bibinfo {volume} {3}},\ \bibinfo {pages} {011016}
  (\bibinfo {year} {2013})}\BibitemShut {NoStop}%
\bibitem [{\citenamefont {Fidkowski}\ and\ \citenamefont
  {Kitaev}(2010)}]{fidkowski10}%
  \BibitemOpen
  \bibfield  {author} {\bibinfo {author} {\bibfnamefont {Lukasz}\ \bibnamefont
  {Fidkowski}}\ and\ \bibinfo {author} {\bibfnamefont {Alexei}\ \bibnamefont
  {Kitaev}},\ }\bibfield  {title} {\enquote {\bibinfo {title} {Effects of
  interactions on the topological classification of free fermion systems},}\
  }\href {\doibase 10.1103/PhysRevB.81.134509} {\bibfield  {journal} {\bibinfo
  {journal} {Phys. Rev. B}\ }\textbf {\bibinfo {volume} {81}},\ \bibinfo
  {pages} {134509} (\bibinfo {year} {2010})}\BibitemShut {NoStop}%
\bibitem [{\citenamefont {Gu}\ and\ \citenamefont {Levin}(2014)}]{gu14b}%
  \BibitemOpen
  \bibfield  {author} {\bibinfo {author} {\bibfnamefont {Z.-C.}\ \bibnamefont
  {Gu}}\ and\ \bibinfo {author} {\bibfnamefont {M.}~\bibnamefont {Levin}},\
  }\bibfield  {title} {\enquote {\bibinfo {title} {Effect of interactions on
  two-dimensional fermionic symmetry-protected topological phases with
  ${Z}_{2}$ symmetry},}\ }\href {\doibase 10.1103/PhysRevB.89.201113}
  {\bibfield  {journal} {\bibinfo  {journal} {Phys. Rev. B}\ }\textbf {\bibinfo
  {volume} {89}},\ \bibinfo {pages} {201113} (\bibinfo {year}
  {2014})}\BibitemShut {NoStop}%
\bibitem [{\citenamefont {Ryu}\ and\ \citenamefont {Zhang}(2012)}]{ryu12}%
  \BibitemOpen
  \bibfield  {author} {\bibinfo {author} {\bibfnamefont {S.}~\bibnamefont
  {Ryu}}\ and\ \bibinfo {author} {\bibfnamefont {S.-C.}\ \bibnamefont
  {Zhang}},\ }\bibfield  {title} {\enquote {\bibinfo {title} {Interacting
  topological phases and modular invariance},}\ }\href {\doibase
  10.1103/PhysRevB.85.245132} {\bibfield  {journal} {\bibinfo  {journal} {Phys.
  Rev. B}\ }\textbf {\bibinfo {volume} {85}},\ \bibinfo {pages} {245132}
  (\bibinfo {year} {2012})}\BibitemShut {NoStop}%
\bibitem [{\citenamefont {Yao}\ and\ \citenamefont {Ryu}(2013)}]{yao13}%
  \BibitemOpen
  \bibfield  {author} {\bibinfo {author} {\bibfnamefont {H.}~\bibnamefont
  {Yao}}\ and\ \bibinfo {author} {\bibfnamefont {S.}~\bibnamefont {Ryu}},\
  }\bibfield  {title} {\enquote {\bibinfo {title} {Interaction effect on
  topological classification of superconductors in two dimensions},}\ }\href
  {\doibase 10.1103/PhysRevB.88.064507} {\bibfield  {journal} {\bibinfo
  {journal} {Phys. Rev. B}\ }\textbf {\bibinfo {volume} {88}},\ \bibinfo
  {pages} {064507} (\bibinfo {year} {2013})}\BibitemShut {NoStop}%
\bibitem [{\citenamefont {Qi}(2013)}]{qi13}%
  \BibitemOpen
  \bibfield  {author} {\bibinfo {author} {\bibfnamefont {X.-L.}\ \bibnamefont
  {Qi}},\ }\bibfield  {title} {\enquote {\bibinfo {title} {A new class of
  (2 + 1)-dimensional topological superconductors with {$\mathbb {Z}_8$}
  topological classification},}\ }\href
  {http://stacks.iop.org/1367-2630/15/i=6/a=065002} {\bibfield  {journal}
  {\bibinfo  {journal} {New Journal of Physics}\ }\textbf {\bibinfo {volume}
  {15}},\ \bibinfo {pages} {065002} (\bibinfo {year} {2013})}\BibitemShut
  {NoStop}%
\bibitem [{\citenamefont {{Wang}}\ \emph {et~al.}(2014)\citenamefont {{Wang}},
  \citenamefont {{Potter}},\ and\ \citenamefont {{Senthil}}}]{wangc-science}%
  \BibitemOpen
  \bibfield  {author} {\bibinfo {author} {\bibfnamefont {C.}~\bibnamefont
  {{Wang}}}, \bibinfo {author} {\bibfnamefont {A.~C.}\ \bibnamefont
  {{Potter}}}, \ and\ \bibinfo {author} {\bibfnamefont {T.}~\bibnamefont
  {{Senthil}}},\ }\bibfield  {title} {\enquote {\bibinfo {title}
  {{Classification of Interacting Electronic Topological Insulators in Three
  Dimensions}},}\ }\href {\doibase 10.1126/science.1243326} {\bibfield
  {journal} {\bibinfo  {journal} {Science}\ }\textbf {\bibinfo {volume}
  {343}},\ \bibinfo {pages} {629--631} (\bibinfo {year} {2014})},\ \Eprint
  {http://arxiv.org/abs/1306.3238} {arXiv:1306.3238} \BibitemShut {NoStop}%
\bibitem [{\citenamefont {Fidkowski}\ \emph {et~al.}(2013)\citenamefont
  {Fidkowski}, \citenamefont {Chen},\ and\ \citenamefont
  {Vishwanath}}]{fidkowski13}%
  \BibitemOpen
  \bibfield  {author} {\bibinfo {author} {\bibfnamefont {L.}~\bibnamefont
  {Fidkowski}}, \bibinfo {author} {\bibfnamefont {X.}~\bibnamefont {Chen}}, \
  and\ \bibinfo {author} {\bibfnamefont {A.}~\bibnamefont {Vishwanath}},\
  }\bibfield  {title} {\enquote {\bibinfo {title} {Non-abelian topological
  order on the surface of a 3d topological superconductor from an exactly
  solved model},}\ }\href {\doibase 10.1103/PhysRevX.3.041016} {\bibfield
  {journal} {\bibinfo  {journal} {Phys. Rev. X}\ }\textbf {\bibinfo {volume}
  {3}},\ \bibinfo {pages} {041016} (\bibinfo {year} {2013})}\BibitemShut
  {NoStop}%
\bibitem [{\citenamefont {{Metlitski}}\ \emph {et~al.}(2014)\citenamefont
  {{Metlitski}}, \citenamefont {{Fidkowski}}, \citenamefont {{Chen}},\ and\
  \citenamefont {{Vishwanath}}}]{metlitski14}%
  \BibitemOpen
  \bibfield  {author} {\bibinfo {author} {\bibfnamefont {M.~A.}\ \bibnamefont
  {{Metlitski}}}, \bibinfo {author} {\bibfnamefont {L.}~\bibnamefont
  {{Fidkowski}}}, \bibinfo {author} {\bibfnamefont {X.}~\bibnamefont {{Chen}}},
  \ and\ \bibinfo {author} {\bibfnamefont {A.}~\bibnamefont {{Vishwanath}}},\
  }\bibfield  {title} {\enquote {\bibinfo {title} {{Interaction effects on 3D
  topological superconductors: surface topological order from vortex
  condensation, the 16 fold way and fermionic Kramers doublets}},}\ }\href@noop
  {} {\bibfield  {journal} {\bibinfo  {journal} {ArXiv e-prints}\ } (\bibinfo
  {year} {2014})},\ \Eprint {http://arxiv.org/abs/1406.3032} {arXiv:1406.3032}
  \BibitemShut {NoStop}%
\bibitem [{\citenamefont {Wang}\ and\ \citenamefont {Senthil}(2014)}]{wangc14}%
  \BibitemOpen
  \bibfield  {author} {\bibinfo {author} {\bibfnamefont {C.}~\bibnamefont
  {Wang}}\ and\ \bibinfo {author} {\bibfnamefont {T.}~\bibnamefont {Senthil}},\
  }\bibfield  {title} {\enquote {\bibinfo {title} {Interacting fermionic
  topological insulators/superconductors in three dimensions},}\ }\href
  {\doibase 10.1103/PhysRevB.89.195124} {\bibfield  {journal} {\bibinfo
  {journal} {Phys. Rev. B}\ }\textbf {\bibinfo {volume} {89}},\ \bibinfo
  {pages} {195124} (\bibinfo {year} {2014})}\BibitemShut {NoStop}%
\bibitem [{\citenamefont {{You}}\ \emph {et~al.}(2014)\citenamefont {{You}},
  \citenamefont {{BenTov}},\ and\ \citenamefont {{Xu}}}]{you14}%
  \BibitemOpen
  \bibfield  {author} {\bibinfo {author} {\bibfnamefont {Y.-Z.}\ \bibnamefont
  {{You}}}, \bibinfo {author} {\bibfnamefont {Y.}~\bibnamefont {{BenTov}}}, \
  and\ \bibinfo {author} {\bibfnamefont {C.}~\bibnamefont {{Xu}}},\ }\bibfield
  {title} {\enquote {\bibinfo {title} {{Interacting Topological Superconductors
  and possible Origin of \$16n\$ Chiral Fermions in the Standard Model}},}\
  }\href@noop {} {\bibfield  {journal} {\bibinfo  {journal} {ArXiv e-prints}\ }
  (\bibinfo {year} {2014})},\ \Eprint {http://arxiv.org/abs/1402.4151}
  {arXiv:1402.4151} \BibitemShut {NoStop}%
\bibitem [{\citenamefont {Morimoto}\ \emph {et~al.}(2015)\citenamefont
  {Morimoto}, \citenamefont {Furusaki},\ and\ \citenamefont
  {Mudry}}]{morimoto15}%
  \BibitemOpen
  \bibfield  {author} {\bibinfo {author} {\bibfnamefont {T.}~\bibnamefont
  {Morimoto}}, \bibinfo {author} {\bibfnamefont {A.}~\bibnamefont {Furusaki}},
  \ and\ \bibinfo {author} {\bibfnamefont {C.}~\bibnamefont {Mudry}},\
  }\bibfield  {title} {\enquote {\bibinfo {title} {Breakdown of the topological
  classification $\mathbb{Z}$ for gapped phases of noninteracting fermions by
  quartic interactions},}\ }\href {\doibase 10.1103/PhysRevB.92.125104}
  {\bibfield  {journal} {\bibinfo  {journal} {Phys. Rev. B}\ }\textbf {\bibinfo
  {volume} {92}},\ \bibinfo {pages} {125104} (\bibinfo {year}
  {2015})}\BibitemShut {NoStop}%
\bibitem [{\citenamefont {{Queiroz}}\ \emph {et~al.}(2016)\citenamefont
  {{Queiroz}}, \citenamefont {{Khalaf}},\ and\ \citenamefont
  {{Stern}}}]{queiroz16}%
  \BibitemOpen
  \bibfield  {author} {\bibinfo {author} {\bibfnamefont {R.}~\bibnamefont
  {{Queiroz}}}, \bibinfo {author} {\bibfnamefont {E.}~\bibnamefont {{Khalaf}}},
  \ and\ \bibinfo {author} {\bibfnamefont {A.}~\bibnamefont {{Stern}}},\
  }\bibfield  {title} {\enquote {\bibinfo {title} {{Classification of
  interacting fermionic phases by dimensional reduction}},}\ }\href@noop {}
  {\bibfield  {journal} {\bibinfo  {journal} {ArXiv e-prints}\ } (\bibinfo
  {year} {2016})},\ \Eprint {http://arxiv.org/abs/1601.01596}
  {arXiv:1601.01596} \BibitemShut {NoStop}%
\bibitem [{\citenamefont {{Witten}}(2015)}]{witten15}%
  \BibitemOpen
  \bibfield  {author} {\bibinfo {author} {\bibfnamefont {E.}~\bibnamefont
  {{Witten}}},\ }\bibfield  {title} {\enquote {\bibinfo {title} {{Fermion Path
  Integrals And Topological Phases}},}\ }\href@noop {} {\bibfield  {journal}
  {\bibinfo  {journal} {ArXiv e-prints}\ } (\bibinfo {year} {2015})},\ \Eprint
  {http://arxiv.org/abs/1508.04715} {arXiv:1508.04715} \BibitemShut {NoStop}%
\bibitem [{\citenamefont {Gu}\ and\ \citenamefont {Wen}(2014)}]{gu-super}%
  \BibitemOpen
  \bibfield  {author} {\bibinfo {author} {\bibfnamefont {Z.-C.}\ \bibnamefont
  {Gu}}\ and\ \bibinfo {author} {\bibfnamefont {X.-G.}\ \bibnamefont {Wen}},\
  }\bibfield  {title} {\enquote {\bibinfo {title} {Symmetry-protected
  topological orders for interacting fermions: Fermionic topological nonlinear
  {$\ensuremath{\sigma}$} models and a special group supercohomology theory},}\
  }\href {\doibase 10.1103/PhysRevB.90.115141} {\bibfield  {journal} {\bibinfo
  {journal} {Phys. Rev. B}\ }\textbf {\bibinfo {volume} {90}},\ \bibinfo
  {pages} {115141} (\bibinfo {year} {2014})}\BibitemShut {NoStop}%
\bibitem [{\citenamefont {{Kapustin}}\ \emph {et~al.}(2014)\citenamefont
  {{Kapustin}}, \citenamefont {{Thorngren}}, \citenamefont {{Turzillo}},\ and\
  \citenamefont {{Wang}}}]{kapustin14}%
  \BibitemOpen
  \bibfield  {author} {\bibinfo {author} {\bibfnamefont {A.}~\bibnamefont
  {{Kapustin}}}, \bibinfo {author} {\bibfnamefont {R.}~\bibnamefont
  {{Thorngren}}}, \bibinfo {author} {\bibfnamefont {A.}~\bibnamefont
  {{Turzillo}}}, \ and\ \bibinfo {author} {\bibfnamefont {Z.}~\bibnamefont
  {{Wang}}},\ }\bibfield  {title} {\enquote {\bibinfo {title} {{Fermionic
  Symmetry Protected Topological Phases and Cobordisms}},}\ }\href@noop {}
  {\bibfield  {journal} {\bibinfo  {journal} {arXiv e-prints}\ } (\bibinfo
  {year} {2014})},\ \Eprint {http://arxiv.org/abs/1406.7329} {arXiv:1406.7329}
  \BibitemShut {NoStop}%
\bibitem [{\citenamefont {{Cheng}}\ \emph {et~al.}(2015)\citenamefont
  {{Cheng}}, \citenamefont {{Bi}}, \citenamefont {{You}},\ and\ \citenamefont
  {{Gu}}}]{cheng15}%
  \BibitemOpen
  \bibfield  {author} {\bibinfo {author} {\bibfnamefont {M.}~\bibnamefont
  {{Cheng}}}, \bibinfo {author} {\bibfnamefont {Z.}~\bibnamefont {{Bi}}},
  \bibinfo {author} {\bibfnamefont {Y.-Z.}\ \bibnamefont {{You}}}, \ and\
  \bibinfo {author} {\bibfnamefont {Z.-C.}\ \bibnamefont {{Gu}}},\ }\bibfield
  {title} {\enquote {\bibinfo {title} {{Towards a Complete Classification of
  Symmetry-Protected Phases for Interacting Fermions in Two Dimensions}},}\
  }\href@noop {} {\bibfield  {journal} {\bibinfo  {journal} {arXiv e-prints}\ }
  (\bibinfo {year} {2015})},\ \Eprint {http://arxiv.org/abs/1501.01313}
  {arXiv:1501.01313} \BibitemShut {NoStop}%
\bibitem [{\citenamefont {Lu}\ and\ \citenamefont {Vishwanath}(2012)}]{lu12}%
  \BibitemOpen
  \bibfield  {author} {\bibinfo {author} {\bibfnamefont {Y.-M.}\ \bibnamefont
  {Lu}}\ and\ \bibinfo {author} {\bibfnamefont {A.}~\bibnamefont
  {Vishwanath}},\ }\bibfield  {title} {\enquote {\bibinfo {title} {Theory and
  classification of interacting integer topological phases in two dimensions: A
  chern-simons approach},}\ }\href {\doibase 10.1103/PhysRevB.86.125119}
  {\bibfield  {journal} {\bibinfo  {journal} {Phys. Rev. B}\ }\textbf {\bibinfo
  {volume} {86}},\ \bibinfo {pages} {125119} (\bibinfo {year}
  {2012})}\BibitemShut {NoStop}%
\bibitem [{\citenamefont {{Lan}}\ \emph {et~al.}(2015)\citenamefont {{Lan}},
  \citenamefont {{Kong}},\ and\ \citenamefont {{Wen}}}]{lan15}%
  \BibitemOpen
  \bibfield  {author} {\bibinfo {author} {\bibfnamefont {T.}~\bibnamefont
  {{Lan}}}, \bibinfo {author} {\bibfnamefont {L.}~\bibnamefont {{Kong}}}, \
  and\ \bibinfo {author} {\bibfnamefont {X.-G.}\ \bibnamefont {{Wen}}},\
  }\bibfield  {title} {\enquote {\bibinfo {title} {{A theory of 2+1D fermionic
  topological orders and fermionic/bosonic topological orders with
  symmetries}},}\ }\href@noop {} {\bibfield  {journal} {\bibinfo  {journal}
  {arXiv e-prints}\ } (\bibinfo {year} {2015})},\ \Eprint
  {http://arxiv.org/abs/1507.04673} {arXiv:1507.04673} \BibitemShut {NoStop}%
\bibitem [{\citenamefont {{Lan}}\ \emph {et~al.}(2016)\citenamefont {{Lan}},
  \citenamefont {{Kong}},\ and\ \citenamefont {{Wen}}}]{lan16}%
  \BibitemOpen
  \bibfield  {author} {\bibinfo {author} {\bibfnamefont {T.}~\bibnamefont
  {{Lan}}}, \bibinfo {author} {\bibfnamefont {L.}~\bibnamefont {{Kong}}}, \
  and\ \bibinfo {author} {\bibfnamefont {X.-G.}\ \bibnamefont {{Wen}}},\
  }\bibfield  {title} {\enquote {\bibinfo {title} {{Classification of 2+1D
  topological orders and SPT orders for bosonic and fermionic systems with
  on-site symmetries}},}\ }\href@noop {} {\bibfield  {journal} {\bibinfo
  {journal} {ArXiv e-prints}\ } (\bibinfo {year} {2016})},\ \Eprint
  {http://arxiv.org/abs/1602.05946} {arXiv:1602.05946} \BibitemShut {NoStop}%
\bibitem [{\citenamefont {Tarantino}\ \emph {et~al.}(2016)\citenamefont
  {Tarantino}, \citenamefont {Lindner},\ and\ \citenamefont
  {Fidkowski}}]{tarantino16}%
  \BibitemOpen
  \bibfield  {author} {\bibinfo {author} {\bibfnamefont {N.}~\bibnamefont
  {Tarantino}}, \bibinfo {author} {\bibfnamefont {N.~H.}\ \bibnamefont
  {Lindner}}, \ and\ \bibinfo {author} {\bibfnamefont {L.}~\bibnamefont
  {Fidkowski}},\ }\bibfield  {title} {\enquote {\bibinfo {title} {Symmetry
  fractionalization and twist defects},}\ }\href
  {http://stacks.iop.org/1367-2630/18/i=3/a=035006} {\bibfield  {journal}
  {\bibinfo  {journal} {New Journal of Physics}\ }\textbf {\bibinfo {volume}
  {18}},\ \bibinfo {pages} {035006} (\bibinfo {year} {2016})}\BibitemShut
  {NoStop}%
\bibitem [{\citenamefont {{Bhardwaj}}\ \emph {et~al.}(2016)\citenamefont
  {{Bhardwaj}}, \citenamefont {{Gaiotto}},\ and\ \citenamefont
  {{Kapustin}}}]{gaiotto16}%
  \BibitemOpen
  \bibfield  {author} {\bibinfo {author} {\bibfnamefont {L.}~\bibnamefont
  {{Bhardwaj}}}, \bibinfo {author} {\bibfnamefont {D.}~\bibnamefont
  {{Gaiotto}}}, \ and\ \bibinfo {author} {\bibfnamefont {A.}~\bibnamefont
  {{Kapustin}}},\ }\bibfield  {title} {\enquote {\bibinfo {title} {{State sum
  constructions of spin-TFTs and string net constructions of fermionic phases
  of matter}},}\ }\href@noop {} {\bibfield  {journal} {\bibinfo  {journal}
  {ArXiv e-prints}\ } (\bibinfo {year} {2016})},\ \Eprint
  {http://arxiv.org/abs/1605.01640} {arXiv:1605.01640} \BibitemShut {NoStop}%
\bibitem [{\citenamefont {Wang}(2016)}]{wangcj16}%
  \BibitemOpen
  \bibfield  {author} {\bibinfo {author} {\bibfnamefont {C.}~\bibnamefont
  {Wang}},\ }\bibfield  {title} {\enquote {\bibinfo {title} {Braiding
  statistics and classification of two-dimensional charge-{$2m$}
  superconductors},}\ }\href {\doibase 10.1103/PhysRevB.94.085130} {\bibfield
  {journal} {\bibinfo  {journal} {Phys. Rev. B}\ }\textbf {\bibinfo {volume}
  {94}},\ \bibinfo {pages} {085130} (\bibinfo {year} {2016})}\BibitemShut
  {NoStop}%
\bibitem [{\citenamefont {Levin}\ and\ \citenamefont {Gu}(2012)}]{levin12}%
  \BibitemOpen
  \bibfield  {author} {\bibinfo {author} {\bibfnamefont {M.}~\bibnamefont
  {Levin}}\ and\ \bibinfo {author} {\bibfnamefont {Z.-C.}\ \bibnamefont {Gu}},\
  }\bibfield  {title} {\enquote {\bibinfo {title} {Braiding statistics approach
  to symmetry-protected topological phases},}\ }\href {\doibase
  10.1103/PhysRevB.86.115109} {\bibfield  {journal} {\bibinfo  {journal} {Phys.
  Rev. B}\ }\textbf {\bibinfo {volume} {86}},\ \bibinfo {pages} {115109}
  (\bibinfo {year} {2012})}\BibitemShut {NoStop}%
\bibitem [{\citenamefont {Wang}\ and\ \citenamefont {Levin}(2015)}]{wangcj15}%
  \BibitemOpen
  \bibfield  {author} {\bibinfo {author} {\bibfnamefont {C.}~\bibnamefont
  {Wang}}\ and\ \bibinfo {author} {\bibfnamefont {M.}~\bibnamefont {Levin}},\
  }\bibfield  {title} {\enquote {\bibinfo {title} {Topological invariants for
  gauge theories and symmetry-protected topological phases},}\ }\href {\doibase
  10.1103/PhysRevB.91.165119} {\bibfield  {journal} {\bibinfo  {journal} {Phys.
  Rev. B}\ }\textbf {\bibinfo {volume} {91}},\ \bibinfo {pages} {165119}
  (\bibinfo {year} {2015})}\BibitemShut {NoStop}%
\bibitem [{\citenamefont {Wang}\ and\ \citenamefont {Levin}(2014)}]{threeloop}%
  \BibitemOpen
  \bibfield  {author} {\bibinfo {author} {\bibfnamefont {C.}~\bibnamefont
  {Wang}}\ and\ \bibinfo {author} {\bibfnamefont {M.}~\bibnamefont {Levin}},\
  }\bibfield  {title} {\enquote {\bibinfo {title} {Braiding statistics of loop
  excitations in three dimensions},}\ }\href {\doibase
  10.1103/PhysRevLett.113.080403} {\bibfield  {journal} {\bibinfo  {journal}
  {Phys. Rev. Lett.}\ }\textbf {\bibinfo {volume} {113}},\ \bibinfo {pages}
  {080403} (\bibinfo {year} {2014})}\BibitemShut {NoStop}%
\bibitem [{\citenamefont {Kitaev}(2006)}]{kitaev06}%
  \BibitemOpen
  \bibfield  {author} {\bibinfo {author} {\bibfnamefont {A.}~\bibnamefont
  {Kitaev}},\ }\bibfield  {title} {\enquote {\bibinfo {title} {Anyons in an
  exactly solved model and beyond},}\ }\href {\doibase
  http://dx.doi.org/10.1016/j.aop.2005.10.005} {\bibfield  {journal} {\bibinfo
  {journal} {Annals of Physics}\ }\textbf {\bibinfo {volume} {321}},\ \bibinfo
  {pages} {2} (\bibinfo {year} {2006})}\BibitemShut {NoStop}%
\bibitem [{\citenamefont {Kitaev}({\natexlab{b}})}]{e8}%
  \BibitemOpen
  \bibfield  {author} {\bibinfo {author} {\bibfnamefont {A.}~\bibnamefont
  {Kitaev}},\ }\href@noop {} {} ({\natexlab{b}}),\ \bibinfo {note}
  {http://online.kitp.ucsb.edu/online/topom at11/kitaev/}\BibitemShut {NoStop}%
\bibitem [{\citenamefont {Ware}\ \emph {et~al.}(2016)\citenamefont {Ware},
  \citenamefont {Son}, \citenamefont {Cheng}, \citenamefont {Mishmash},
  \citenamefont {Alicea},\ and\ \citenamefont {Bauer}}]{ware16}%
  \BibitemOpen
  \bibfield  {author} {\bibinfo {author} {\bibfnamefont {B.}~\bibnamefont
  {Ware}}, \bibinfo {author} {\bibfnamefont {J.~H.}\ \bibnamefont {Son}},
  \bibinfo {author} {\bibfnamefont {M.}~\bibnamefont {Cheng}}, \bibinfo
  {author} {\bibfnamefont {R.~V.}\ \bibnamefont {Mishmash}}, \bibinfo {author}
  {\bibfnamefont {J.}~\bibnamefont {Alicea}}, \ and\ \bibinfo {author}
  {\bibfnamefont {B.}~\bibnamefont {Bauer}},\ }\bibfield  {title} {\enquote
  {\bibinfo {title} {Ising anyons in frustration-free majorana-dimer models},}\
  }\href {\doibase 10.1103/PhysRevB.94.115127} {\bibfield  {journal} {\bibinfo
  {journal} {Phys. Rev. B}\ }\textbf {\bibinfo {volume} {94}},\ \bibinfo
  {pages} {115127} (\bibinfo {year} {2016})}\BibitemShut {NoStop}%
\bibitem [{\citenamefont {Read}\ and\ \citenamefont {Green}(2000)}]{read00}%
  \BibitemOpen
  \bibfield  {author} {\bibinfo {author} {\bibfnamefont {N.}~\bibnamefont
  {Read}}\ and\ \bibinfo {author} {\bibfnamefont {D.}~\bibnamefont {Green}},\
  }\bibfield  {title} {\enquote {\bibinfo {title} {Paired states of fermions in
  two dimensions with breaking of parity and time-reversal symmetries and the
  fractional quantum hall effect},}\ }\href {\doibase
  10.1103/PhysRevB.61.10267} {\bibfield  {journal} {\bibinfo  {journal} {Phys.
  Rev. B}\ }\textbf {\bibinfo {volume} {61}},\ \bibinfo {pages} {10267--10297}
  (\bibinfo {year} {2000})}\BibitemShut {NoStop}%
\bibitem [{\citenamefont {Ivanov}(2001)}]{ivanov01}%
  \BibitemOpen
  \bibfield  {author} {\bibinfo {author} {\bibfnamefont {D.~A.}\ \bibnamefont
  {Ivanov}},\ }\bibfield  {title} {\enquote {\bibinfo {title} {Non-abelian
  statistics of half-quantum vortices in $\mathit{p}$-wave superconductors},}\
  }\href {\doibase 10.1103/PhysRevLett.86.268} {\bibfield  {journal} {\bibinfo
  {journal} {Phys. Rev. Lett.}\ }\textbf {\bibinfo {volume} {86}},\ \bibinfo
  {pages} {268--271} (\bibinfo {year} {2001})}\BibitemShut {NoStop}%
\bibitem [{\citenamefont {Wen}(2004)}]{wen-book}%
  \BibitemOpen
  \bibfield  {author} {\bibinfo {author} {\bibfnamefont {X.-G.}\ \bibnamefont
  {Wen}},\ }\href@noop {} {\emph {\bibinfo {title} {Quantum Field Theory of
  Many-Body Systems}}}\ (\bibinfo  {publisher} {Oxford University Press},\
  \bibinfo {address} {Oxford},\ \bibinfo {year} {2004})\BibitemShut {NoStop}%
\bibitem [{\citenamefont {Bais}\ and\ \citenamefont
  {Slingerland}(2009)}]{bais09}%
  \BibitemOpen
  \bibfield  {author} {\bibinfo {author} {\bibfnamefont {F.~A.}\ \bibnamefont
  {Bais}}\ and\ \bibinfo {author} {\bibfnamefont {J.~K.}\ \bibnamefont
  {Slingerland}},\ }\bibfield  {title} {\enquote {\bibinfo {title}
  {Condensate-induced transitions between topologically ordered phases},}\
  }\href {\doibase 10.1103/PhysRevB.79.045316} {\bibfield  {journal} {\bibinfo
  {journal} {Phys. Rev. B}\ }\textbf {\bibinfo {volume} {79}},\ \bibinfo
  {pages} {045316} (\bibinfo {year} {2009})}\BibitemShut {NoStop}%
\bibitem [{\citenamefont {{de Wild Propitius}}(1995)}]{propitius95}%
  \BibitemOpen
  \bibfield  {author} {\bibinfo {author} {\bibfnamefont {M.}~\bibnamefont {{de
  Wild Propitius}}},\ }\bibfield  {title} {\enquote {\bibinfo {title}
  {{Topological interactions in broken gauge theories}},}\ }\href@noop {}
  {\bibfield  {journal} {\bibinfo  {journal} {Arxiv e-prints}\ } (\bibinfo
  {year} {1995})},\ \Eprint {http://arxiv.org/abs/arXiv:hep-th/9511195}
  {arXiv:hep-th/9511195} \BibitemShut {NoStop}%
\bibitem [{\citenamefont {Essin}\ and\ \citenamefont
  {Hermele}(2013)}]{essin13}%
  \BibitemOpen
  \bibfield  {author} {\bibinfo {author} {\bibfnamefont {A.~M.}\ \bibnamefont
  {Essin}}\ and\ \bibinfo {author} {\bibfnamefont {M.}~\bibnamefont
  {Hermele}},\ }\bibfield  {title} {\enquote {\bibinfo {title} {Classifying
  fractionalization: Symmetry classification of gapped ${\mathbb{z}}_{2}$ spin
  liquids in two dimensions},}\ }\href {\doibase 10.1103/PhysRevB.87.104406}
  {\bibfield  {journal} {\bibinfo  {journal} {Phys. Rev. B}\ }\textbf {\bibinfo
  {volume} {87}},\ \bibinfo {pages} {104406} (\bibinfo {year}
  {2013})}\BibitemShut {NoStop}%
\bibitem [{\citenamefont {Mesaros}\ and\ \citenamefont
  {Ran}(2013)}]{mesaros13}%
  \BibitemOpen
  \bibfield  {author} {\bibinfo {author} {\bibfnamefont {A.}~\bibnamefont
  {Mesaros}}\ and\ \bibinfo {author} {\bibfnamefont {Y.}~\bibnamefont {Ran}},\
  }\bibfield  {title} {\enquote {\bibinfo {title} {Classification of symmetry
  enriched topological phases with exactly solvable models},}\ }\href {\doibase
  10.1103/PhysRevB.87.155115} {\bibfield  {journal} {\bibinfo  {journal} {Phys.
  Rev. B}\ }\textbf {\bibinfo {volume} {87}},\ \bibinfo {pages} {155115}
  (\bibinfo {year} {2013})}\BibitemShut {NoStop}%
\bibitem [{\citenamefont {{Barkeshli}}\ \emph {et~al.}(2014)\citenamefont
  {{Barkeshli}}, \citenamefont {{Bonderson}}, \citenamefont {{Cheng}},\ and\
  \citenamefont {{Wang}}}]{barkeshli14}%
  \BibitemOpen
  \bibfield  {author} {\bibinfo {author} {\bibfnamefont {M.}~\bibnamefont
  {{Barkeshli}}}, \bibinfo {author} {\bibfnamefont {P.}~\bibnamefont
  {{Bonderson}}}, \bibinfo {author} {\bibfnamefont {M.}~\bibnamefont
  {{Cheng}}}, \ and\ \bibinfo {author} {\bibfnamefont {Z.}~\bibnamefont
  {{Wang}}},\ }\bibfield  {title} {\enquote {\bibinfo {title} {{Symmetry,
  Defects, and Gauging of Topological Phases}},}\ }\href@noop {} {\bibfield
  {journal} {\bibinfo  {journal} {ArXiv e-prints}\ } (\bibinfo {year}
  {2014})},\ \Eprint {http://arxiv.org/abs/1410.4540} {arXiv:1410.4540}
  \BibitemShut {NoStop}%
\bibitem [{\citenamefont {Chen}\ \emph {et~al.}(2015)\citenamefont {Chen},
  \citenamefont {Burnell}, \citenamefont {Vishwanath},\ and\ \citenamefont
  {Fidkowski}}]{chen14}%
  \BibitemOpen
  \bibfield  {author} {\bibinfo {author} {\bibfnamefont {X.}~\bibnamefont
  {Chen}}, \bibinfo {author} {\bibfnamefont {F.~J.}\ \bibnamefont {Burnell}},
  \bibinfo {author} {\bibfnamefont {A.}~\bibnamefont {Vishwanath}}, \ and\
  \bibinfo {author} {\bibfnamefont {L.}~\bibnamefont {Fidkowski}},\ }\bibfield
  {title} {\enquote {\bibinfo {title} {Anomalous symmetry fractionalization and
  surface topological order},}\ }\href {\doibase 10.1103/PhysRevX.5.041013}
  {\bibfield  {journal} {\bibinfo  {journal} {Phys. Rev. X}\ }\textbf {\bibinfo
  {volume} {5}},\ \bibinfo {pages} {041013} (\bibinfo {year}
  {2015})}\BibitemShut {NoStop}%
\bibitem [{\citenamefont {Wang}\ \emph {et~al.}(2016)\citenamefont {Wang},
  \citenamefont {Lin},\ and\ \citenamefont {Levin}}]{bbc}%
  \BibitemOpen
  \bibfield  {author} {\bibinfo {author} {\bibfnamefont {C.}~\bibnamefont
  {Wang}}, \bibinfo {author} {\bibfnamefont {C.-H.}\ \bibnamefont {Lin}}, \
  and\ \bibinfo {author} {\bibfnamefont {M.}~\bibnamefont {Levin}},\ }\bibfield
   {title} {\enquote {\bibinfo {title} {Bulk-boundary correspondence for
  three-dimensional symmetry-protected topological phases},}\ }\href {\doibase
  10.1103/PhysRevX.6.021015} {\bibfield  {journal} {\bibinfo  {journal} {Phys.
  Rev. X}\ }\textbf {\bibinfo {volume} {6}},\ \bibinfo {pages} {021015}
  (\bibinfo {year} {2016})}\BibitemShut {NoStop}%
\bibitem [{\citenamefont {{Heinrich}}\ \emph {et~al.}(2016)\citenamefont
  {{Heinrich}}, \citenamefont {{Burnell}}, \citenamefont {{Fidkowski}},\ and\
  \citenamefont {{Levin}}}]{heinrich16}%
  \BibitemOpen
  \bibfield  {author} {\bibinfo {author} {\bibfnamefont {C.}~\bibnamefont
  {{Heinrich}}}, \bibinfo {author} {\bibfnamefont {F.}~\bibnamefont
  {{Burnell}}}, \bibinfo {author} {\bibfnamefont {L.}~\bibnamefont
  {{Fidkowski}}}, \ and\ \bibinfo {author} {\bibfnamefont {M.}~\bibnamefont
  {{Levin}}},\ }\bibfield  {title} {\enquote {\bibinfo {title} {{Symmetry
  enriched string-nets: Exactly solvable models for SET phases}},}\ }\href@noop
  {} {\bibfield  {journal} {\bibinfo  {journal} {ArXiv e-prints}\ } (\bibinfo
  {year} {2016})},\ \Eprint {http://arxiv.org/abs/1606.07816}
  {arXiv:1606.07816} \BibitemShut {NoStop}%
\bibitem [{\citenamefont {{Cheng}}\ \emph {et~al.}(2016)\citenamefont
  {{Cheng}}, \citenamefont {{Gu}}, \citenamefont {{Jiang}},\ and\ \citenamefont
  {{Qi}}}]{cheng16}%
  \BibitemOpen
  \bibfield  {author} {\bibinfo {author} {\bibfnamefont {M.}~\bibnamefont
  {{Cheng}}}, \bibinfo {author} {\bibfnamefont {Z.-C.}\ \bibnamefont {{Gu}}},
  \bibinfo {author} {\bibfnamefont {S.}~\bibnamefont {{Jiang}}}, \ and\
  \bibinfo {author} {\bibfnamefont {Y.}~\bibnamefont {{Qi}}},\ }\bibfield
  {title} {\enquote {\bibinfo {title} {{Exactly Solvable Models for
  Symmetry-Enriched Topological Phases}},}\ }\href@noop {} {\bibfield
  {journal} {\bibinfo  {journal} {ArXiv e-prints}\ } (\bibinfo {year}
  {2016})},\ \Eprint {http://arxiv.org/abs/1606.08482} {arXiv:1606.08482}
  \BibitemShut {NoStop}%
\bibitem [{\citenamefont {{Hung}}\ and\ \citenamefont {{Wan}}(2014)}]{hung14}%
  \BibitemOpen
  \bibfield  {author} {\bibinfo {author} {\bibfnamefont {L.-Y.}\ \bibnamefont
  {{Hung}}}\ and\ \bibinfo {author} {\bibfnamefont {Y.}~\bibnamefont {{Wan}}},\
  }\bibfield  {title} {\enquote {\bibinfo {title} {{Symmetry-enriched phases
  obtained via pseudo anyon condensation}},}\ }\href {\doibase
  10.1142/S0217979214501720} {\bibfield  {journal} {\bibinfo  {journal}
  {International Journal of Modern Physics B}\ }\textbf {\bibinfo {volume}
  {28}},\ \bibinfo {eid} {1450172} (\bibinfo {year} {2014})},\ \Eprint
  {http://arxiv.org/abs/1308.4673} {arXiv:1308.4673} \BibitemShut {NoStop}%
\bibitem [{\citenamefont {Jiang}\ \emph {et~al.}(2014)\citenamefont {Jiang},
  \citenamefont {Mesaros},\ and\ \citenamefont {Ran}}]{ran14}%
  \BibitemOpen
  \bibfield  {author} {\bibinfo {author} {\bibfnamefont {S.}~\bibnamefont
  {Jiang}}, \bibinfo {author} {\bibfnamefont {A.}~\bibnamefont {Mesaros}}, \
  and\ \bibinfo {author} {\bibfnamefont {Y.}~\bibnamefont {Ran}},\ }\bibfield
  {title} {\enquote {\bibinfo {title} {Generalized modular transformations in
  $(3+1)\mathrm{D}$ topologically ordered phases and triple linking invariant
  of loop braiding},}\ }\href {\doibase 10.1103/PhysRevX.4.031048} {\bibfield
  {journal} {\bibinfo  {journal} {Phys. Rev. X}\ }\textbf {\bibinfo {volume}
  {4}},\ \bibinfo {pages} {031048} (\bibinfo {year} {2014})}\BibitemShut
  {NoStop}%
\bibitem [{\citenamefont {Cheng}\ and\ \citenamefont {Wang}()}]{cheng-wang}%
  \BibitemOpen
  \bibfield  {author} {\bibinfo {author} {\bibfnamefont {M.}~\bibnamefont
  {Cheng}}\ and\ \bibinfo {author} {\bibfnamefont {C.}~\bibnamefont {Wang}},\
  }\href@noop {} {}\bibinfo {note} {In preparation.}\BibitemShut {Stop}%
\end{thebibliography}%

\end{document}